\documentclass[11pt,titlepage]{report}

\usepackage{graphicx}
\usepackage{rkstops}

\begin{document}
\let\footnotesize\normalsize
\addtolength{\baselineskip}{2ex}
\renewcommand{\thepage}{\roman{page}}
\setcounter{page}{1}

\begin{titlepage}
\thispagestyle{empty}
\addtolength{\baselineskip}{-2ex}

\vskip 10mm
\begin{center}
\large{TOPICS IN SUPERSYMMETRY \\
PHENOMENOLOGY  AT \\
THE LARGE HADRON COLLIDER}
\end{center}

\vskip 30mm
\begin{center}
A DISSERTATION SUBMITTED TO THE GRADUATE DIVISION OF THE \\
UNIVERSITY OF HAWAI`I AT M\={A}NOA IN PARTIAL FULFILLMENT OF \\ THE 
REQUIREMENTS FOR THE DEGREE OF \\
\vskip 2.5ex
DOCTOR OF PHILOSOPHY \\
\vskip 2.5ex
IN \\
\vskip 2.5ex
PHYSICS \\
\vskip 5ex
DECEMBER 2011
\vskip 19ex
By \\
\vskip .5ex
{\large Roger HK Kadala}
\vskip 5ex
Dissertation Committee:
\vskip 2.5ex
Xerxes Tata, Chairperson\\
Sandip Pakvasa\\
John Learned\\
Pui Lam\\
Marvin Ortel
\end{center}
\addtolength{\baselineskip}{2ex}
\end{titlepage}

\newpage
\thispagestyle{empty}

\paragraph{Acknowledgements}
Many people have had a positive influence throughout my life, either by word or action, through shared experiences or knowledge, assistance, or advice. To all, my heartfelt thanks.\\
To my friends who had a hand in my developing and completing my research, Javier Ferrandis, Jose Kenichi Mizukoshi, Srikanth Hundi, and Roman Nevzorov, I thank you for your time, your willingness to help, and your friendship. Mahalo. \\
To my advisor and friend, Xerxes Tata, thank you. Always patient, always giving and forgiving, always teaching. You have helped me achieve my childhood dream, and I am forever grateful.\\
To my family, I am truly blessed. Beatriz, my wife, my soulmate, my companion, your love and spirit are my strength, my bright day. My daughter, Dalybeth, and my son, John Roger, your boundless faith in me makes all challenges surmountable. My granddaughters, Isabel, my best buddy, my little guardian angel, you  taught me how to really love, and Infinity and Victoria, you are all a Blessing from God. I love you.\\
\underline{\it Gracias $\aleph$ Dios}, thank you for being there during my years growing up alone in Miami and New York, my countless flights throughout the world. Thank you for being with me unconditionally always, eternally. I could not have done this without you. Mahalo and Aloha!\\

\newpage
\setcounter{page}{2}
\addcontentsline{toc}{chapter}{Abstract}
\begin{abstract}
This dissertation focuses on phenomenological studies for possible signals for supersymmetric events at the Large Hadron Collider (LHC).  We have divided our endeavours into three separate projects. First, considering that the branching fraction for the decays of gluinos to third generation squarks is expected to be enhanced in classes of supersymmetric models where either third generation fermions are lighter than other squarks, or models of mixed higgsino dark matter which are constructed in agreement with the measured density of cold dark matter(\textbf{CDM}), the gluino production in such scenarios at the LHC should be rich in top and bottom quark jets. Requiring $b$-jets  in addition to missing energy $\eslt$ should, therefore, enhance the supersymmetry signal relative to Standard Model backgrounds. We quantify the increase in the supersymmetry reach of the LHC from $b$-tagging in a variety of well-motivated models of supersymmetry. We also explore $top$-tagging at the LHC.
Second, we explore the prospects for detecting the direct production of third generation squarks in models with an inverted squark mass hierarchy. This is signalled by $b$-jets + $\eslt$ events harder than in the Standard Model, but softer than those from the production of gluinos and heavier squarks.  We find that these events can be readily separated from SM background (for third generation squark masses in the $200-400$ GeV range), and the contamination from the much heavier gluinos and squarks although formidable can effectively be suppressed.
Third, we attempt to extract model-independent information about neutralino properties from LHC data. assuming only the particle content of the MSSM and that all two-body neutralino decays are kinematically suppressed, with the neutralino inclusive production yielding a sufficient cross section. We show that the Lorentz invariant dilepton mass distribution encodes clear information about the relative sign of the mass eigenvalues of the parent and daughter neutralinos. We show that we can extract most neutralino mass matrix parameters if there is a double mass edge.
\end{abstract}
\addtocounter{page}{2}
\def\contentsname{Table of Contents}
\addtolength{\baselineskip}{1ex}
\tableofcontents
\addtolength{\baselineskip}{-1ex}
\clearpage\addcontentsline{toc}{chapter}{List of Tables}\listoftables
\clearpage\addcontentsline{toc}{chapter}{List of Figures}\listoffigures
\clearpage

\renewcommand{\thepage}{\arabic{page}}
\setcounter{page}1
\widowpenalty 150
\clubpenalty 150
\setlength{\parindent}{3em}

\chapter{Introduction to Supersymmetry}
\label{chap: susy101}
\section{Introduction}
Supersymmetry (\textbf{SUSY}) \cite{wss,manuel,martin,weinberg,mkw} is one of the more peculiar theoretical discoveries in the history of physics, since, despite the enormous effort invested in its study (its discovery dates over 30 years ago), there is no experimental evidence of SUSY. \\
Additionally, in the past decades, the Standard Model (\textbf{SM}) has been verified with great precision by numerous experiments. When discrepancies have been encountered, these vanish with the increased precision in measurements, and the greater the precision that is achieved, the more precisely is the SM confirmed. From the viewpoint of precision measurements, there is little need for new physics beyond the SM\footnote{Exceptions to this are the $g-2$ \cite{gmenos2a},\cite{gmenos2b} experiment and the proton size anomaly~\cite{protanomal}}.\\
The experimental evidence of the need for new physics beyond the SM comes from neutrino physics and observations supporting the existence of Dark Matter in the Universe without viable candidates in the SM. Also gravity interactions are not part of the SM. From the theoretical perspective there are also good reasons for going beyond the SM: the solution of the hierarchy problem, or the desire to find a new unified theory, or a much simpler one, that offers an explanation of the symmetries, the spectrum, or the parameters of the SM. SUSY is one of the best candidates we have to this date of new physics beyond the SM. Besides offering a natural solution to the hierarchy problem, it allows for the unification of the gauge couplings. \\
We must keep in mind that the hierarchy problem was not the primary motive behind the invention of SUSY in the 1970's. This is why it is surprising that although the initial SUSY models were quite different from the current Minimal Supersymmetric Standard Model (\textbf{MSSM}), with time this has become the principal candidate to succeed the SM. With the advent of the new generation of colliders, such as the Large Hadron Collider (\textbf{LHC}) coming online, a more definite test of SUSY is viable. \\
\section{SUSY Theory}
The construction of a SUSY theory would have as its underlying algebraic structure that of a graded Lie Algebra (\textbf{gLA}). These are extensions of the Lie Algebras, in which a distinction is established between elements of odd and even nature. Those of even nature obey commutation rules (Lie Algebra), while those of odd nature obey anti-commutation rules amongst them, and commutation rules with the even ones, i.e. the elements of odd nature constitute a representation of the gLA, so that for $A_m$ and $Q_{\alpha}$ being the even and odd elements of this gLA respectively, we would have\\
\begin{eqnarray}
&[ A_m , A_n ]=&f^l_{mn}\,A_l \nonumber \\
&[ A_m , Q_{\alpha} ]=& S^{\beta}_{ma}\,Q_{\beta}\nonumber\\
&\{ Q_{\alpha} , Q_{\beta} \}=& F^{m}_{\alpha\beta}\,A_m
\end{eqnarray}
where the repeated indices on the right-hand side are summed over. In the context of the extensions of the Poincare group by one self-conjugate spinor charge $Q$ the even generators are the generators of the Poincare group and the odd generator is the generator of SUSY. 
A supersymmetry transformation will turn a bosonic state into a fermionic state, and viceversa, with the generator given by the operator $Q$ as an anticommuting spinor, so that,
\[
fermion \stackrel{Q}{\longleftrightarrow} boson
\]
An extension of the Coleman-Mandula theorem ~\cite{cmthm}, by  Haag-Lopuszanski-Sohnius  \cite{haag}, restricts the possible supersymmetries acceptable in a Quantum Field Theory (\textbf{QFT}) with interactions. Only theories with one spinorial charge $Q_{\alpha}$, known as $N=1$ SUSY, allow for chiral fermions, i.e., fermions whose left-handed and right-handed pieces transform differently under symmetry transformations, theories crucial for phenomenology. For this reason, we restrict our focus to $N=1$ SUSY. We can then write the algebra as,
\begin{eqnarray}
&[ P_{\mu} , Q_{\alpha} ]=&0 \nonumber \\
&[ Q_{\alpha} , M^{\mu\nu} ]=&\frac{1}{2}\,(\sigma^{\mu\nu})^{\beta}_{\alpha}\,Q_{\beta}\nonumber \\
&\{ Q_{\alpha} , \overline{Q_{\beta}}\}=&2\,(\gamma^{\mu})_{\alpha\beta}\,P_{\mu}
\end{eqnarray}
The algebra closes to yield the generators of the Poincare group, $P_{\mu}$~\footnote{the $M^{\mu\nu}$ are the generators of Lorentz transformations}, so these show that supersymmetry is a spacetime symmetry.The irreducible representations of the SUSY Algebra are labeled supermultiplets, each containing both bosonic and fermionic states having the exact same number of degrees of freedom. The supermultiplets of the SUSY Algebra utilized in the construction of the MSSM are: 
\begin{itemize}
\item ( $ \Phi , \Psi$ )  chiral superfield (or scalar , or matter) consisting of one Weyl fermion ($n_f = 2$) and two real scalars ($n_b = 2\times 1$).
\item ( $V^\mu , \lambda$ )  vector superfield (or gauge) consisting of one spin-1 massless boson ($n_b = 2$) and one Weyl fermion ($n_f = 2$).
\end{itemize}
The operator $P^2$ commutes with all generators, so that all particles occurring in a supermultiplet will have the same eigenvalues of $P^2$, and therefore the same mass. The supersymmetry generators also commute with the genrators of gauge transformations, which means that members of a supermultiplet belong to the same representation of the gauge group, thus having the same electric charge, weak isospin, and color degrees of freedom.\\
The superfield formalism ~\cite{salstra} provides a convenient tool for studying supersymmetric theories including studying the multiplet structure of these theories, the unitary supersymmetric representations of the particle states, the construction of supersymmetric invariants, amongst others. Most importantly, it provides a recipe for the construction of a Lagrangian density of a supersymmetric Yang-Mills theory in terms of the ordinary boson and fermion fields of QFT.  Assume that the chiral supermultiplets ($\phi_i, \psi_i$) transform under a gauge group representations and that ($V_{a}^\mu, \lambda_{a}$) are the gauge supermultiplets, with $a$ as the gauge group index. Then the Lagrangian density can be written as,
\be
\lagr = \lagr_K + \lagr_{M\lambda} + \lagr_Y + \lagr_S+\lagr_{MG}+\lagr_{\lambda G}+\lagr_{GG}
\ee
where $\lagr_K$ contains the kinetic term,
\be
\lagr_K = \sum_{j}\,\,|\partial_{\mu}\phi_{j}|^{2} + \frac{i}{2}\,\overline{\psi}_{j}\,\not\partial\,\psi_{j} -\frac{1}{4}\,V^{a}_{\mu\nu}V^{\mu\nu}_{a} +\frac{i}{2}\,\overline{\lambda}^{a}\not\partial\lambda_{a} + h.c.
\ee
$\lagr_{M\lambda}$ gives the interactions of gauginos with the scalars and fermions of the chiral multiplets
\be
\lagr_{M\lambda} = -ig\,\sqrt{2}(\phi^{\ast}_j\,(t_a)\overline{\lambda}^a\,\psi_{L_{k}} + h.c.
\ee
$\lagr_Y$ yields the fermion mass terms and the Yukawa-type interactions,
\be
\lagr_Y = -\frac{1}{2}\,[\sum_{l,k}\,\frac{\partial W[\hat{\phi}]}{\partial\phi_l\,\partial\phi_k}\,\overline{\psi}_l\psi_{Lk} + h.c.]
\ee
$\lagr_S$ includes interactions between scalar fields , known as F-terms and D-terms , which have an important role in the breaking of SUSY
\be
\lagr_S =  -\frac{1}{2}\,|g\,\phi_{i}^{\ast}\,(t_a)\,\phi_j\,|^2 - \sum_i\,|\frac{\partial W[\hat{\Phi}]}{\partial\phi_l}|^2 
\ee
$\lagr_{MG}$ gives us the interactions between each particle and the gauge fields,
\begin{eqnarray}
\lagr_{MG} &=&-g\,\overline{\psi}_{i}\,\gamma^{\mu}\,V^{a}_{\mu}\,(t_{a})_{ij}\,\psi_{j} - ig\,\phi^{\ast}_{i}\,V^{a}_{\mu}\,(t_{a})_{ij} \stackrel{\leftrightarrow}{{\partial}^{\mu}}\phi_{j} + \nonumber\\
&&+ g^{2}V^{a}_{\mu}V^{b\mu}\,\phi^{\ast}_{i}\,(t_{a}t_{b})_{ij}\,\phi_{j} + h.c.
\end{eqnarray}
$\lagr_{\lambda G}$ includes the interactions between the gauge fields and the gauginos,
\be
\lagr_{\lambda G} = ig\,f_{abc}\,\lambda^a\,\gamma^{\mu}\,\overline{\lambda}^b\,V^c_{\mu} + h.c.
\ee
and $\lagr_{GG}$ contains the self-interactions of the gauge fields where the $t^a$ are the matrices of the Lie Algebra associated with the gauge group, 
\be
\lagr_{GG} = -g\,f^a_{bc}\,V^b_{\mu}V^c_{\nu}\,\partial^{\mu}V^{\nu}_a - \frac{1}{4}g^2\,f^a_{bc}\,f_{ade}\,V^b_{\mu}V^c_{\nu}V^{d\mu}V^{e\nu}
\ee
which satisfy\\
\be
[ t_a , t_b ] = i f^{c}_{ab} t_c
\ee
The complete lagrangian may be written in a more compact form if covariant derivatives are used, as illustrated in eqs. (6.44) and (6.45a-d) in ~\cite{wss}. \\
\smallskip
The model is completely specified once the superpotential  $W[\hat{\phi}_j]$ is specified. For chiral superfields renormalizability requires $W$ to be the most general gauge invariant function that is at most cubic in the fields, i.e.
\be
W[\hat{\phi}_j] = \sum_j\,k_i\hat{\phi}_i + \frac{1}{2}\,\sum_{i.j}\,m_{ij}\,\hat{\phi}_i\hat{\phi}_j + \frac{1}{3}\,\sum_{i,j,k}\,\lambda_{ijk}\hat{\phi}_i\hat{\phi}_j\hat{\phi}_k
\ee
\section{Spectrum of the MSSM}
\label{sec:spectrum}
To create a SUSY version of the SM we must pair up the SM fields in supermultiplets and introduce the SUSY partners of all the SM fields. For the 12 gauge bosons of  $SU(3)_{C} \times SU(2)_{L} \times U(1)_{Y}$ there are no available partners so we must introduce 12 fermions: 8 gluinos ($\tg$), 1 bino ($\tilde{\lambda}_{0}$), and 3 winos ($\tw_{j}$), whose definition is a generalization of the one in the SM. For the chiral fermions of the SM we need new complex scalar partners known as squarks and sleptons. One Higgs doublet is not enough to provide mass to both of the charge (-1/3 , 2/3 ) quarks and satisfy invariance under SUSY transformations, so we must introduce an additional Higgs doublet, defined in Table~\ref{tab:partic}. This is also, what is required to cancel the chiral anomaly that would otherwise arise. The resulting spectrum, as the simplest SUSY generalization of the Glashow-Weinberg-Salam model then consists of the fields listed in Table~\ref{tab:partic}\\
\begin{table}[htdp]
\begin{center}
\begin{tabular}{|c||c|}
\hline
\hline
Field&$SU(3)_C\times SU(2)_L\times U(1)_Y$\\
\hline
&\\
$\hat{L}$ =$ \left(
\begin{array}{c}
\hat{\nu}_{eL}\\
\hat{e}_L
\end{array}
\right)$
&$(\mathbf{1} , \mathbf{2} , -1)$\\
&\\
$\hat{E}^c$&$(\mathbf{1} , \mathbf{1} , 2 )$\\
&\\
$\hat{Q}$ = $\left(
\begin{array}{c}
\hat{u}_L\\
\hat{d}_L
\end{array}
\right)$
&$(\mathbf{3} , \mathbf{2} , \frac{1}{3} )$\\
&\\
$\hat{U}^c$&$(\mathbf{3^{\ast}} , \mathbf{1} , -\frac{4}{3} )$\\
&\\
$\hat{D}^c$&$(\mathbf{3^{\ast}} , \mathbf{1} , \frac{2}{3} )$\\
&\\
$\hat{H}_{u}$ = $\left(
\begin{array}{c}
\hat{h}^{+}_{u}\\
\hat{h}^{0}_{u}
\end{array}
\right)$
&$(\mathbf{1} , \mathbf{2} , 1)$\\
&\\
$\hat{H}_{d}$ = $\left(
\begin{array}{c}
\hat{h}^{-}_{d}\\
\hat{h}^{0}_{d}
\end{array}
\right)$
&$(\mathbf{1} , \mathbf{2^{\ast}} , -1)$\\
&\\
\hline
\end{tabular}
\end{center}
\caption{\label{tab:partic}MSSM particle content. Only the first generation of matter particles is shown; the second and third generations are replicas of this.}%
\end{table}%
\section{A SUSY Toy Model}
As an illustration consider a field theory ~\cite{wss,wzmod} with Lagrangian given by,
\[
\lagr = \lagr_{kin}+\lagr_{mass}
\]
with 
\be
\lagr_{kin}=\frac{1}{2}(\partial_{\mu}A)^2+\frac{1}{2}(\partial_{\mu}B)^2+\frac{i}{2}\overline{\psi}\not\partial \psi + \frac{1}{2}(F^2+G^2)
\ee
\be
\lagr_{mass}=-m[\frac{1}{2}\overline{\psi}\psi-GA-FB]
\ee
where A and B are real scalar fields with mass dimension  $[A]=[B]=1$, $\psi$ is a 4-component $Majorana$ spinor field with mass dimension $[\psi]=3/2$, and F and G are also real scalar fields with dimension $[F]=[G]=2$. Both F and G have no kinetic terms, so their equations of motion are algebraic and can be used to eliminate these fields from the Lagrangian. This yields,
\be
\lagr= \frac{1}{2}(\partial_{\mu}A)^2+\frac{1}{2}(\partial_{\mu}B)^2+\frac{i}{2}\overline{\psi}\not\partial \psi-\frac{1}{2}m^2(A^2+B^2)-\frac{1}{2}m\frac{i}{2}\overline{\psi} \psi
\label{wzlagrfree}
\ee
As a note, we see that the number of bosonic and fermionic degrees of freedom in the Lagrangian exactly balance: without the equations of motion, the four real components for the Majorana spinor field are balanced by the four real scalar fields. This is the  Lagrangian  for free fields $A, B$ and $\psi$. After applying the respective equations of motion, their quanta correspond to two spin zero particles $A$ and $B$ and a self-conjugate, spin $\frac{!}{2}$ particle, all with the same mass. We can add interactions to our Lagrangian~(\ref{wzlagrfree}), assuming renormalizability,
\begin{eqnarray}
\lagr_{int}=-\frac{g}{\sqrt{2}}A\overline{\psi}\psi+\frac{ig}{\sqrt{2}}B\overline{\psi}\gamma_{5}\psi-gm\sqrt{2}AB^2 -\frac{gm}{\sqrt{2}}A(A^2-B^2) \nonumber \\
-g^2 A^2B^2-\frac{g^2}{4}(A^2-B^2)^2
\label{intlagr}
\end{eqnarray}
Note that the Lagrangian~(\ref{intlagr}) has just one mass and one coupling parameter for \underline{all} fields. \\
It is important to note that in the SM radiative corrections to the mass of the fundamental scalars of the SM will be quadratically divergent. If the SM is coupled to new physics at a scale $M$, the quadratic divergence manifests itself as corrections that grow as $M^2$, and so destabilize the weak scale if $M$ is much larger than the Fermi scale, which is the natural scale of the SM. This is known as the {\it hierarchy problem}. The expectation is, that the new physics will have inherent a symmetry which will induce the cancellation of the quadratically divergent contributions. SUSY is just such a symmetry, and as long as the new energy scale is $< 1-2$ TeV, the fine-tuning is ameliorated. \\
To illustrate the cancellation of quadratic divergences, we can use our toy theory and show the cancellation of the quadratically divergent contributions that would destabilize the scalar sector.
Before doing so, we need to evaluate the following quadratically divergent integral up to some momentum cutoff value $\Lambda$,
\be
I_{qd}\equiv \int \frac{d^4q}{(2\pi)^4}\, \frac{i}{q^2-m^2+i\epsilon}
\label{quadint1}
\ee 
Noting the poles of the integrand and choosing an appropiate contour, and setting limits of integration,
\begin{eqnarray}
\int \frac{d^4q}{(2\pi)^4}\, \frac{i}{q^2-m^2+i\epsilon}=\frac{1}{4\pi^2}\int_{0}^\Lambda dq \frac{q^2}{\sqrt{q^2+m^2}}
\nonumber \\
I_{qd}\approx \frac{1}{8\pi^2}[ \Lambda^2 - m^2 ln(\frac{\Lambda}{m}) + const\times m^2]
\label{qdint2}
\end{eqnarray}
Proceeding as in ~\cite{wss}, we consider first the one-point function of the field $A$ to first order in the coupling $g$. The relevant interaction Hamiltonian from ~(\ref{intlagr}) is
\be
\mathcal{H}_{int} = -\lagr_{int} \ni \frac{g}{\sqrt{2}}A\overline{\psi}\psi+\frac{g}{\sqrt{2}}mAB^2+\frac{g}{\sqrt{2}}mA^3
\label{lagr1point}
\ee
If we expand the matrix element $\langle \Omega | TA(x)|\Omega \rangle$, where $|\Omega \rangle$ is the ground state of the interacting theory, perturbatively to order $G$, we get,
\be
-i\frac{g}{\sqrt{2}}\int d^{4}y  D_{F}^{A}(x-y)[(-1)Tr S_{F}(y-y)+mD_{F}^{B}(y-y)+3mD_{F}^{A}(y-y)]
\ee
where $D_{F}$ is the Fourier transform of the scalar field propagator in momentum space given by,
\be
D_{F}(x-y)=\int \frac{d^4q}{(2\pi)^4}\,e^{-iq\cdot (x-y)} \frac{i}{q^2-m^2+i\epsilon}
\ee
The factor in square brackets above is proportional to,
\begin{eqnarray}
Tr\int \frac{d^{4}p}{\not{p}-m_{\psi}} - m\int \frac{d^{4}p}{p^{2}-m_{B}^2}-3m\int \frac{d^{4}p}{p^{2}-m_{A}^2} \nonumber \\
= 4m_{\psi}\int \frac{d^{4}p}{p^{2}-m_{\psi}^2}  -m\int \frac{d^{4}p}{p^{2}-m_{B}^2}-3m\int \frac{d^{4}p}{p^{2}-m_{A}^2}
\label{1ptqd}
\end{eqnarray}
where $m_{\psi}, m_{A}, m_{B}$ are exactly the same as mass parameter $m$ in the trilinear scalar   couplings in eq.~(\ref{intlagr}). Since these masses are exactly equal in a supersymmetric theory, the three contributions in ~(\ref{1ptqd}) add to zero. So although each contribution is separately quadratically divergent, the divergence due to the fermionic term cancels the sum of the divergences from the bosonic terms. In order for this to happen it is necessary that the couplings are exactly those in ~(\ref{intlagr}). Also, the quadratic divergence in ~(\ref{1ptqd}) is independent of $m_{A}$ and $m_{B}$, however the fermion mass must be equal to $m$. \\
If we look at the lowest order quadratic divergences in the two-point function of $A$,  $\langle \Omega | TA(x)A(y)|\Omega \rangle$, once again the quadratic divergences cancel out between fermionic and bosonic contributions, and the cancellation occurs for \underline{all} values pf particle masses. It is again crucial that couplings are as in ~(\ref{intlagr}). We thus see that as long as the dimensionless couplings are as given by supersymmetry, the quadratic divergences cancel even if supersymmetry is broken by scalar masses different from fermion masses. This is an example of Soft Supersymmetry Breaking (\textbf{SSB}) discussed in the next section.
\section{SUSY Breaking}
If SUSY were an exact symmetry of nature, SUSY particles would have the same mass as their SM partners. This is not so, otherwise discovery of the spartners of the known particles should have been possible at the accelerators available prior to the LHC operation. Therefore, SUSY is a broken symmetry at the Fermi scale.  Having a simple model to describe the breaking of SUSY which we could connect to the MSSM would make our endeavours much easier, but unfortunately, such models are far from being simple. 
As previously mentioned during the discussion of our toy model, we can add to $\lagr_{SUSY}$ terms which violate supersymmetry but which are of little importance at high energies. The complete list of possible terms, all of which are required to have mass dimension $< 4$, which may be added to $\lagr_{SUSY}$ \underline{without} altering the cancellation of quadratic divergences in the radiative correction to the SM Higgs mass is as follows:\\
\begin{itemize}
\item mass scalar terms: $\phi_{i}^{\ast}\phi_j$ , $\phi_i\phi_j$
\item trilinear scalar interactions: $a_{ijk}\phi_i\phi_j\phi_k$, $c_{ijk}\phi_i^{\ast}\phi_j\phi_k$ and their h.c.
\item gaugino masses: $\frac{1}{2} M_l\lambda_l\lambda_l$ + h.c.
\item linear terms: $C_i\phi_i$
\end{itemize}
These all are known as Soft-SUSY-Breaking terms (\textbf{SSB}). We then distinguish two separate components in the complete Lagrangian density: \\
\[
\lagr = \lagr_{SUSY} + \lagr_{SOFT}
\]
$\lagr_{SOFT}$ terms parameterize the fundamental mechanisms of SUSY breaking and include the majority of the parameters appearing in the Lagrangian. This complete Lagrangian is what we have introduced before as the MSSM. Any sensible phenomenological study is impaired by the very huge number of  parameters which are present in $\lagr_{SOFT}$. It would be helpful to have a theory capable of predicting the soft parameters, and effectively there are several such theories.\\
Two fundamental mechanisms exist which attempt to explain how MSSM superpartners acquire their masses. Common to both of them  is the existence of a hidden sector responsible for SUSY breaking and an interaction responsible for transmitting the breaking to the visible sector where we find the particles which constitute the MSSM. These fundamental models are known as,
\begin{enumerate}
\item Gauge Mediated SUSY Breaking (\textbf{GMSB}): where the transmission interaction is the same gauge interaction of the SM ~\cite{gsbreak}.
\item Gravity Mediated SUSY Breaking (\textbf{SUGRA}): where it is gravity which acts as the messenger for SUSY breaking ~\cite{sugraref}.
\end{enumerate}
In these models the parameters of the MSSM are determined in terms of a handful of parameters at specified high energy scales. A consequence of this  is that the Higgs mass parameters acquire negative values and produce the ElectroWeak Symmetry Breaking \textbf(EWSB). We see then that EWSB is intimately related to SUSY breaking.\\
The minimal SUGRA (\textbf{mSUGRA}) model has been extensively studied phenomenologically, as well as the GMSB model.  Within mSUGRA, the soft parameters acquire a simple structure at the unification scale (or Planck scale) in which,\\
\begin{enumerate}
\item Scalar masses are universal (diagonal)
\[
m_{Q}^2 = m_{D}^2 = m_{U}^2 = m_{L}^2 = m_{E}^2=m_{H}^2 = m_{0}^2
\]
\item Gaugino masses are universal
\[
M_{1} = M_{2} = M_{3} = m_{\frac{1}{2}}
\] 
\item The cube terms of the soft potential are proportional to the superpotential Yukawas\\
\[
[a_f] = A_{0} [h_f]
\]
\end{enumerate}
where $A_0$ is a common parameter. At low energies, parameters are determined by the renormalization group equations (\textbf{RGE}) from their high energy values. For mSUGRA, the model is fixed by the 18 parameters from the SM with five additional parameters
\[ \langle m_0 , m_{1/2} , A_{0} , B_{0} , \mu \rangle \]
Radiative EWSB (Electroweak Symmetry Breaking) determines $\mu^2$ and it is traditional to eliminate $B_0$ in favor of tan$\beta$ leaving the often used parameter set 
\[ \langle m_0 , m_{1/2} , A_{0} , tan\beta , sgn(\mu) \rangle \]
If the sparticle masses are $\sim 10^2-10^3$ GeV, then extrapolating to high energies by way of the RGE's, with the three SM gauge couplings measured at the weak scale, these very nearly meet at a point under MSSM evolution, suggesting physics at scales $M_{GUT}\sim 2\times10^{16}$ GeV is described by a SUSY GUT. Below $M_{GUT}$, the correct effective field theory is provided by the MSSM.\\
We would like to introduce a symmetry which acts differently on the component fields of the superfields so as to distinguish the SM particles from their superpartners.  This is known as $R$-symmetry, consisting of a $U(1)$ phase applied to the components of the superfields. When the phase is restricted to the value $\pi$, the $R$-symmetry is referred to as $R$-parity, with the phase being either $+1$ or $-1$.  All SM particles are even under $R$-parity while superpartners are odd under $R$-parity. If we now restrict $\lagr_{SUSY}$ to be invariant under $R$-parity, a consequence will be that there cannot be interactions coupling a single superpartner to two SM particles. This implies that all superpartners will ultimately decay to an sparticle, {\it the lightest supersymmetric particle} or \textbf{LSP}, which will be stable. Then the Universe must be filled with these sparticles, and from experiments on the charge-to-mass ratio of matter, the possibility of it being electrically charged has been ruled out. Thus, the LSP is electrically neutral. The LSP would be a viable candidate for Dark Matter. In our work, we assume $R$-parity invariance. A formula to calculate $R$-parity is given by
\[
R= (-1)^{3(B-L)+2s}
\]
where $B, L$ are the baryon (lepton) quantum numbers and $s$ is the spin. The factor $(-1)^{2s}$ guarantees that particles and their superpartners will have opposite $R$-parity. Imposing $R$-parity invariance eliminates all baryon (lepton) violating interactions, if interactions are renormalizable. An important phenomenological consequence of $R$-parity invariance is that sparticles can be produced only in pairs at colliders, and must decay to SM particles plus an odd number of sparticles. \\   
Summarizing, the MSSM is a quantum field theory with  supersymmetry relating bosonic and fermionic degrees of freedom. Providing us with a solution to the SM fine-tuning problem by eliminating the quadratic dependence on the cut-off scale $\Lambda$, it is perturbatively calculable for all energies up to M$_{Planck}$ or M$_{GUT}$ scales without requiring huge fine tuning. It is less $UV$ divergent than corresponding non-SUSY theory due to cancellation of the leading quadratic  divergence of fermionic loops with those of bosonic loops. When elevated to a local supersymmetry gravity is automatically introduced leading  gravitational interactions along with strong, weak and electromagnetic interactions in an effective field theory. With conservation of $R$-parity it includes a stable massive particle which is usually electrically and color neutral, providing us with an excellent candidate for the observed cold dark matter in the Universe. \\

\chapter{SUSY Phenomenology at the LHC}
\label{chap: collide}
\section{Large Hadron Collider}
\subsection{Overview}
\begin{figure}[ht]
\begin{center}
\includegraphics[width=10cm]{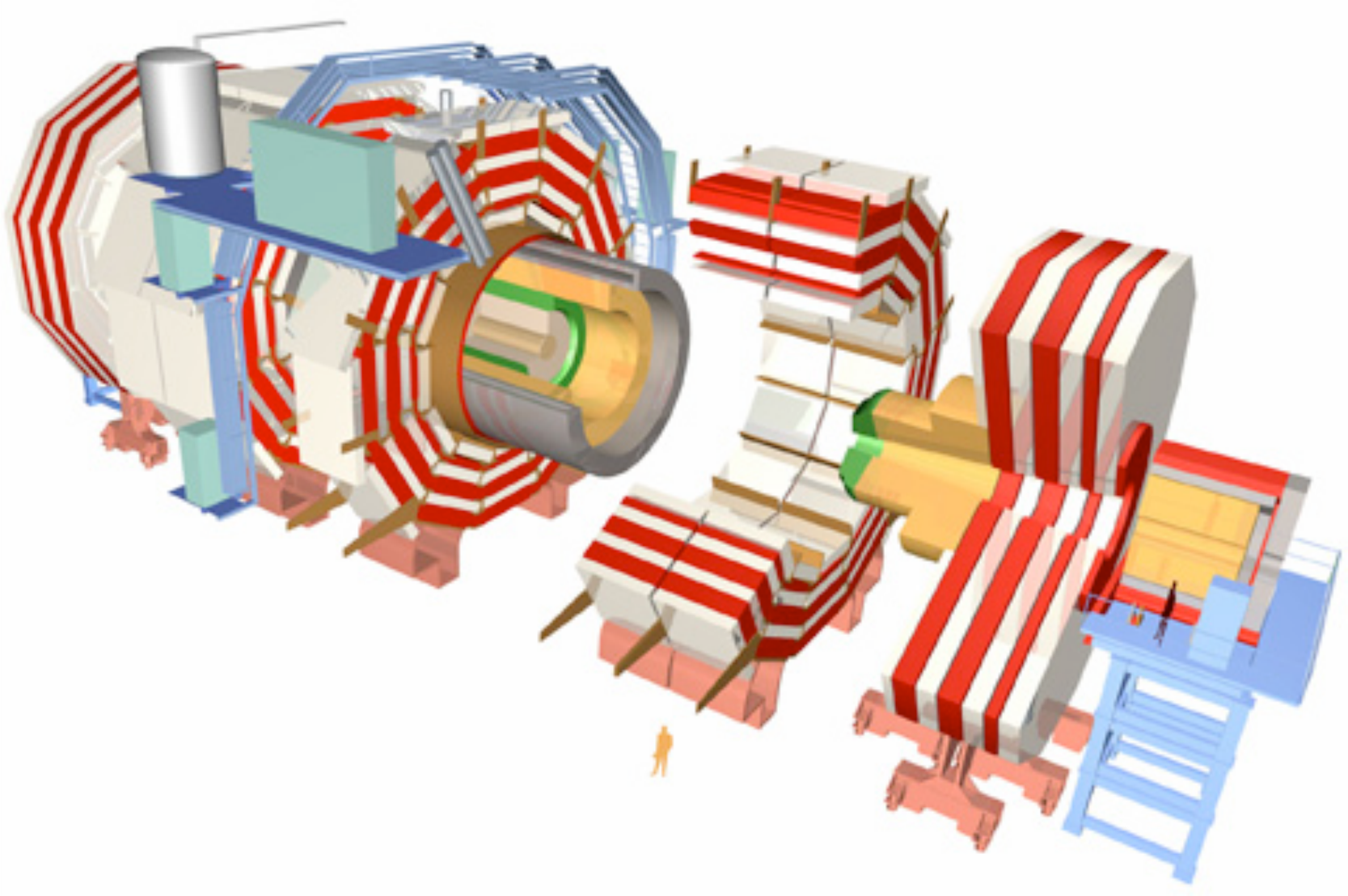}
 \end{center}
 \caption{ \label{cmsdet}CMS detector at the LHC }
 \end{figure}

\begin{figure}[ht]
\begin{center}
\includegraphics[width=10cm]{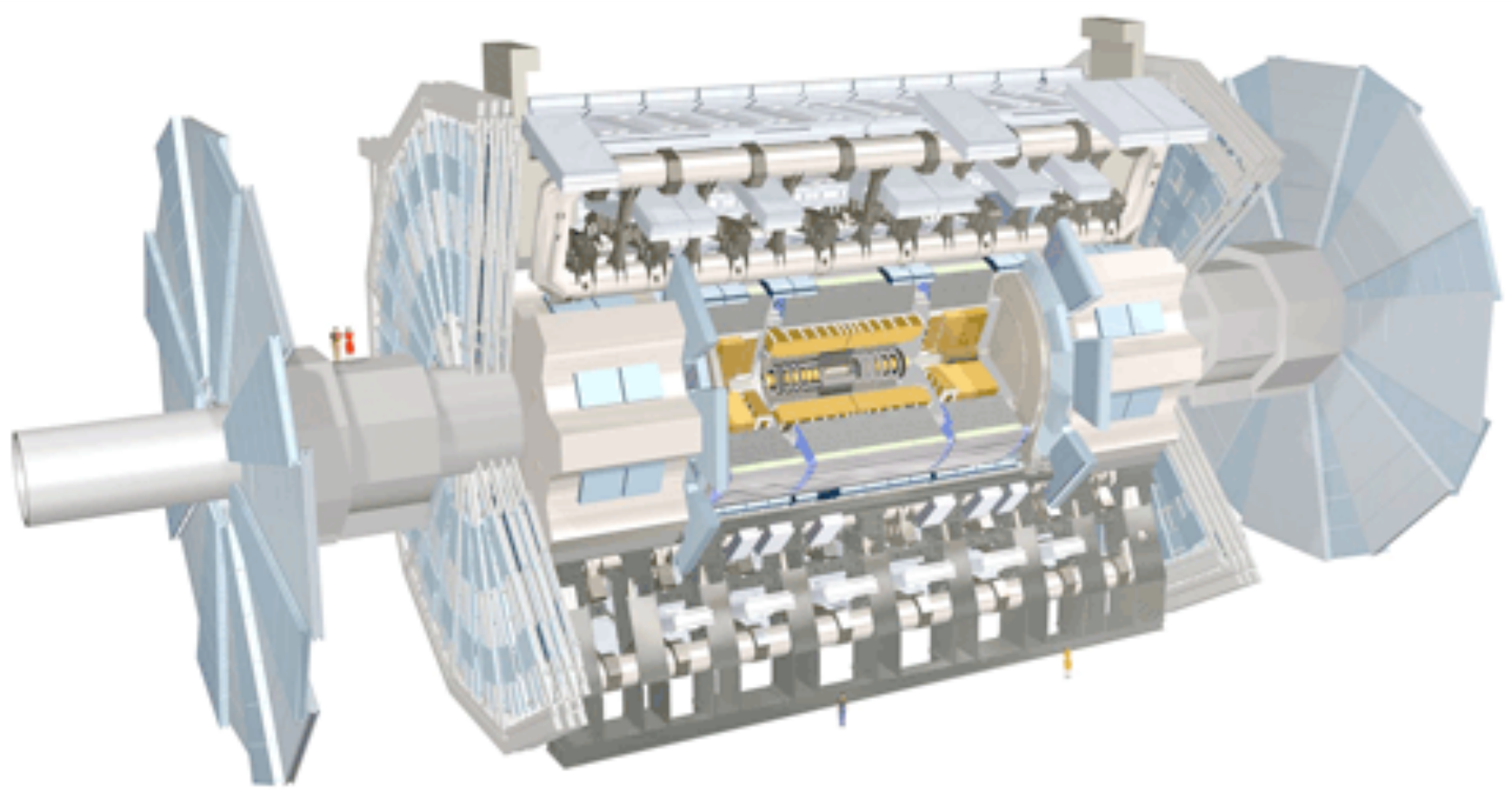}
 \end{center}
 \caption{ \label{atlasdet}ATLAS detector at the LHC }
 \end{figure}

Located in Geneva, Switzerland, the Large Hadron Collider \textbf{LHC} is a $pp$  collider, built to operate initially at a C.M. energy of  $\sqrt{s} = 7$ TeV, with plans to eventually reach its design energy of $\sqrt{s} = 14$ TeV. At these energy scales, the LHC is capable of creating the conditions essential for discovery of new physics at the weak scale \cite{lhc1,lhc2}.\\
The discovery of a single SUSY particle would be as groundbreaking as that of finding the elusive Higgs boson. The available energy would facilitate the discovery of a superpartner particle,  allowing these hypothetical particles to manifest themselves over the background from SM sources. Such an event  holds promise of explaining one of the most persistent mysteries in physics and astronomy, the existence of dark matter, first theorized in the 1930's. \\
The LHC work may also reveal the existence of additional dimensions of space, if nature really has hidden dimensions, over the known $3 + 1$, allowing for a structure of the Universe more complex than our current knowledge describes. The LHC has a length of $26.659$km, with detectors located at several points along the circumference.
The proton beams are obtained by ionizing the Hydrogen gas travel in opposite directions while being accelerated to speeds very close to the speed of light. This happens in a successive series of smaller accelerators, before the beam is finally  injected into the LHC for the last  stage of acceleration. Here, powerful frequency devices provide a kick to the particles each time they pass by. There are over 10-thousand superconducting electromagnets, supercooled by liquid Helium to 1.9K.\\
The detectors are:
\begin{itemize}
\item Compact Muon Solenoid (\textbf {CMS}) \cite{cms,cms1}.
\item A Toroidal LHC Apparatus (\textbf  {ATLAS})\cite{atlas,atlas1,atlas2} .
\item A Large Ion Collider Experiment (\textbf{ALICE})\cite{alice1,alice2,alice3}.
\item LHCb\cite{lhcb1,lhcb2}.
\end{itemize}
CMS and ATLAS are general purpose detectors and the analysis of data of these experiments will be of primary interest to us in this dissertation.\\ 
With the LHC using approximately $2\times 10^{-9}$ grams of Hydrogen per day, the ultimate collisions emerge after a succession of processes:
\begin{enumerate}
\item Protons are given an initial boost in the small \underline{linear} accelerator known as Linac2. to about $0.314c$.
\item Protons then move into CERN's old \underline{circular} accelerator, the Proton Synchroton (\textbf{PS})Booster, where they can boost their speed every lap until reaching speeds of about $ 0.916c$.
\item The next boost happens at the Proton Synchroton, to about $ 0.9993c$.
\item Protons are now funneled into the Super Proton Synchroton (\textbf{SPS}), where in 1983 both the W and the Z were first detected. Here protons reach $\approx 0.99998c$ which is equivalent to a C.M. energy of about $ 450$GeV.
\item Protons are led into the LHC, where for a C.M. energy of $ 7$ TeV at current operation, the speeds are $\approx 0.999999991c$ . One beam consists of some 2808 bunches, with $\approx 10^9 $ protons in each bunch. 
\item Collisions now occur at ATLAS (point 1), CMS (point 5), LHCb (point 8), and ALICE (point 2).
\end{enumerate}
The ability of a detector to find and measure particle momenta with high accuracy is propotional to the strength of the magnetic field $B \times$ the distance travelled inside the detector. For the CMS design, as shown in Fig.~\ref{cmsdet} the choice was to build a compact instrument offering a relatively short path for the muons inside the detector, but using a high magnetic field $B \approx 4  Tesla$.\\
 \smallskip
The ATLAS detector, as shown in Fig.~\ref{atlasdet} was designed with the alternate choice, a bigger instrument offering a larger path, but using a smaller magnetic field $B \approx 1$Tesla, thus achieving the same capability. The pre-assembled piece of the CMS detector, containing the giant electromagnet weighs in as much as 5 Boeing 747 airplanes.\\
 \subsection{Techniques for LHC Searches}
 From the billions of $pp$ collisions, some 10-15 petabytes ($10^{15}$bytes) of data are generated per year. In the CMS, from this huge amount of collisions, maybe 1 in $10^5$ is of interest, 300 are permanently recorded for complete reconstruction and analysis, and 1 is placed on screen every second.\\
 Four kinds of emissions follow proton collisions in the LHC:\\
 \begin{enumerate} 
 \item Jets: streams of quarks and gluons that ultimately convert to hadrons emanating at various angles, depending on the energy and type of reactions produced in the collision. 
 \item Discrete emissions of isolated leptons : $e, \mu, \tau.$
 \item Missing transverse energy $\mathbf{E_T^{\rm miss}}$: energy of particles that are undetected and moving in directions that have a transverse component to the direction of the colliding beam of protons.
 \item Photon emission.
 \end{enumerate}
The total transverse momentum of the final products  in the center of mass frame of the collision  should be zero. Then the difference between the measured amount and zero yields the  $E_T^{\rm miss}$(or alternatively, $\NEG{E}\mathstrut _{\top }^{\text{ }}$). In hadronic collisions, the partons which participate in the hard process carry a fraction of the beam energy. The remnants of the beam associated with the remaining partons mostly escape undetected in the beam pipe. Thus, only conservation of the momentum in the direction transverse to the beams is relevant, making the missing transverse energy $\eslt$ the important quantity, rather than the total missing energy. \\
 
\begin{figure}[ht]
\begin{center}
\includegraphics[width=7.7cm]{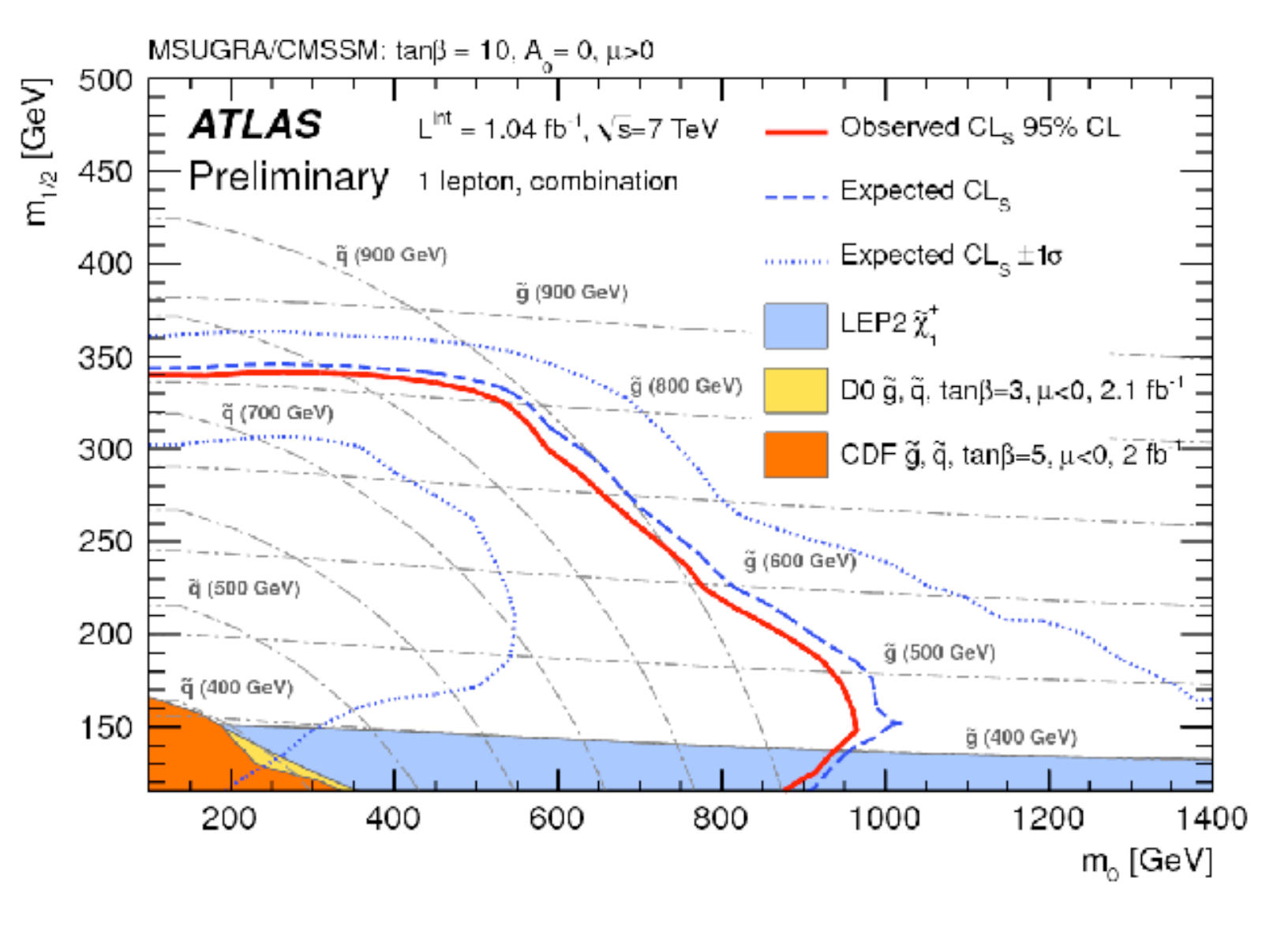}
 \caption{ \label{excl1}Exclusion region in the mSUGRA/CMSSM  $(m_0,m_{1/2})$ plane for $\tan\beta=10$, $A_0=0$ and $\mu>0$ for an integrated luminosity of 1.04 fb$^{-1}$ at ATLAS ~\cite{exclude}}
 \end{center}
 \end{figure}

 February 2010 provided the first report of a collision taking place inside the CMS detector at the LHC late in 2009. On March 19, 2010 LHC reached its  target energy for the next two years : $3.5$ TeV per beam, $7$ TeV total. After this period the LHC will undergo a year of maintainance, following which it expects to operate at its design energy  of $14$ TeV  for both beams. 
After more than a year of operation, its total integrated luminosity for 2010-11 is 3.3 fb$^{-1}$, as of this writing. Recently recorded events are viewable at ~\cite{atlaslive}.\\
The ATLAS and CMS have searched for an excess of events above SM expectations in channels that would be populated by the production of gluinos and squarks at the LHC. Unfortunately no such excess was found. This has then been translated to upper limits on the cross section for particle production. This exclusion is illustrated within the mSUGRA model, introduced in Chapt.~\ref{chap: susy101}, in Fig~\ref{excl1} for ATLAS data and in Fig~\ref{excl1} for CMS data where a composite is made for an integrated luminosity  of about 1 fb$^{-1}$ that was analyzed in Summer, 2011.\\
 \begin{figure}[ht]
\begin{center}
\includegraphics[width=7.7cm]{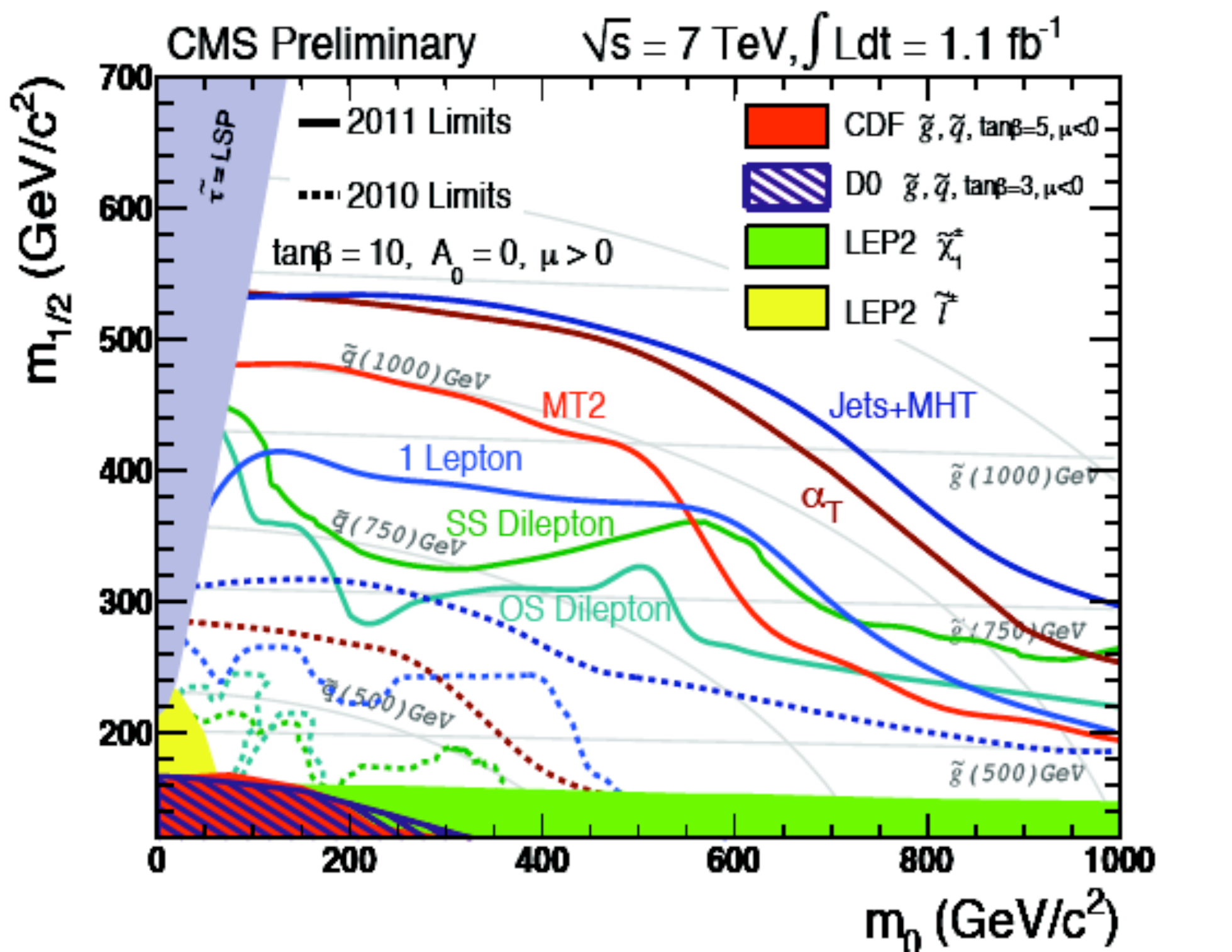}
\caption{ \label{excl2}Exclusion region in the mSUGRA/CMSSM  $(m_0,m_{1/2})$ plane for $\tan\beta=10$, $A_0=0$ and $\mu>0$ for an integrated luminosity of 1.04 fb$^{-1}$ at CMS.}
\end{center}
 \end{figure}
\section{SUSY Event Simulation}
The key link between theoretical predictions of SUSY or other new physics and the actual experimental observations of particle tracks and calorimeter depositions in collider detectors is provided  through event generator programs. These allow us to compute how a theory would manifest in actual collider experiments. With the LHC operating at energy scales of $\sqrt{s}= 7-14$ TeV, this should be enough to produce superpartners and provide evidence for viable  particle models, such as weak scale SUSY. It is possible  that discovery of new physics beyond the SM can result from indirect searches, but for SUSY at the weak scale it is widely accepted that evidence for it will come from direct creation of supersymmetric matter in colliding beam experiments., and the detailed analysis of the resultant scattering events.\cite{wss}\\
\subsection{Event generator}
Different models describing supersymmetry exist, which are used to predict sparticle production rates and subsequent decay patterns into final states of SM fundamental particles. Some of these, such as quarks and gluons, cannot be detected directly in a collider detector. The detectors will measure tracks and momenta of quasi-stable charged particles bending in a magnetic field, in addition to energy deposited in calorimeter cells by charged leptons, hadrons, and photons. We realize then that a gap exists between the predictions of SUSY theories for final states of fundamental particles, and the actual detection at the experimental level. It is the existence of this gap that necessitates the development of event generator programs\cite{hbaer}. Currently available general purpose event generator programs that incorporate SUSY include {\small ISAJET \cite{isajet}, PYTHIA \cite{pythia}, HERWIG \cite{herwig}, SUSYGen \cite{sgen}, and SHERPA \cite{sherpa}.} Once a SUSY theory and collider type are specified, the event generator program produces a full simulation of types of scattering events to be expected. The final states are completely specified including detailed kinematics of each particle in the event.\\
The present work studies prospects for physics at the LHC operating at its design energy of $14$ TeV in the center of mass. Then, for an input of an MSSM set of parameters, the generator, in our case {\small ISAJET} version 7.74, generates the sparticle pair production events according to the ratio of their production cross sections. The sparticles will then undergo a decay into a partonic final state, according to branching ratios specified by the model. Then the partonic state final state is converted to one composed of particles which are detected experimentally.\\
The fraction of the hadron's longitudinal momentum carried by the initial hard scattering partons is unknowable, so there is an irreducible uncertainty in the longitudinal boost of the center-of-momentum frame for the colliding system. However, by forming the vector sum of all the energy deposited in the transverse direction, we get an important quantity called missing transverse energy, $\NEG{E}\mathstrut _{\top }^{\text{ }}$, as mentioned previously. A certain amount of $\NEG{E}\mathstrut _{\top }^{\text{ }}$ is due to jet and lepton mismeasurement from imperfect energy resolution, particles going into cracks in the detector, and other ÔÔnon-physicsÕÕ causes. A large $\NEG{E}\mathstrut _{\top }^{\text{ }}$, however, generally indicates the
production of one or more high-energy  weakly-interacting particles that escape the experimental apparatus without depositing energy. In the Standard Model these would be neutrinos. In SUSY searches, large $\NEG{E}\mathstrut _{\top }^{\text{ }}$ is the signature of escaping LSPs.
Indeed, since a pair of LSPs are always produced in a SUSY reaction where $R$-parity is conserved, large $\NEG{E}\mathstrut _{\top }^{\text{ }}$ is the hallmark of a supersymmetric reaction.\\
\subsection{Detector simulation}\label{sec:detsim}
As each event is generated, it is processed through a toy detector simulator consisting of the following elements (where $\eta $ is the pseudorapidity, $\phi $ is the azimuthal angle, and $\Delta R=\sqrt{(\Delta \eta)^{2}+(\Delta \phi )^{2}}$). The toy detector captures the salient features of the LHC detectors.
\begin{itemize}
\item  Calorimeter simulator: We implement a toy calorimeter based on the {\small ISAJET}\ \texttt{CALSIM} subroutine. The segmentation is $\Delta\eta \times \Delta \phi =0.05\times 0.05$ extending to a rapidity of $|\eta|=5$. There is a hadronic calorimeter, into which hadrons deposit their energy with a resolution given in Table~\ref{simdef} for different ranges of $|\eta|$, and an electromagnetic calorimeter which captures electrons and photons with resolution also listed in Table~\ref{simdef}. We do not attempt to simulate effects of cracks or dead regions that are specific to particular detectors.
\item  Isolated lepton identification: We sum the hadronic transverse energy in a cone of $\Delta R<0.3$ around each lepton. If this hadronic energy is less than 50\% of the lepton's transverse energy, then the lepton is declared isolated. The $p\mathstrut _{\top }^{\text{ }}$ thresholds for isolated leptons are given in Table~\ref{simdef} for each case.
\item  Jet identification: We use {\small ISAJET}'s \texttt{GETJET} jet-finding algorithm. Jets are defined as hadronic clusters with total $E_{\mathrm{\top }}^{{}}>  50$ GeV falling within a cone of radius $\Delta R<0.5$ and subject to $|\eta |<3$. We do not correct jet energy.
\item  Silicon vertex detector (SVX): We simulate a SVX detector for tagging $b$-jets. We identify each weakly-decaying $B$ hadron in an event with $E_{\mathrm{\top }}^{{}}>15\mathrm{\ GeV}$ and $|\eta |<3$. If $\Delta R(B,\text{jet})<0.5$ for some jet then that jet is tagged as a $b$-jet. with an efficiency of 50\% \cite{brej} at the LHC design luminosity of 100~fb$^{-1}$/y, and assume that gluon and light quark jets can be rejected as $b$ jets by a factor $R_b= 150$ (50) if $E_T < 100$~GeV ($E_T > 250$~GeV) and a linear interpolation in between \cite{brej}.\\
\end{itemize}
In Table~\ref{simdef}  we summarize  the  basic parameters used to define jets, b-jets, and isolated  leptons. \\
\begin{table}[htdp]
\begin{center}
\begin{tabular}{lcc}
calorimeter $|\eta|$ & $<$ 5\\
cell size & $\Delta\eta$ x $\Delta\phi$ = 0.05 x 0.05\\
hadronic resolution & \\
$|\eta| < $ 3 &${ 0.5/\sqrt{E}}\bigoplus 0.03$\\
3$< |\eta| < $ 5 & ${1/\sqrt{E}}\bigoplus 0.07$\\
$\bigoplus$ & addition in quadrature\\
EM resolution & ${0.1/\sqrt{E}}\bigoplus 0.01$\\
Jets are & hadronic clusters\\
$E_T >$50 GeV & $\Delta R = \sqrt{\Delta \eta^2 + \Delta \phi^2} = 0.5$\\
$|\eta| < $3.0 &\\
B-Jets are & tagged at 50$\%$ eff \\
$E_T >$ 50 GeV & $|\eta| < 3$ \\
B-hadron & $p_{T} > $ 15 GeV\\
Isolated Leptons &\\
in LHC reach studies &$p_{T} > $ 10 GeV\\
in LHC dilepton studies&  $p_{T} > $ 6 GeV   
\end{tabular}
\end{center}
\caption{simulation initial set of parameters defining jets bjets and isolated leptons }
\label{simdef}
\end{table}%
\section{Phenomenological goals}\label{sec:goals}
Given the opportunity to make a wish list of measurements that may shed light on the properties of the sparticles, which would establish a roadmap to the high energy scale physics responsible for the breaking of SUSY, one would set apart as being of upmost importance the  following:\\
\begin{itemize}
\item The discovery of sparticles, whose properties would be indicative of the validity of existing SUSY models.
\item The measurement of the masses of as many superpartners as possible.
\item The relevance of signals from third generation squarks, possibly shedding some light on an Inverted Mass Hierarchy (\textbf{IMH}) in SUSY, where the sparticles belonging to the third generation  have masses lower than the corresponding ones in the first/second generation.
\item the determination of the parameters involved in higgsino-gaugino mixing
\end{itemize}
In view of the above as motivation for this dissertation, we have divided our endeavours into three separate projects.\\
\begin{enumerate}
\item {\large\textbf{Heavy-flavor tagging and the SUSY reach at the LHC:}}\\
Considering that the branching fraction for the decays of gluinos to third generation squarks is expected to be enhanced in classes of SUSY models where either third generation squarks are lighter than other squarks, or models of mixed higgsino dark matter which are constructed in agreement with the measured density of cold dark matter(\textbf{CDM}), the gluino production in such scenarios at the LHC should be rich in top and bottom quark jets.  Requiring $b$-jets \cite{MMT} in addition to $\eslt$ should, therefore, enhance the supersymmetry signal relative to Standard Model backgrounds from $V$ + jet, $VV$ and
QCD backgrounds ($V=W, Z$). We quantify the increase in the supersymmetry reach of the LHC from $b$-tagging in a variety of well-motivated models of supersymmetry. We also explore ``top-tagging'' at the LHC. We find that while the efficiency for this turns out to be too low to give an increase in reach beyond that obtained via $b$-tagging, top-tagging can indeed provide a confirmatory signal if gluinos are not too heavy \cite{rk01}.
\item {\large\textbf{Signals for light third generation squarks (stops) at the LHC:}}\\
We explore the prospects for detecting the direct production of third generation squarks in models
with an inverted squark mass hierarchy. This is signalled by $b$-jets + $\eslt$ events harder than in the Standard Model, but softer than those from the production of gluinos and heavier squarks.  We find that
these events can be readily separated from SM background (for third generation squark masses $\lesssim 500$~GeV), and the contamination from the much heavier gluinos and squarks although formidable can effectively be suppressed \cite{rk01}.
\item {\large\textbf{Neutralino mass reconstruction and MSSM parameter determination:}}\\
We attempt to extract model-independent information about neutralino properties from LHC data. assuming only the particle content of the MSSM and that all two-body neutralino decays are kinematically suppressed, with the neutralino inclusive production yielding a sufficient cross section. We show that the Lorentz invariant dilepton mass distribution encodes clear information about the relative sign of the mass eigenvalues of the parent and daughter neutralinos. We attempt to answer question as to whether  from the dilepton distribution we can establish if the decay is the result of a virtual $Z$-boson or a virtual slepton $\tilde{l}_{L,R}$ exchange. We attempt to extract information as to the values of the MSSM parameters that determine the mass of the neutralinos.
\end{enumerate}
The first two items listed above make up the contents of Chap.~\ref{chap:btag}, while the third item is elaborated upon in Chap.~\ref{chap:zslep}. We end in Chap.~\ref{chap:epilog} with a brief outlook for the future.\\



\chapter{Heavy Flavor Tagging and the LHC Reach }
\label{chap:btag}
\section{\textbf{Introduction}}\label{sec:inicial}
In the previous chapter, in Sec.~\ref{sec:goals} we presented three projects which comprise the focus of the work presented in this dissertation. Now, in this chapter, we detail our work with respect to the first two of these projects mentioned in Sec.~\ref{sec:goals} whose primary goal is to provide answers to the following two questions:\\
\bi
\item Q1: What information obtained from  LHC events can we use to develop techniques which will allow us to extend the SUSY reach projections using $\eslt$, jets and leptons, at the LHC? Extending the SUSY reach encompasses extending the region of MSSM parameter space where SUSY signals can be distinguished from SM background events, based on specific observability requirements.\\
Additionally, we would like these techniques not to be restricted to a specific SUSY model, but rather to be applicable to  classes of models. We consider models that: first, they conform to Dark Matter constraints, specifically CDM, which we describe below. Second, we study models that exhibit an Inverted Mass Hierarchy (\textbf{IMH}), whereby the third generation scalar sparticles are lighter than their corresponding $1^{st}$ and  $2^{nd}$ generation counterparts. IMH models are interesting because serves  third generation sfermion mass parameters are driven to sub-TeV values, leaving first and second generation scalars as heavy as 2--3~TeV. The multi-TeV values of first and second generation scalar masses ameliorate the SUSY $CP$ and flavour problems without destroying the SUSY resolution of the gauge hierarchy problem, since the fields with substantial direct couplings to the Higgs sector (gauginos and third generation scalars) have masses below the TeV scale.  
\item Q2: Can we use these same techniques to isolate signals corresponding to direct production of third generation sparticles, not only from SM background events, but Additionally from other SUSY events which would now be considered an added contamination to the background? This would provide unequivocal evidence for the production of third generation squarks.
\ei
It has been shown that both squark and gluino masses are smaller than $2-3$ TeV and their production will be observable above SM backgrounds via signals consisting of multi-jet plus multi-lepton events with large amounts of $\eslt$ carried off by the escaping LSPs.
We will assume that the lightest neutralino is the LSP as is the case in many models.
Remarkably, SUSY models with a stable neutralino LSP naturally lead to the right magnitude for the measured relic density of thermally produced cold dark matter \cite{wmap}, if superpartner masses are $\sim 100$~GeV.\\
Defining $\Omega$ as the total matter/energy density of the Universe as a fraction of the critical closure density $\rho_{c} \simeq 1.88 \times 10^{-29} h^{2} gcm^{-3}$ where $h$ is the Hubble parameter in units of $100 km/sec / Mpc$, then the component arising from non-relativistic and non-radiating matter is labeled as $\Omega_{\rm CDM}$ with an inferred value of ~\cite{wmap2011},
\be
\Omega_{\rm CDM}h^2=0.1120 \pm 0.0056 \;, \ \ (2\sigma)
\label{wmap1}
\ee
\\
Assuming thermal production and standard Big Bang cosmology, the upper limit from (\ref{wmap1})  provides a stringent constraint  on any theory with stable weakly interacting particles, in particular on weak scale SUSY theories. Since the dark matter may well consist of several components, the contribution from any single component may well not saturate the observed value, so that strictly speaking the relic density measurement serves as an upper bound, 
\be
\Omega_{\tz_1}h^2 < 0.12\;,
\label{wmap}
\ee
on the relic density of neutralinos, or for that matter, on the density of any other stable particle.\\

Direct searches for charged sparticles at LEP~2 have resulted in lower limits of about 100~GeV on chargino and selectron masses, and slightly lower on the masses of smuons and staus \cite{lep2}. Since neutralinos can annihilate via $t$-channel sfermion exchange, the measured value of the relic density, on the other hand, favours sfermions lighter than about 100~GeV, resulting in some tension with the LEP~2 bounds. In many constrained models where all sparticle masses and couplings are fixed by just a few parameters, such light sparticles often also lead to measurable deviations in {\it other} observables, and hence are disfavoured. If the SUSY mass scale is raised to avoid these constraints, the annihilation cross-section which is proportional to $\frac{1}{M_{\rm SUSY}^2}$ is correspondingly reduced, and the neutralino relic density turns out to be too large.  One way to fix this is by invoking non-thermal relics or non-standard cosmology to dilute the relic density. However, it seems much more economical to invoke SUSY mechanisms that enhance the neutralino annihilation rate to bring their thermal relic density in line with (\ref{wmap}).\\

The primary reason for the low neutralino annihilation rate lies in the fact that the LSP is dominantly a bino in many models with assumed gaugino mass unification, where the bino and wino masses are related by $M_1 \simeq {1\over 2} M_2$. The annihilation of bino pairs to gauge bosons is forbidden because $SU(2)\times U(1)$ precludes the couplings of binos to the gaugino-gauge boson system, while annihilation to fermions may be suppressed by large sfermion masses and the relatively small hypercharge coupling. Finally, annihilation to Higgs boson pairs is suppressed by the (usually large) higgsino mass, as well as by the small hypercharge gauge coupling.  This then suggests several ways in which the neutralino annihilation rate may be enhanced to bring their thermal relic density in accord with (\ref{wmap}).
\begin{itemize}
\item We can arrange the mass of a charged or coloured sparticle to be  close to that of the LSP. Since these coloured/charged sparticles can   annihilate efficiently, interactions between them and the neutralino  which maintain thermal equilibrium will necessarily also reduce the neutralino relic density \cite{GS}. Within the mSUGRA model, the   co-annihilating sparticle is usually either the scalar tau \cite{stau} or the scalar top \cite{stop}, but different  choices are possible in other models.
  
\item We can arrange $2m_{\tz_1} \simeq m_A \simeq m_H$, so that neutralino annihilation is resonantly enhanced through $s$-channel  heavy Higgs boson exchange \cite{Afunnel}. The large widths of $A$ and  $H$ together with the thermal motion of the LSPs in the early universe then enhances the annihilation cross section over a considerable range of parameters. Within the mSUGRA model, this is possible only if  $\tan\beta$ is very large. However, in models with non-universal Higgs mass (NUHM) parameters, where the Higgs scalar mass parameters do not unify with matter scalar parameters as in mSUGRA \cite{NUHMgen,NUHM}, agreement with (\ref{wmap}) may be obtained via resonant $A/H$ annihilation for any value of $\tan\beta$.  We mention that resonantly enhanced 
annihilation may also occur via $h$ exchange, albeit for a much smaller range of parameters \cite{hfunnel}.

\item It is also possible to obtain an enhanced neutralino annihilation  rate if the light top squark, $\tst_1$, is relatively light so that  neutralinos efficiently annihilate via $\tz_1\tz_1 \to t\bar{t}$ \cite{compressed}, or in  NUHM models via $\tz_1\tz_1 \to u\bar{u}$ or $c\bar{c}$, via $t$-channel top- or right-squark exchanges, respectively \cite{NUHM}.  
\end{itemize}

Instead of adjusting sparticle masses, we can also adjust the composition of the neutralino. More specifically, 
\begin{itemize}
\item We can increase the higgsino content of the neutralino so that its couplings to the gaugino-gauge boson pairs are increased, leading to  mixed higgsino dark matter (MHDM). Within the mSUGRA framework, we  can only do so in the so-called hyperbolic branch/focus point (HB/FP) region where $m_0$ takes on multi-TeV values \cite{focus}, but  in NUHM models this is possible for {\it all} values of $m_0$  \cite{NUHM}. The higgsino content may also be increased by relaxing the assumed high scale universality between gaugino masses. The  usually assumed universality of gaugino masses follows if the auxiliary field that breaks supersymmetry does not break the underlying grand unification symmetry; if this is not the case,  non-universal gaugino masses can result. It has been shown that if the GUT scale gluino mass is smaller than the other gaugino masses,  $m_{H_u}^2$ does not run as negative as usual, yielding a smaller  value of $\mu^2$, resulting in an increased higgsino content of $\tz_1$
 \cite{lm3dm}.  This has been dubbed as low $M_3$ dark matter (LM3DM). It has been pointed out \cite{hm2dm} that  increasing the GUT scale wino mass parameter from its unified value also results in a low value of $|\mu|$, resulting in consistency with  (\ref{wmap}) via MHDM.
\item Finally, depending on the gauge transformation property of the  SUSY breaking auxiliary field, it may also be possible to enhance the wino content of the neutralino leading to mixed wino dark matter (MWDM) \cite{mwdm}. This requires that the weak scale values of bino  and wino masses to be approximately equal. If instead these are  roughly equal in magnitude but differ in sign, bino-wino mixing is suppressed, but agreement with the observed relic density is possible via bino-wino co-annihilation (BWCA) \cite{bwca}.
\end{itemize}

Of interest to us here is the potential for an enhanced rate for bottom quark production in SUSY events that occurs for MHDM, as exemplified by (but not limited to) the HB/FP region of the mSUGRA model \cite{DDDR,MMT}, or models where third generation squarks are significantly lighter than other squarks as, for instance, in the stop co-annihilation region of mSUGRA, in  inverted hierarchy models where third generation sfermions are much lighter than those of the first two generations \cite{bagger,imh}, or in the framework suggested in Ref.~\cite{compressed}. \\
It has been shown previously \cite{MMT}  that using $b$-jet tagging techniques that are available at the LHC, the SUSY reach may be enhanced by as much as 20\% for parameters in the HB/FP of the mSUGRA model. Toward this end, we examine the reach of the LHC with and without $b$-jet tagging, in several models motivated by the relic density measurement just discussed as well as by other considerations, to precisely delineate the circumstances under which $b$-jet tagging will significantly enhance the LHC reach. Since SUSY events may also be enriched in $t$-jets, we also examine prospects for  top jet tagging in SUSY events at the LHC.\\
Having completed this introduction, for the benefit of the reader we provide a summarized version of the chapter's structure, which we have divided as follows:
\bi
\item Sec~\ref{sec:models}: we introduce the various models chosen for the study.
\item Sec~\ref{sec:sim}: event simulation and calculational details are discussed.
\item Sec~\ref{sec:btaggs}: we discuss the simulation of the signal and the analysis cuts.
\item Sec~\ref{sec:relts}: the results for the different models are presented.
\item Sec~\ref{sec:third}: results specific to isolating third generation squarks are discussed.
\item Sec~\ref{sec:toptag}: top-tagging is introduced, and its relative efficiency compared to b-tagging is discussed.
\item Sec~\ref{sec:ctag}: charm tagging is discussed, including its relative merits in previous studies in comparison to the LHC.
\item Sec~\ref{sec:resumen}: we summarize our work presented in this chapter.
\ei

\section{Models} \label{sec:models}
Here we discuss several models in which we may expect third generation fermions to be preferentially produced in SUSY models. We begin with the familiar mSUGRA model, and work our way through various other models motivated either by the relic density observation discussed in Sec.~\ref{sec:inicial}, or by other considerations.

\subsection{The mSUGRA model} The mSUGRA model \cite{msugra} was already introduced in Chap.~\ref{intro}. We remind the reader that this model is completely specified by the parameter set,
\be
m_0, m_{1/2}, \tan\beta, A_0 \ \ {\rm and} \  {\rm sign}(\mu)\;.
\ee
\\
Typically, the weak scale value of $|\mu|$ is similar in magnitude to $m_{\tg}$, and the bino is the LSP. However, for any chosen value of $m_{1/2}$, the requirement that electroweak symmetry be correctly broken imposes an upper bound on $m_0$, since the value of $\mu^2$ becomes negative for yet larger values of $m_0$. There is thus a contour in the $m_0-m_{1/2}$ plane where $\mu^2=0$. For values of $m_0$ just below this bound, $\mu^2 \ll m_{\tg}^2$ and can be comparable to the SSB bino mass parameter, $M_1$ so that the lightest neutralino is a mixed bino-higgsino state that can annihilate rapidly in the early universe, mainly via its higgsino content. This is the celebrated HB/FP region of the mSUGRA model \cite{focus}, one of the regions of mSUGRA parameter space where the expected neutralino relic density is consistent with (\ref{wmap}) \cite{wilczek}. For parameters in this region, squark masses are in the multi-TeV range, and the reach of the LHC is determined by final states from gluino pair production: although the higgsino-like chargino may be light, the mass difference $m_{\tw_1}-m_{\tz_1}$ is small so that leptons from its decays are too soft to increase the reach beyond that obtained via the $\eslt$ signal from gluino pair production \cite{howiefocus}. Since the LSP couples
preferentially to the third family via its higgsino component, cascade decays of the gluino to third generation fermions tend to be enhanced. As a result, the requirement of a $b$-tagged jet in SUSY events reduces SM backgrounds and enhances the LHC reach by 15--20\% beyond the reach via the inclusive $\eslt$ channel in the HB/FP region of the mSUGRA model~\cite{MMT}. We should also mention that the $b$-jet multiplicity may also be enhanced in the mSUGRA model if third generation squarks happen to be light, either because of large bottom quark Yukawa couplings when $\tan\beta$ is large, or because the $A_t$ parameter happens to be just right so that $m_{\tst_1}\ll m_{\tq}$, and $\tst_1$ mainly decays via $\tst_1 \to b\tw_1 \ {\rm and} \  t\tz_1$, or $\tst_1 \to bW\tz_1$.\\

\subsection{Inverted mass hierarchy models} 

The evidence for neutrino oscillations \cite{neutrino} and its interpretation in terms of neutrino masses provides strong motivation for considering $SO(10)$ SUSY grand unified theories (GUTS) \cite{soten}. Each generation of matter (including the sterile neutrino) can be unified into a single ${\bf 16}$ dimensional representation of $SO(10)$ while the Higgs superfields $\hat{H}_u$ and $\hat{H_d}$ are both contained in a single ${\bf 10}$ dimensional representation, allowing for the unification of both gauge (and separately) Yukawa couplings.\\
 $SO(10)$ may either be directly broken to the SM gauge group, or by a two step process via an intermediate stage of $SU(5)$ unification. The spontaneous breakdown of $SO(10)$ with the concomitant reduction of rank leaves an imprint on the SSB masses which is captured by one additional parameter $M_D^2$ with a weak scale magnitude but which can take either sign \cite{dterm}. The model is then completely specified by the parameter set,
\be
m_{16}, m_{10}, m_{1/2}, M_D^2, \tan\beta, A_0 \ {\rm and} \ {\rm sign}(\mu)\;.
\ee
where we have assumed a common SSB mass parameter $m_{16}$ and a different parameter $m_{10}$ for matter and Higgs fields in the ${\bf 16}$ and ${\bf 10}$ dimensional representations, respectively. The GUT scale SSB masses for MSSM fields then take the form \cite{dterm},
$$ m_Q^2 = m_E^2 = m_U^2 = m_{16}^2 + M_D^2\;,$$ 
\be m_D^2 = m_L^2 = m_{16}^2 - 3 M_D^2\;, \ee
$$ m_{N}^2 = m_{16}^2 + 5 M_{D}^2\;, $$
$$m_{H_{u,d}}^2 = m_{10}^2 \mp 2 M_D^2\;. $$   
Unification of Yukawa couplings is possible for very large values of $\tan\beta$ \cite{raby,yukunif}. 
\\
The $SO(10)$ framework that we have just introduced naturally allows a phenomenologically interesting class of models in which the matter sfermion mass order is inverted with the order for the corresponding
fermions \cite{bagger}. Specifically, in models with Yukawa coupling unification, the choice
\be A_0^2 = 2m_{10}^2 = 4m_{16}^2 
\label{so10rel}
\ee 
for the SSB parameters serves to drive third generation sfermion mass parameters to sub-TeV values, leaving first and second generation scalars as heavy as 2--3~TeV. A positive value of $M_D^2 \lesssim
(m_{16}/3)^2$ is necessary to obtain radiative electroweak symmetry breaking \cite{imh}. The multi-TeV values of first and second generation scalar masses ameliorate the SUSY $CP$ and flavour problems without destroying the SUSY resolution of the gauge hierarchy problem, since the fields with substantial direct couplings to the Higgs sector (gauginos and third generation scalars) have masses below the TeV scale.  Because third generation sfermions are significantly lighter than their first/second generation cousins, we may expect that SUSY events are enriched in $b$- (and possibly $t$-) quark jets in this scenario.\\

\subsection{Non-Universal Higgs Mass Models} Within the mSUGRA model, if $m_0^2=m_{H_u}^2({\rm GUT})$ is smaller than or comparable to $m_{1/2}^2$,  $m_{H_u}^2$ runs to a large negative value at the weak scale. The minimization condition for the (tree level) Higgs scalar potential which reads
\be
\mu^2 =\frac{m_{H_d}^2-m_{H_u}^2\tan^2\beta}{\tan^2\beta-1} 
-\frac{M_Z^2}{2} \simeq -m_{H_u}^2 -\frac{M_Z^2}{2}
\label{lowmu}
\ee
(where the last approximation is valid for modest to large values of $\tan\beta$), then implies that 
$|\mu| > |M_{1,2}|$ so that the LSP is essentially a bino, while the heavier -inos are mainly higgsino-like. A way of avoiding this conclusion is to choose $m_{H_u}^2({\rm GUT})$ to be so  large that $m_{H_u}^2$ runs to small negative values at the weak scale. Within the mSUGRA model, this can only be realized by choosing $m_0 \gg m_{1/2}$ which gives us the well studied HB/FP region with MHDM discussed above. \\
A different way would be to relax the assumed universality \cite{NUHMgen} between the matter scalar and Higgs boson SSB mass parameters in what has been dubbed as non-universal Higgs mass (NUHM) models, and adopt a large value for $m_{H_u}^2({\rm GUT})$. In order to avoid unwanted flavour changing neutral currents, we maintain a universal value $m_0$ for matter scalars. The GUT scale value of the SSB down Higgs mass parameter may (may not) be equal to $m_{H_u}^2$ leading
to a one (two) parameter extension of the mSUGRA framework that we will refer to as the NUHM1 (NUHM2) model \cite{NUHM}. The NUHM1 model is thus completely specified by the mSUGRA parameter set together with $m_{\phi}={\rm sign}(m_{H_{u,d}}^2)\sqrt{|m_{H_{u,d}}^2|}$, {\it i.e.}
by,
\be m_0, m_{\phi}, m_{1/2}, A_0, \tan\beta \ {\rm and} \ {\rm sign}(\mu)
\ \ ({\rm NUHM1})\;.  
\ee 
If $m_\phi$ is chosen to be sufficiently larger than $m_0$, the parameter $m_{H_u}^2$ runs down to negative values but remains small in magnitude so that we obtain MHDM {\it for any value of
$m_0$ and $m_{1/2}$}.\footnote{Of course, if $m_{\phi}$ is chosen to be  too large then $m_{H_u}^2$ does not run to negative values and  electroweak symmetry breaking is no longer obtained.} Curiously, the NUHM1 model accommodates another possibility of getting agreement with (\ref{wmap}). If $m_{\phi} < 0$, $m_{H_u}^2$ and $m_{H_d}^2$ both run to large, negative values at the weak scale so that 
\be
m_A^2 = m_{H_u}^2 + m_{H_d}^2 +2\mu^2 \simeq m_{H_d}^2-m_{H_u}^2 - M_Z^2
\ee
may be small enough for neutralinos to annihilate via the $A$ and $H$ resonances. Within the NUHM1 framework, the Higgs funnel thus occurs for {\it all values of $\tan\beta$.} Since the Higgs bosons $A$ and $H$ with relatively small masses are expected to be produced via cascade decays of gluinos and squarks, and since these decay preferentially to third generation fermions,  we may once again expect an enhancement of the $b$- and, perhaps also, $t$-jet multiplicity. 
\\
The NUHM2 model requires two more parameters than the mSUGRA framework for its complete specification. While these may be taken to be the GUT scale values of $m_{H_u}^2$ and $m_{H_d}^2$, it is customary and more convenient to eliminate these in favour of $m_A$ and $\mu$, and work with the hybrid parameter set,
\be 
m_0, m_{1/2}, m_A, \mu, A_0, \tan\beta  \ \ ({\rm NUHM2})\;.  
\ee 
This then allows us to adjust the higgsino content of charginos and neutralinos at will, and furthermore allows as much freedom in the (tree-level) Higgs sector as in the unconstrained MSSM.

\subsection{Low ${\bf |M_3|}$ Dark Matter Model} Instead of relaxing the universality between scalar masses as in the NUHM model, we can also relax the universality between the gaugino mass parameters. If we adjust the GUT scale value of $M_1/M_2$ so that $M_1\simeq M_2$ at the weak
scale, we obtain mixed wino DM \cite{mwdm}. Since there is no principle that forces $M_1/M_2$ to be positive, we can instead adjust this ratio so that $M_1 \simeq -M_2$ at the weak scale. In this case the LSP remains a bino with charged and neutral winos close in mass to it and agreement with (\ref{wmap}) is obtained via bino-wino co-annihilation \cite{bwca}. Although collider signatures are indeed altered from mSUGRA expectations, we do not expect any enrichment of $b$-jet multiplicity in this case.
\\
Although not obvious, agreement with (\ref{wmap}) is also obtained if we maintain $M_1=M_2$ at $Q=M_{\rm GUT}$, but instead {\it reduce} the value of $|M_3|$. Specifically, for smaller values of $|M_3|$, the (top)-squark mass parameters and also $A_t^2$ are driven to smaller values at the weak scale. These smaller values of top-squark masses and of $A_t^2$, in turn, slow down the evolution of $m_{H_u}^2$ so that it runs to negative values more slowly than in the mSUGRA model. As a result, the weak scale value of $m_{H_u}^2$ though negative, has a smaller magnitude than in the mSUGRA case, so that the value of $\mu^2$ is correspondingly reduced [see Eq.~(\ref{lowmu})] and the LSP becomes
MHDM \cite{lm3dm}. This is referred to as the low $|M_3|$ DM (LM3DM) model, and the corresponding parameter space is given by,
\be
m_0, m_{1/2}, M_3, A_0, \tan\beta, {\rm sign}(\mu) \ \ ({\rm LM3DM})\;.
\ee
Here $m_{1/2}> 0$ denotes the GUT scale value of $M_1=M_2$, while $M_3$ (which is either positive or negative) denotes the corresponding value of $M_3$ at the GUT scale. For $m_0 \sim m_{1/2} \lesssim 1$~TeV, the GUT scale value of $|M_3|$ must be {\it reduced} from its mSUGRA value in
order to obtain MHDM as discussed above. In contrast, if we fix $m_{1/2}\simeq 1$~TeV, and take $m_0$ to be multi-TeV, MHDM is obtained for values $|M_3|/m_{1/2} > 1$. To simplify fine tuning issues, we will confine ourselves to $m_0 \lesssim 1$~TeV where we can obtain agreement with (\ref{wmap}) by reducing the value of $|M_3|$. We may expect an enhancement in the $b$-multiplicity from SUSY events at the LHC because of the enhanced higgsino content of the LSP.

\subsection{High ${\bf{M_2}}$ Dark Matter Model}

Very recently, it has been pointed out \cite{hm2dm} that raising the GUT scale value of $M_2$ from its unified value of $m_{1/2}$ to about (2.5--3)$m_{1/2}$ for $M_2 > 0$, or to between $-2$ and $-2.5$ times $m_{1/2}$ for $M_2< 0$, also leads to a small value of $|\mu|$, giving rise to a relic density in agreement with (\ref{wmap}).\\
The parameter space of this high $|M_2|$ dark matter (HM2DM) model is given by, 
\be
m_0, m_{1/2}, M_2, A_0, \tan\beta, {\rm sign}(\mu) \ \ ({\rm HM2DM})\;.
\ee
where $|M_2|$, the GUT scale value of the wino mass parameter, is dialled to large magnitudes to obtain MHDM. The large value of $|M_2|$ causes the Higgs SSB $m_{H_u}^2$ to initially increase from its GUT scale value of $m_0^2$ as $Q$ is reduced from $M_{\rm GUT}$. Ultimately, however, the usual top quark Yukawa coupling effects take over, causing $m_{H_u}^2$ to evolve to negative values resulting in the well-known radiative breaking of electroweak symmetry. However, because of its initial upward evolution, the weak scale value of $m_{H_u}^2$ is not as negative as in models with unified gaugino masses, and the value of $\mu^2$ is correspondingly smaller. The neutralino LSP then has a
significant higgsino component, and we may expect an enhancement of $b$-jets in SUSY events at the LHC.

\section{Event simulation and calculational details} \label{sec:sim}

We use ISAJET 7.74 \cite{isajet} with the toy calorimeter described in Chap.~\cite{chap:LHC} for the calculation of the SUSY signal as well as of SM backgrounds in the experimental environment of the LHC. We use parameters as described in Sec.~\ref{sec:detsim}  We conservatively take the tagging efficiency  $\epsilon_b=0.5$ at the LHC design luminosity of 100~fb$^{-1}$/y, and assume that gluon and light quark jets can be rejected as $b$ jets by a factor $R_b= 150$ (50) if $E_T < 100$~GeV ($E_T > 250$~GeV) and a linear interpolation in between \cite{brej}. For jets not tagged as a $b$-jet, we require $E_T(j) \ge 50$~GeV.\\

Gluino and squark production is the dominant sparticle production mechanism at the LHC for gluino and squark masses up to about 1.8~TeV, if $m_{\tq} \simeq m_{\tg}$. If instead squarks are very heavy, gluino pair production will dominate the sparticle production rate up to about $m_{\tg}\sim 0.8$~TeV. Cascade decays of the parent gluinos and squarks then lead to signals in various multi-jet plus multi-lepton plus $\eslt$ topologies \cite{cascade}. \\

Since SUSY particles are expected to be heavy (relative to SM particles) sparticle production is expected to be signalled by events with hard jets, possibly with hard, isolated leptons and large $\eslt$.  The dominant physics backgrounds to these events with hard jets come from $t\bar{t}$ production, $V + j$ production ($V=W, Z$), $VV$ production and QCD production of light jets, where the $\eslt$ comes from neutrinos produced by the decays of $W$ or $Z$ bosons or of heavy flavours. Missing $E_T$ may also arise from mismeasurement of jet or lepton transverse momenta and from uninstrumented regions of the detector. These non-physics sources of $\eslt$ are detector-dependent, and only qualitatively accounted for in our simulation with the toy calorimeter. With the hard cuts that we use to obtain the reach, we expect that the physics backgrounds will dominate the difficult-to-simulate detector-dependent backgrounds, and the results of our analyses of the SUSY reach will be reliable. This expectation
is indeed borne out since results of previous theoretical analyses of the SUSY reach \cite{bcpt,update} compare well with the projected reaches obtained by the CMS \cite{cms} and ATLAS \cite{atlas}
collaborations. The {\it gain in reach}, if any, that we obtain from $b$-jet tagging, should if anything be more reliable than the absolute value of the reach.\footnote{The absolute reach may also suffer from the
fact that SM backgrounds may be somewhat larger than those obtained using shower Monte-Carlo programs when proper matrix elements for jet production are included. We expect though that the gain in the reach from $b$-tagging may again be less sensitive to the inclusion of the proper matrix elements.}
\\
In the analysis detailed in the next chapter, we have examined the reach of the LHC for a wide range of sparticle masses, for the different models introduced in Sec.~\ref{sec:models}. To facilitate this, we generate
signals and backgrounds (calculational details are described below) and only write out events that include at least two jets with $E_T(j) \ge 100$~GeV and $\eslt \ge 100$~GeV, which we refer to as our basic cuts. The corresponding cross sections for SM events are shown in the second column of Table~\ref{tab:bkg}.  For low to medium values of sparticle masses, the sparticle production cross sections are large enough for us to extract the signal above SM backgrounds with relatively soft analysis cuts. For very heavy sparticles, however, the production rate is small, but essentially all events contain very energetic jets and large $\eslt$. The detection of the signal is then optimized by using very hard cuts that strongly suppress SM backgrounds while retaining bulk of the SUSY signal. Since our aim is to develop a strategy that can be applied to essentially the entire interesting mass range of a wide variety of models, we are led to evaluate the signal together with the SM background for a wide range of cuts, detailed in the next section. \\
To understand the relative importance of the different background sources, in the last three columns of Table~\ref{tab:bkg} we list the corresponding cross sections for the {\it softest} set of cuts that we use in our analysis detailed in Sec.~\ref{sec:btaggs}.
\begin{table}[htdp]
\begin{center}
\begin{tabular}{lcccc}
\hline\hline
Source & $\sigma_{\rm basic}$  & $\sigma_{\rm cut}(0b)$  
&$\sigma_{\rm cut}(1b)$  & $\sigma_{\rm cut}(2b)$\\ \hline
$t\bar{t}$ &19900 &2.16 & 1.41 & 0.365 \\
$W+j$ &21400 & 12.0 & 1.36 & 0.133 \\
$Z+j$ & 8850 & 5.11 & 0.059 & 0.0052 \\
$VV$  &89.8 & 0.0248 & 0.0020 & 0.0001 \\
QCD & 93700 &  11.6 & 3.11 & 0.467\\
Total &$1.44\times 10^5$ & 30.9 & 5.94 & 0.97 \\
mSUGRA1 & 261 & 12.0 &9.26 & 3.86\\
mSUGRA2 & 48.4 & 2.44 &1.95 & 0.87\\
\hline\hline
\end{tabular}
\end{center}
\caption{Cross sections in fb for the SM production of $t\bar{t}$,  $W+j$, $Z+j$, $VV$, and QCD jet events that form the dominant backgrounds. The second column gives the cross section for events with the basic requirements of two jets with $E_T(j) \ge 100$~GeV and $\eslt \ge 100$~GeV. The last three columns give the corresponding cross sections for the softest set of cuts listed in Table~\ref{tab:cuts1} and Table~\ref{tab:cuts2} that we actually use in our analysis, with no requirement of $b$-jet tagging (column 3), requiring at least one
tagged $b$-jet (column 4) and at least two tagged $b$-jets (column 5). For illustration, we also list the corresponding signal cross sections for two points in the HB/FP region of the mSUGRA model, with $A_0=0$, $\tan\beta=10$ and $m_{\tg}\simeq 1$~TeV, $m_{\tq}\sim 3$~TeV (mSUGRA1) and $m_{\tg}\simeq 1.5$~TeV and $m_{\tq}\sim 3.9$~TeV (mSUGRA2). }
\label{tab:bkg}
\end{table}
\\
In the last two rows we also list the corresponding signal cross sections for two WMAP-consistent cases in the HB/FP region of the mSUGRA model. Several comments are worth noting.
\begin{itemize}
\item We see that with the basic requirements of two jets with $E_T \ge 100$~GeV and $\eslt \ge 100$~GeV, the background is two (three) orders of magnitude larger than the signal  for $m_{\tg} \simeq 1$ (1.5)~TeV; however, the analysis cuts very efficiently reduce the background, while reducing the signal by a much smaller factor. 
\item After these analysis cuts we see that QCD, followed by $V+j$ production, are the leading backgrounds to the inclusive $\eslt$ signal. Top pair production, while significant, is considerably
smaller. Since we do not require the presence of leptons, the background from $VV$ production is negligible.

\item The backgrounds from QCD and $V+j$ production can be sharply reduced by the use of $b$-jet tagging with relatively small loss of the signal. In contrast, since top events necessarily contain $b$-jets, $b$-tagging reduces the $t\bar{t}$ background only by a modest amount.
\end{itemize}

Table~\ref{tab:bkg} highlights the importance of a careful evaluation of the QCD and the $V+j$ backgrounds. This is technically complicated because the large size of the cross sections necessitates simulations of very large number of events to obtain a reliable estimate for the backgrounds after the very hard cuts that are needed for optimizing the reach of the LHC.\footnote{Of course, the fact that we are far into the tails of these backgrounds where the simulations (which will be tuned to data when these become available) require possibly unjustified extrapolations is a different matter.} Moreover, since the cross section is a rapidly falling function of the centre of mass energy, or equivalently, the hard scattering $p_T$ of the initial partons, we must ensure that our procedure generates events even for very large values of $P_T^{HS}$ where the matrix element is very small, resulting in a much smaller weight. To facilitate this, we have generated the various backgrounds using different numbers, $N_i^{HS}$, of hard scattering bins: the bin intervals are finely spaced for low values of $P_T^{HS}$ where event weights are very large. We choose $N_i^{HS}=53, 13, 8$ and 7 for $i =$ QCD, $V+j$, $t\bar{t}$ and $VV$, respectively, where the choice $N_{HS}^{\rm QCD} = 53$ reflects the largeness of the QCD cross
section. We have generated a total of about 10M QCD events, about 1M $W+j$ events and about 500K-700K events for each of the other backgrounds. If, for any set of cuts, we find zero events in our simulation of a particular background, we set this background cross section to a value corresponding to the one event level in the bin with the smallest weight in our simulation.\\
\section{Bottom jet tagging and the reach of the LHC} \label{sec:btaggs}

\subsection{Simulation of the Signal and the LHC reach} 
Simulation of the signal events is technically much easier than that of the background. This is largely because the signal typically originates  in heavy sparticles, and so passes the hard analysis cuts with relative ease compared to the background. To assess how much $b$-jet tagging extends the SUSY reach of any particular model, rather than perform extensive and time-consuming scans of the parameter space, we have defined ``model lines'' along which the sparticle mass scale increases.  We then choose parameters along these lines, and for every such parameter set use ISAJET 7.74 to generate a SUSY event sample. Next, we pass this event sample through the set of analysis cuts defined below, and define the signal to be observable at the LHC if for {\it any}
choice of cuts 
\bi
\item the signal exceeds 10 events, assuming an integrated luminosity of  100~fb$^{-1}$, 
\item the statistical significance of the signal $N_{\rm  signal}/\sqrt{N_{\rm back}}\ge 5$,
  and 
 \item the signal to background ratio, $N_{\rm  signal}/N_{\rm back} \ge 0.25$.  
\ei 
We also require a minimum of 15 events after cuts in our simulation of the signal. We obtain the reach for each model line by comparing the corresponding signal with the background, and ascertaining where the signal just fails our observability criteria for {\it the entire set of
cuts in Table~\ref{tab:cuts1} and Table~\ref{tab:cuts2}}. 
An important part of our work involves tagging of b-jets at the LHC, using both the presence or complete absence of these tagged jets to achieve our results. \\

\subsection{Analysis cuts} \label{sec:cuts}

The inverted mass hierarchy model based on $SO(10)$ SUSY GUTs, whose hallmark is the light third generation, serves as the prototypical case  where we expect enhanced $b$-jet multiplicity in SUSY events. We have used this framework to guide us to the set of analysis cuts that can be used for the optimization of the SUSY signal for a wide range of sparticle masses in a wide class of models. Toward this end, we fix $\mu < 0$, $A_0<0$, and $\tan\beta=47$ (a large value is needed for the unification of Yukawa couplings) and choose $m_{10} =\sqrt{2}m_{16}$, $A_0=-2m_{16}$ to
obtain the hierarchy between the first/second and third generation scalars as discussed above. The choice $M_D = 0.25m_{16}$ facilitates electroweak symmetry breaking. We vary the gluino mass along the ``model line'' with $m_{1/2}=0.36m_0+48$~GeV which maintains a hierarchy between
the generations.\\
The value of
$$ S \equiv \frac{3(m_{\tu_L}^2 + m_{\td_L}^2 + m_{\tu_R}^2 +
m_{\td_R}^2) + m_{\te_L}^2 + m_{\te_R}^2 + m_{\tnu_e}^2}{3(m_{\tst_1}^2
+ m_{\tb_1}^2 + m_{\tst_2}^2 + m_{\tb_2}^2) + m_{\ttau_1}^2 +
m_{\ttau_2}^2 + m_{\tnu_\tau}^2} 
$$
is typically around 3.5-4.1 along this model line. \\
The optimal choice of cuts depends on the ({\it a priori} unknown) sparticle spectrum, and to a smaller extent on their decay patterns. While hard cuts optimize the signal if sparticles are heavy, these would drastically reduce (or even eliminate) the signal if sparticles happen to be light. In order to obtain a general strategy that can be used for a wide variety of models, we have used the $SO(10)$ model with $\mu < 0$ to define a universal set of cuts that can be used for SUSY discovery in any of the various models that we have introduced, and likely, also for a wider class of models.\\
Toward this end, we generate a sample of signal events for this ``test model line'' and run this, as well as the SM backgrounds that we discussed above, through each one of the large set ({\it i.e.} the complete set that includes the numbers listed in the parentheses) of analysis cuts detailed in the nine rows  in both Table~\ref{tab:cuts1} and Table~\ref{tab:cuts2}. Here, $m_{\rm eff}$ is the scalar sum of the transverse energies of the four hardest jets in the event combined with the missing transverse energy, $\Delta\phi$ is the transverse plane opening angle between the two hardest jets, and $\Delta\phi_b$ the corresponding angle between the two tagged $b$-jets in events with
$n_b\ge 2$. To clarify, the softest set of cuts that we use for the $0b$ signal has [$\eslt$, $E_T(j_1)$, $E_T(j_2)$, $E_T(b_1)$, $m_{\rm eff}$] $\ge (300, 300, 200, 40, 1500)$~GeV, $n_j \ge 4$ and transverse sphericity $S_T >0.1$, with no restriction on jet opening angles. Note that because $E_T(j_1)> E_T(j_2)$, there are 21 combinations for the minimum values of  [$E_T(j_1), E_T(j_2)$] that we have used. Next, we harden the cut on one of these observables to the next level, keeping the others at the same value, {\it etc.} until the complete set of $6\times 21\times 5\times 6\times 3\times 3\times 5\times 2$ combinations has been examined for $n_b\ge 2$. Since there are (is) no (just one) tagged $b$ jets in the $n_b=0$ ($n_b=1$) case, there are correspondingly fewer combinations for these analyses. 
%
\\

\begin{table}[htdp]
\begin{center}
\begin{tabular}{lc} \hline \hline
 Variable  & $0b$, $1b$  \\  \hline
$\eslt$~(GeV) $>$ & $(300), 450,..., 900, (1050)$  \\ $E_T(j_1)$~(GeV) $>$ & $300, 400,..., 
800$  \\ $E_T(j_2)$~(GeV) $>$ & $200, 300,..., E_T^{\rm
min}(j_1)-100$~GeV  \\
$E_T(b_1)$~(GeV) $>$ & $40, 100, 200, (300, 400)$  \\
$m_{\rm eff}$~(GeV) $>$ & $1500, 2000, 2500,..., 4000$ \\ 
$\Delta\phi < $ & $180^\circ$, $160^\circ$,
$140^\circ$  \\
$\Delta\phi_{b}< $ & N/A  \\
$n_j\ge$ & $4, 5,..., 8$   \\ 
$S_T\ge$ & $0.1, (0.2)$ \\ \hline 
\hline \hline
\end{tabular}
\end{center}
\caption{The complete set of cuts examined for extraction of the SUSY signal over the SM backgrounds. The $0b$ and $1b$ entries  denote requirements for events without any restriction $b$-jet tagging, or with at least one tagged $b$-jet. }
\label{tab:cuts1}
\end{table}
\begin{table}[htdp]
\begin{center}
\begin{tabular}{lc} \hline \hline
 Variable  & $2b$ \\  \hline
$\eslt$~(GeV) $>$  &
$(300), 450, 600, 750, (900, 1050)$ \\ 
$E_T(j_1)$~(GeV) $>$  & $300, 400,..., 800$ \\
 $E_T(j_2)$~(GeV) $>$  & $200, 300,..., E_T^{\rm min}(j_1)-100$~GeV \\
$E_T(b_1)$~(GeV) $>$  & $40, 100, 200, 300, (400)$\\
$m_{\rm eff}$~(GeV) $>$ &$1500, 1750, 2000, 2250, (2500, 2750)$\\ 
$\Delta\phi < $  & $180^\circ$, ($160^\circ$, $140^\circ$) \\
$\Delta\phi_{b}< $  & $180^\circ$, $150^\circ$, $120^\circ$\\
$n_j\ge$ & $4, 5,..., 7, (8)$ \\ 
$S_T\ge$ & $0.1, (0.2)$\\ \hline 
\hline \hline
\end{tabular}
\end{center}
\caption{The complete set of cuts examined for extraction of the SUSY signal over the SM backgrounds. The $2b$ entries  denote requirements for events  with at least two tagged $b$-jets.
For the final analysis of the reach in the various models, we dropped the cut values within the parenthesis, and replaced the 21 combinations for the minimum values of [$E_T(j_1),E_T(j_2)$] with the 11 combinations in the last two rows (below the horizontal lines) in the table. }
\label{tab:cuts2}
\end{table}
\begin{table}[htdp]
\begin{center}
\begin{tabular}{lcc} \hline \hline
$[E_T(j_1), E_T(j_2)]$~(GeV) $>$ &
\multicolumn{2}{c} 
{$(300, 200), (400, 200),(500, 200), (500, 300),$}\\
&\multicolumn{2}{c} {$(500, 400), (600, 200), (600, 500),(700, 300),$}\\
& \multicolumn{2}{c}{$(700, 600),(800, 300), (800,600) $}\\
\hline \hline
\end{tabular}
\end{center}
\caption{For the final analysis of the reach in the various models, we dropped the cut values within the parenthesis in Table~\ref{tab:cuts1} and Table~\ref{tab:cuts2}, and replaced the 21 combinations for the minimum values of [$E_T(j_1),E_T(j_2)$] with the 11 combinations in the table above. }
\label{tab:cuts3}
\end{table}

For each of these cut choices, we analysed the observability and statistical significance of the LHC signal for our test $SO(10)$ model line for an integrated luminosity of 100~fb$^{-1}$. We found that a subset of cuts was sufficient to ensure the observability of the SUSY signal over the entire mass range. Specifically, restricting the minimum values of the transverse energies of the two hardest jets to the eleven combinations shown in Table~\ref{tab:cuts3}, and dropping the cut values shown in parenthesis for the other variables in both Table~\ref{tab:cuts1} and Table~\ref{tab:cuts2}  had no impact upon the observability (and the statistical significance) of the signal over the entire sparticle mass range.\\

In the remainder of this project we, therefore, confine ourselves to this limited subset of cuts, as this speeds up the analysis considerably.\\

\section{\textbf{Results for LHC reach using $b$-jet tagging}}\label{sec:relts}
In this section, we evaluate prospects for increasing the reach of the LHC by the use of $b$-tagging to reduce SM backgrounds, thereby increasing the statistical significance of the SUSY signal, for each of the models introduced in Sec.~\ref{sec:models}. We confine ourselves to various 1-parameter model lines (introduced below) along which sparticle masses increase and run the signal and backgrounds through each of the final set of cuts in Table~\ref{tab:cuts1} and Table~\ref{tab:cuts2}, and optimize the signal by
selecting the cut choice that yields an observable signal with the highest statistical significance.  To assess the gain from $b$-tagging, for each model line we first do so without any requirement on
$b$-tagging, and then repeat it requiring, in addition, at least one and at least two tagged $b$-jets.

\subsection{The HB/FP region of the mSUGRA model} 

The possibility of increasing the LHC reach was first studied in the HB/FP region of the mSUGRA framework \cite{MMT}, where it was found that the reach could be increased by up to 15-20\%. We have repeated this study, albeit with a somewhat different model line with $$m_{1/2}= 0.295 m_0 - 507.5~{\rm GeV}, \tan\beta=30, A_0=0,$$ in the HB/FP region that saturates the relic density in (\ref{wmap}) and of course, with the different set of cuts that we use here. We  find an increased reach from $b$-tagging in qualitative agreement with Ref.~\cite{MMT}. 

\subsection{Inverted mass hierarchy model} \label{imhmodel}

As discussed in Sec.~\ref{sec:cuts}, we have already used the $SO(10)$ model with $\mu <0$ and parameters related by (\ref{so10rel}) where we obtain an  inverted mass hierarchy to choose the final set of cuts for our analysis. Here, we show results for the reach of the LHC with and without requirements of $b$-jet tagging for two model lines with a significant inversion of the sfermion mass hierarchy, one for each sign of $\mu$. For both of these, we choose
\begin{equation}
-A_0=2m_{16}=\sqrt{2}m_{10}, \tan\beta=47, \label{above} \end{equation} 
with
\begin{eqnarray}
M_{D} = 0.25m_{16} \ {\rm and} \ m_{1/2} = 0.36m_{16} + 48~{\rm GeV}
\ {\rm for} \ 
\mu <0, \label{imhneg}  \\
M_{D} = 0.20m_{16}  \ {\rm and} \ m_{1/2} = 0.30m_{16} + 39~{\rm GeV} \
{\rm for} \
\mu >0. \label{imhpos}
\end{eqnarray}

\begin{figure}[ht]
\begin{center}
\includegraphics[width=5cm]{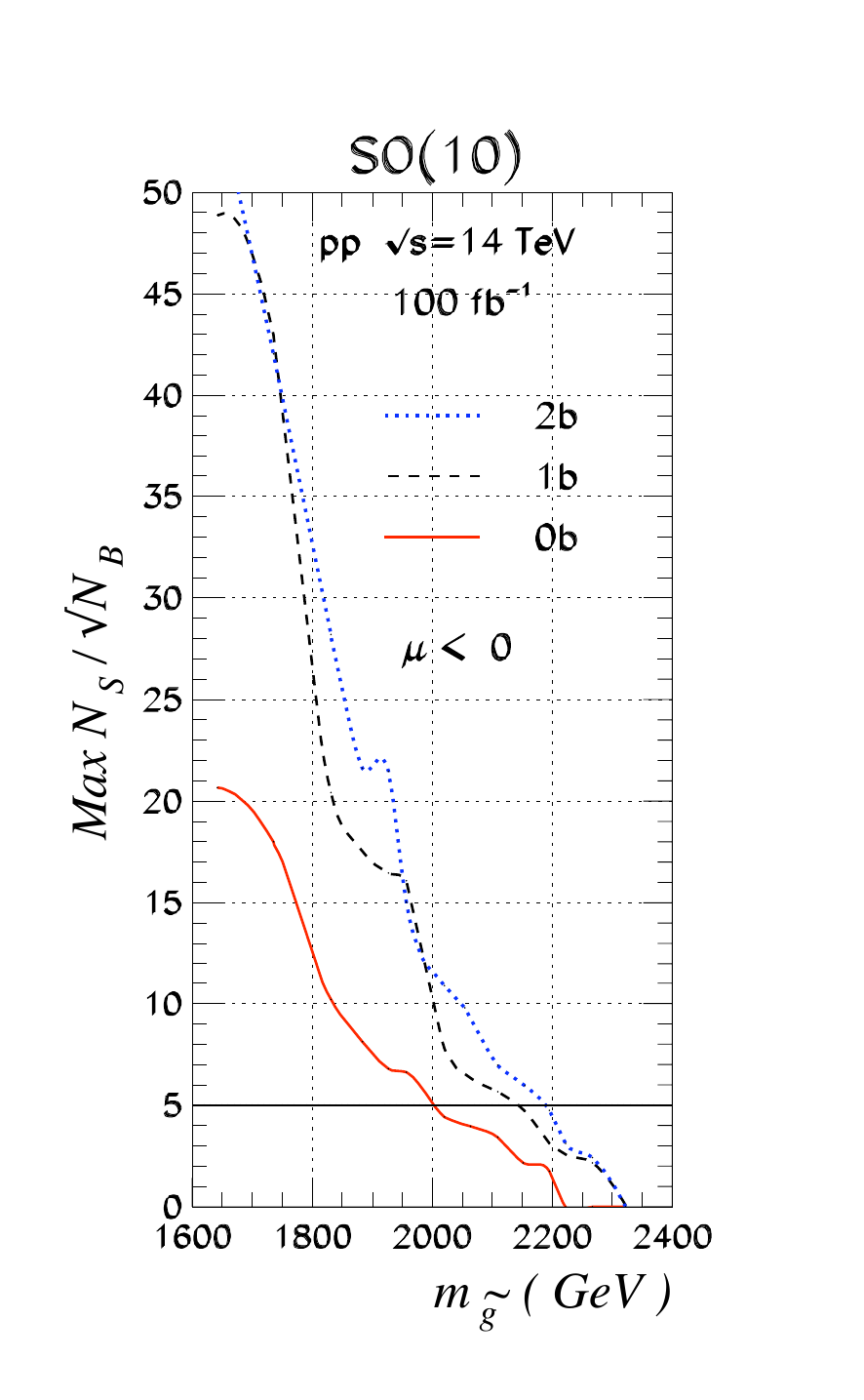}
\includegraphics[width=5cm]{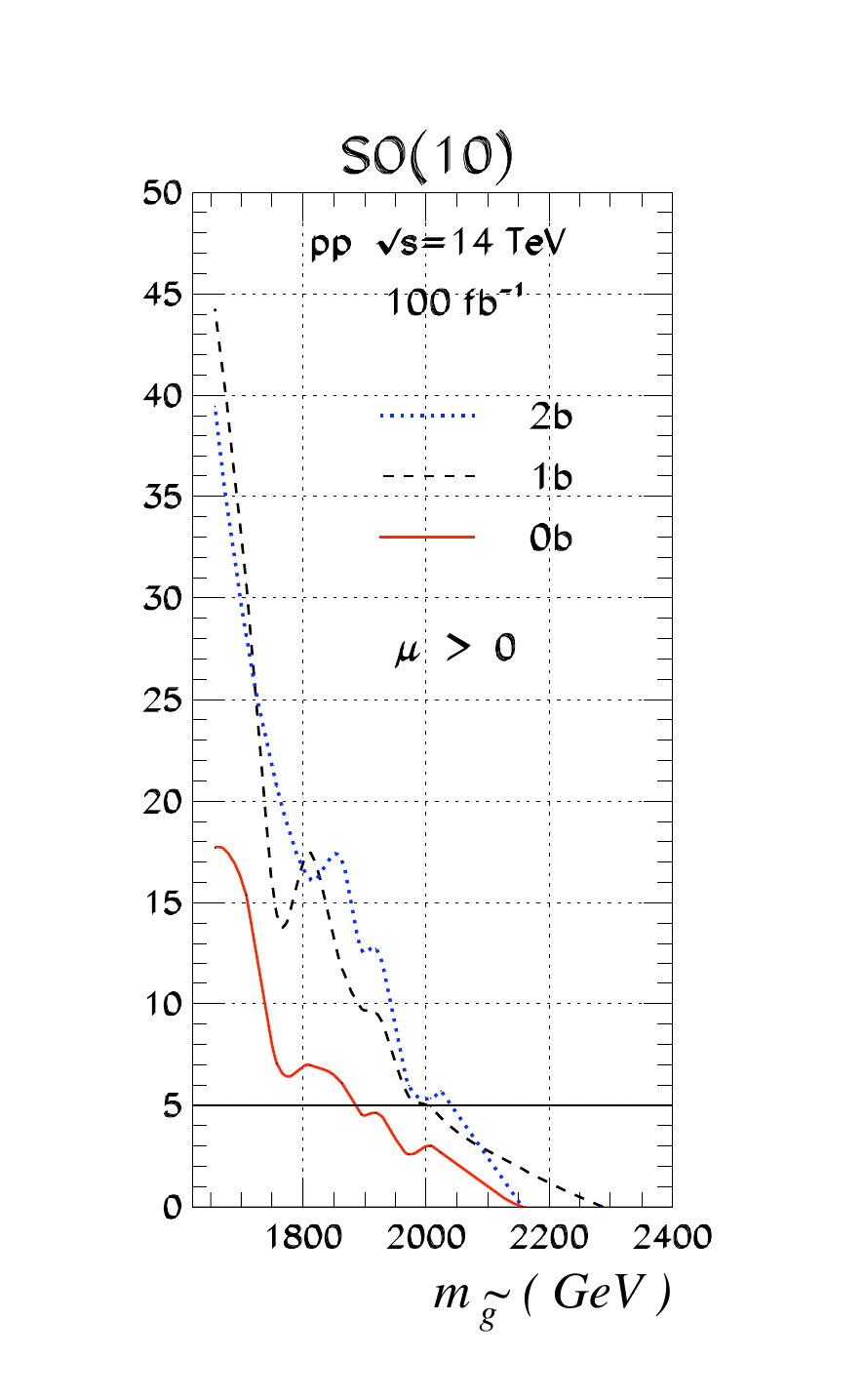}
 \caption{ \label{reach:so10}The statistical significance  for the  inverted hierarchy $SO(10)$ model lines introduced in the text, assuming an  integrated luminosity of 100~fb$^{-1}$ for  ({\it a})~$\mu <0$, and ({\it b})~$\mu>0$. The solid (red) line is for the signal with no  requirement on $b$-tagging, the dashed (black) line is with the  requirement of at least one tagged $b$-jet, and the dotted (blue) line is with at least two tagged $b$-jets. The signal is observable if the  statistical significance is above the horizontal line at $N_{\rm  signal}/\sqrt{N_{\rm back}}=5$.  }
 \end{center}
 \end{figure}

Our results are shown in Fig.~\ref{reach:so10}, where we plot the largest statistical significance of the signal, $N_{\rm signal}/\sqrt{N_{\rm back}}$, versus the corresponding gluino mass for ({\it a})~$\mu <0$, and ({\it b})~$\mu >0$, assuming an integrated luminosity of 100~fb$^{-1}$. The maximal
$N_{\rm signal}/\sqrt{N_{\rm back}}$ was obtained running over all the cuts in Table~\ref{tab:cuts1} and Table~\ref{tab:cuts2},  subjecting it to the requirement that the $N_{\rm signal}/N_{\rm back}>0.25$ and $N_{\rm signal}>10$  event criteria are satisfied. The solid (red) curves show this significance for the inclusive $\eslt$ signal with no requirement of $b$-jet tagging, while the dashed (black) curve and the dotted (blue) curves correspond to cases where we require at least one and two tagged $b$-jets, respectively. The wiggles in these curves reflect the statistical errors in our simulation.\\ 
We attribute the somewhat larger reach in the left frame to the fact that the mass hierarchy (as measured by the value of $S$) is somewhat smaller for $\mu<0$, so that $\tq\tg$ makes a larger contribution in this case.  We also see that for $\mu <0$, $b$-tagging leads to an increase of the LHC reach by $\sim 200$~GeV, or about 10\%, while the corresponding increase is somewhat
smaller for the model line with positive $\mu$. \\
This difference (which may well not be very significant in view of the wiggles) is evidently due to the increased reach in the $2b$ channel, and could arise from a complicated interplay between the effect of cuts and the sparticle spectrum: for instance, for $m_{\tg}\sim 1960$~GeV, $m_{\tb_1}$ is significantly lighter in the $\mu<0$ case, while $m_{\tst_1}$ is considerably heavier. As a result, the branching fraction for the decays $\tg \to b\tb_i$, which likely leads to a harder spectrum for $b$-jets (compared to $\tg \to t\tst_1$, which constitutes the bulk of the remaining decays of the gluino), falls from 38\% for negative $\mu$ to 28\% for positive $\mu$.\\

\subsubsection{Non-universal Higgs mass models} 

Next, we turn to the impact of $b$-tagging on the reach in NUHM models with just one additional parameter $m_\phi$ that is adjusted so that agreement with the observed relic density is obtained either by tempering the LSP content so that it is MHDM ($m_\phi>m_0$), or by adjusting the masses
so that the LSP annihilation rate is resonantly enhanced by the exchange of neutral $A$ or $H$ bosons in the $s$-channel ($m_\phi<0$). We did not study the NUHM model where both Higgs SSB mass parameters are arbitrary -- the so-called NUHM2 models in the nomenclature of
Ref.~\cite{NUHM} -- because this meant that both $m_A$  and $\mu$ are arbitrary, resulting in too much freedom for definitive analysis. Beginning with the MHDM cases of the LSP where sparticle decays to third generation quarks are enhanced by the higgsino content of the LSP, we introduce two model lines with $A_0=0$, $\tan\beta=10$ and $\mu>0$, with (1)~$m_0= m_{1/2}$, and (2)~$m_0=3m_{1/2}$. In the former case, the squarks of the first two generations are roughly degenerate with gluinos, whereas in the latter case $m_{\tq}\sim 1.6 m_{\tg}$. \\
\begin{figure}[ht]
\begin{center}
\includegraphics[width=5cm]{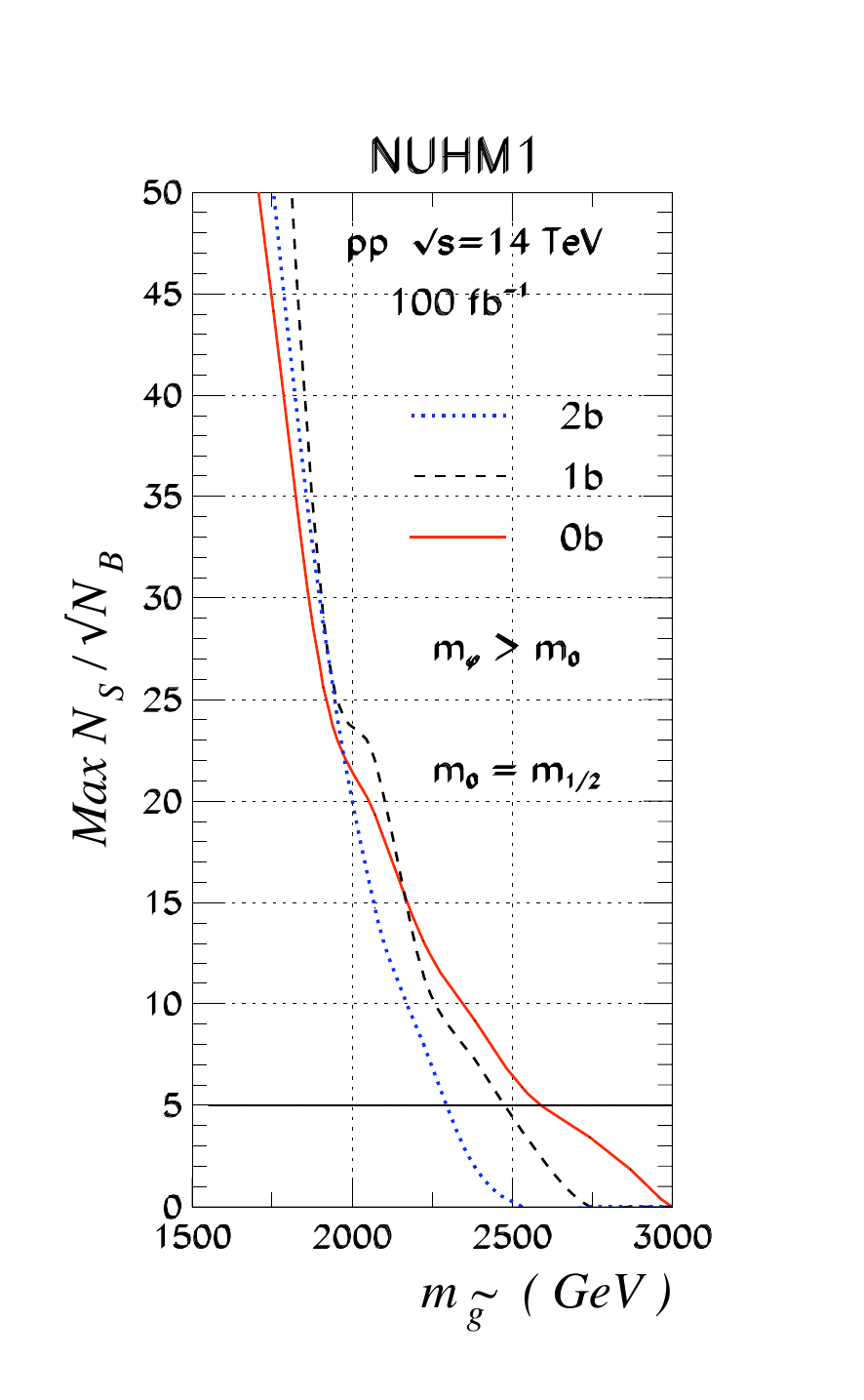}
\includegraphics[width=5cm]{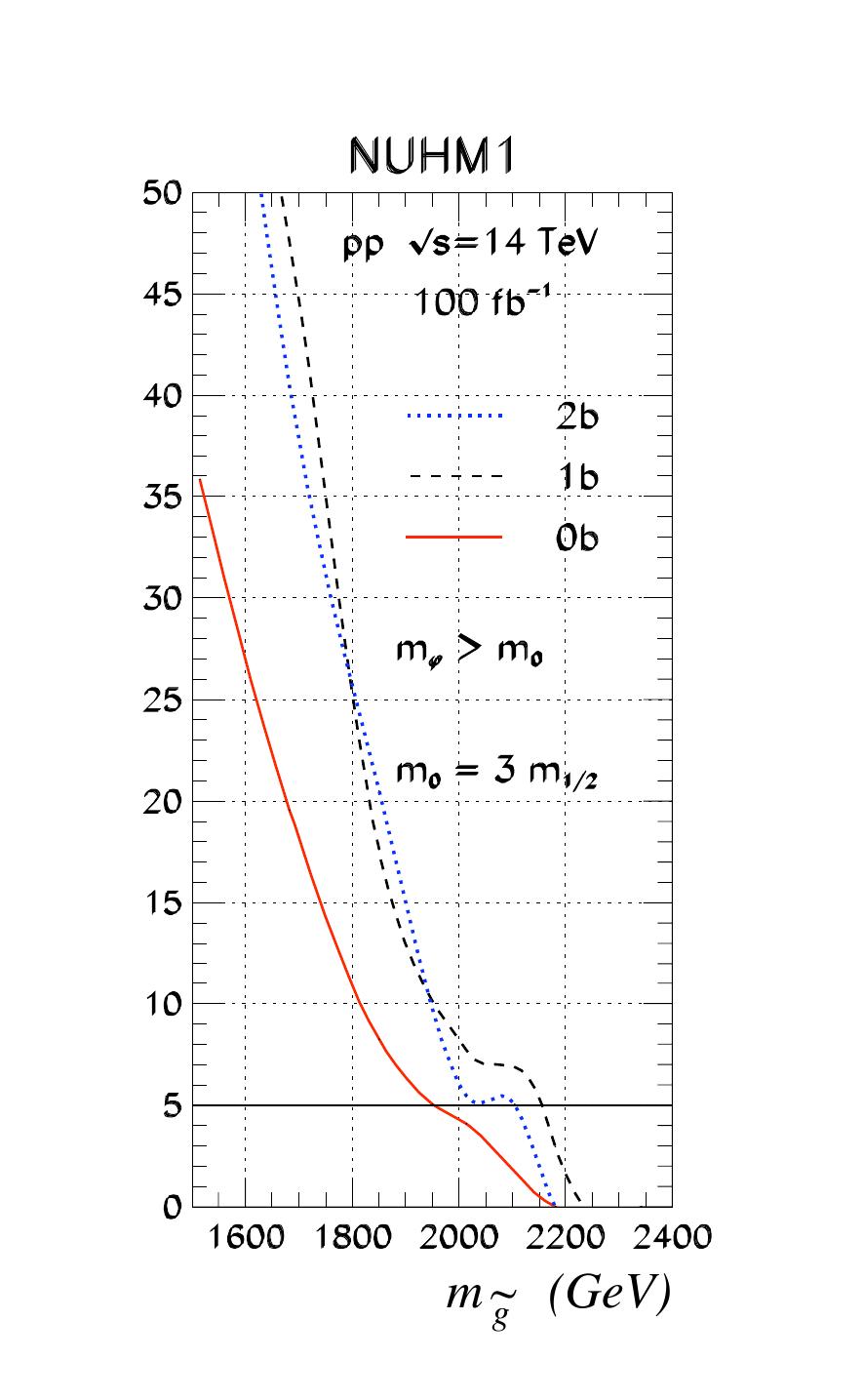}
\includegraphics[width=5cm]{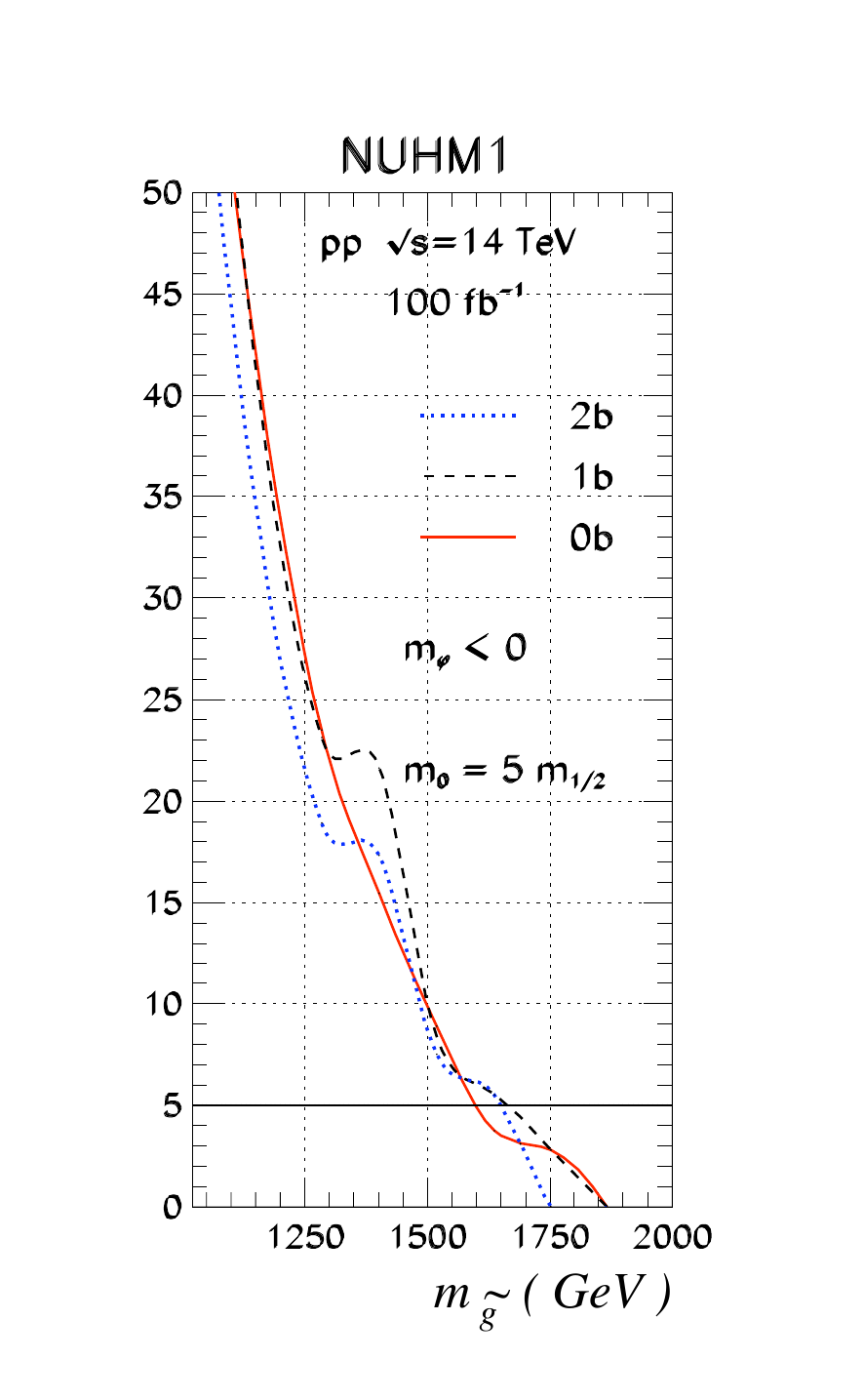}
 \caption{ \label{reach:nuhm}The statistical significance  for the  three NUHM  model lines introduced in the text,
  assuming an integrated luminosity of 100~fb$^{-1}$. All the model  lines have $A_0=0$ and $\mu>0$,  with  ({\it  a})~$m_\phi >0, \tan\beta=10, m_0=m_{1/2}$,  ({\it b})~$m_\phi >0, \tan\beta=10, m_0=3m_{1/2}$,  and ({\it c})~$m_\phi <0, \tan\beta=20, m_0=5m_{1/2}$. The solid (red)  line is for the signal with no requirement on $b$-tagging, the dashed  (black) line is with the requirement of at least one tagged $b$-jet,  and the dotted (blue) line is with at least two tagged $b$-jets. The signal is observable if the statistical significance is above the  horizontal line at $N_{\rm signal}/\sqrt{N_{\rm back}}=5$.  }
 \end{center}
 \end{figure}

Our results for the statistical significance of the LHC SUSY signal, with and without $b$-jet tagging are shown in Fig.~\ref{reach:nuhm} for ({\it a})~$m_0=m_{1/2}$, and ({\it b})~$m_0=3m_{1/2}$. We see that while $b$-tagging clearly improves the reach by $\sim 10\%$ in the case shown in frame ({\it b}), it leads to a {\it degradation} of the reach in frame ({\it a}). \\
We have traced this to the fact that for this case where squark and gluino masses are comparable, squark production (particularly first generation squark production) makes a significant contribution to the signal after the hard cuts. Then, since unlike gluinos which decay ``democratically'', these squarks decay to charginos and neutralinos (remember that because $m_{\tq}\sim m_{\tg}$, the decay $\tq \to q\tg$ is suppressed by phase-space) plus  {\it quarks of their own generation}, a sizeable fraction of the inclusive $\eslt$ signal is actually cut out by any $b$-tagging requirement.\\
In frame ({\it b}), the squarks are much heavier than gluinos and so contribute a smaller fraction of the signal, but more relevantly, $\tq \to q\tg$ with a large branching fraction, so that $b$-tagging helps in this case. These considerations also explain why the increase in reach from $b$-tagging is not as large as in the case of the HB/FP region of the mSUGRA model where $m_{\tq} \gg m_{\tg}$ \cite{MMT}. \\
We now turn to the $m_\phi <0$ model line shown in Fig.~\ref{reach:nuhm}{\it c} for which we have chosen $m_0=5m_{1/2}$ (to ensure squark contributions to the signal do not dilute the effect of
$b$ tagging as in the case that we just discussed), $A_0=0$, $\tan\beta=20$ and $\mu>0$, and $m_{\phi}<0$ is adjusted to give agreement with (\ref{wmap}) via resonant annihilation of LSPs through $A/H$ exchanges in the $s$-channel. This means that $A$ and $H$ must be relatively light and accessible in cascade decays of gluinos and squarks.  However, we see {\it no enhancement} of the LHC reach in this case.\\
We understand this in hindsight. In this case $|\mu|$ is large so the lighter neutralinos produced in gluino cascade decays are gaugino-like, with $m_{\tw_1}\simeq m_{\tz_2}\simeq 2m_{\tz_1}$. Then the very condition $2m_{\tz_1}\sim m_A$ that makes the LSP anihilation cross section resonant suppresses the phase space for the decays of $\tz_2 \to A \ {\rm or} \ H + \tz_1$, so that these are not significantly produced in cascade decays of gluinos. Since squarks are very heavy, they are essentially irrelevant to this discussion.\\

\subsubsection{Low ${\bf M_3}$ dark matter model} 

As explained above, we can also obtain MHDM, and hence a potential increase in reach via $b$-tagging, in models with non-universal gaugino
\begin{figure}[ht]
\begin{center}
\includegraphics[width=9cm]{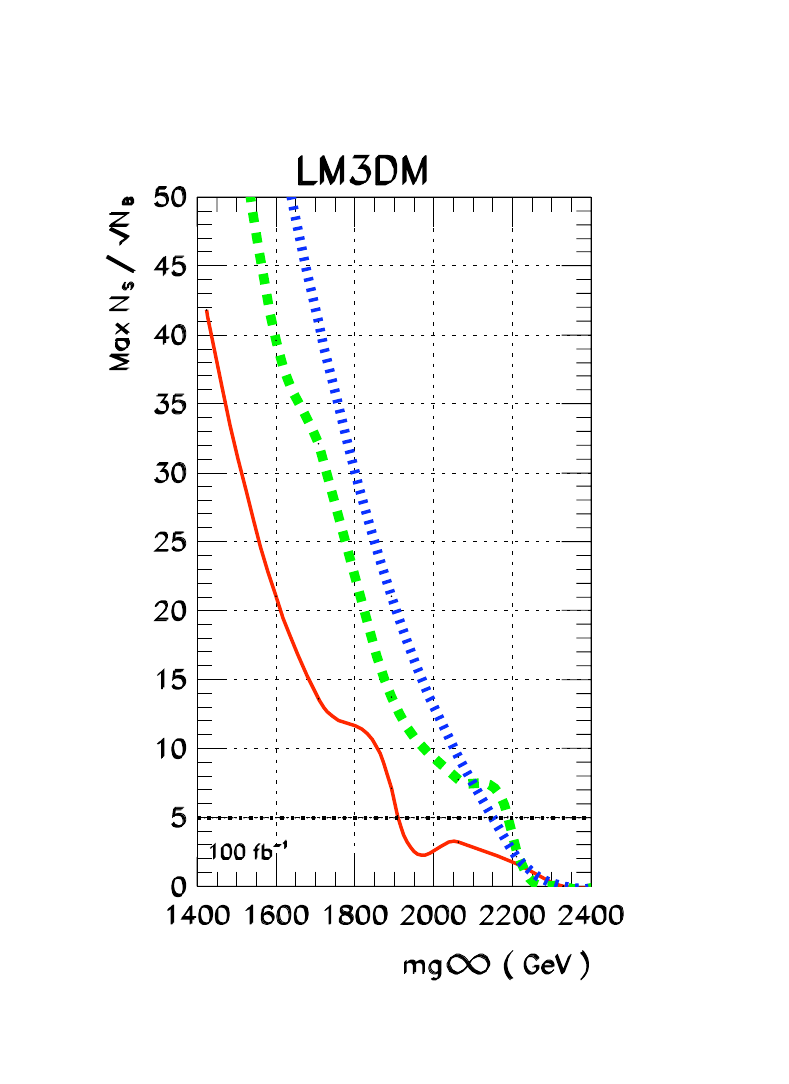}
 \caption{ \label{reach:lm3dm}The statistical significance for the  LM3DM  model line with $m_0=m_{1/2}, A_0=0, \tan\beta=10$ and  $\mu>0$, where $M_3({\rm GUT})$ is adjusted to saturate the measured CDM relic density,  assuming an integrated luminosity of 100~fb$^{-1}$. The solid (red)  line is for the signal with no requirement on $b$-tagging, the dashed  (black) line is with the requirement of at least one tagged $b$-jet,  and the dotted (blue) line is with at least two tagged $b$-jets. The  signal is observable if the statistical significance is above the  horizontal line at $N_{\rm signal}/\sqrt{N_{\rm back}}=5$. } 
 \end{center}
 \end{figure}
mass parameters where $|M_3({\rm GUT})|$ is taken to be reduced compared to its value in models with gaugino mass unification. To study the gain in the reach that we may obtain in this case, we have explored an LM3DM model line with  $$m_0=m_{1/2}, A_0=0, \tan\beta=10, \mu >0, $$ where the GUT scale value of $M_3$ (which we take to be positive) is adjusted to saturate the measured CDM relic density.\footnote{Roughly   speaking, for $m_0=m_{1/2}=700$~GeV, $M_3({\rm GUT})=277$~GeV, and for   an increase of $\delta m_0$ in $m_0=m_{1/2}$, the GUT scale value of   $M_3$ has to be raised by about $\delta M_3 \sim \delta m_0/2.25$. } The corresponding dependence of the statistical significance of the SUSY signal on $m_{\tg}$ is shown in Fig.~\ref{reach:lm3dm}. We see that in this case $b$-tagging leads to an increase in reach close to 15\%. This is because though gluinos and squarks are both reduced in mass relative to their uncoloured cousins, the reduced value of the gluino mass parameter leads to $m_{\tq}\sim (1.4-1.5)m_{\tg}$ even for $m_0=m_{1/2}$, to be compared to $m_{\tq}\sim m_{\tg}$ that we obtained for models with unified  gaugino masses as {\it e.g.} in  the NUHM case just discussed. The large value of $m_{\tq}$ relative to $m_{\tg}$ then leads to an enhanced reach via $b$-tagging just as before.  \\
\begin{figure}[ht]
\begin{center}
\includegraphics[width=5cm]{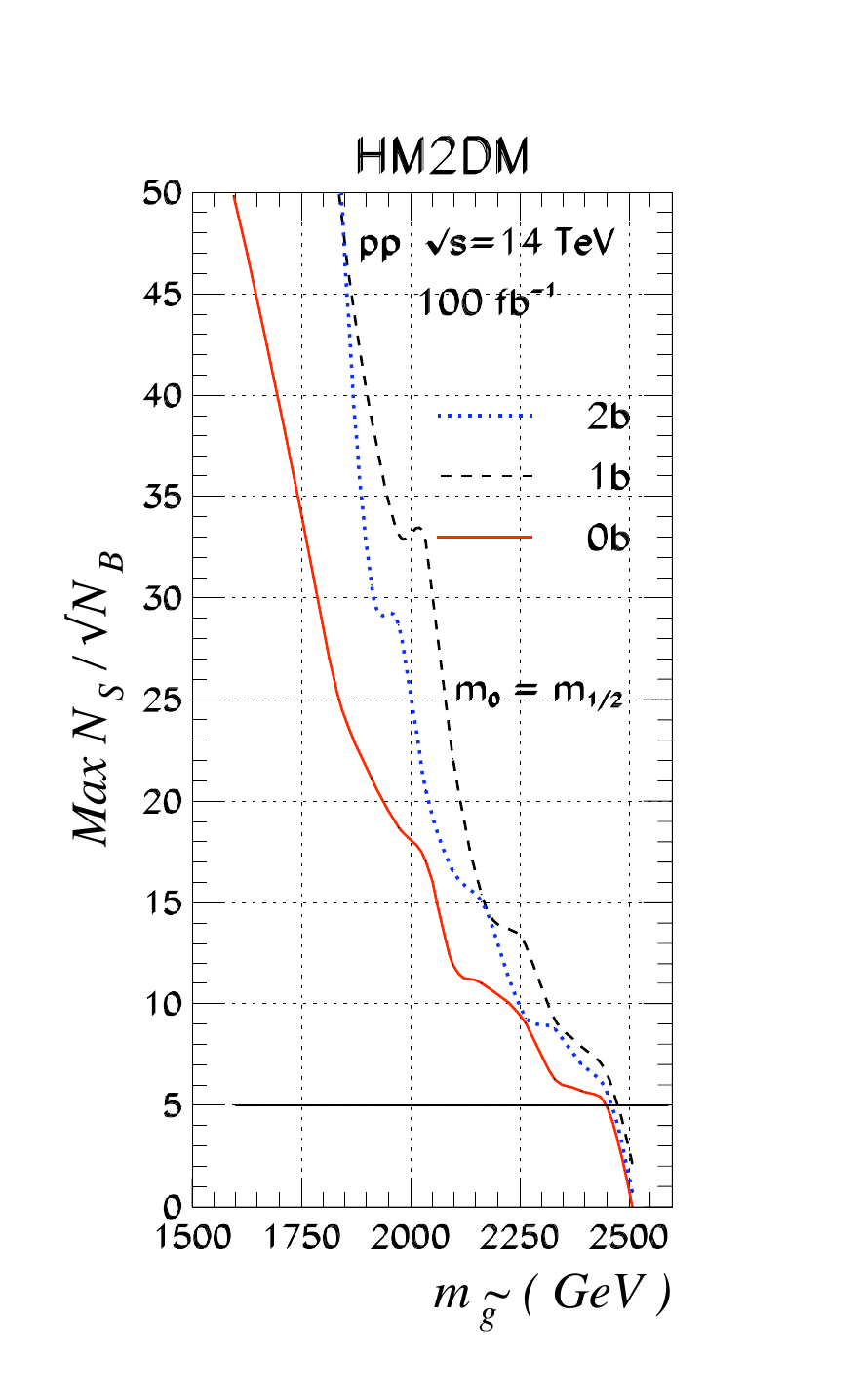}
\includegraphics[width=5cm]{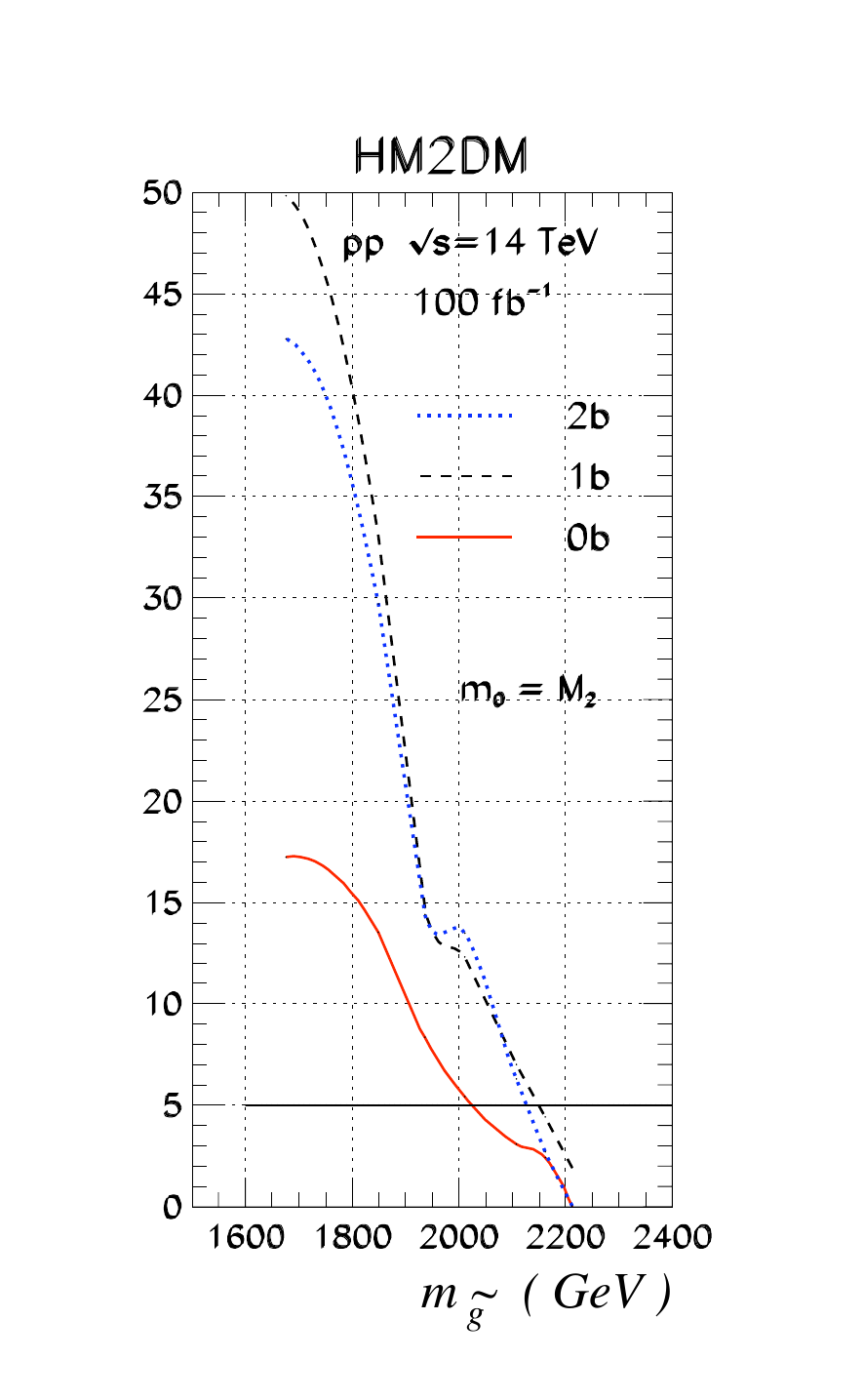}
 \caption{ \label{reach:hm2dm}The statistical significance  for the  HM2DM  model line with $A_0=0, \tan\beta=10$,
  $\mu>0$ and ({\it a})~$m_0=m_{1/2}$, and ({\it b})~$m_0=M_2({\rm  GUT})$. In both frames, $M_2({\rm GUT})$ is adjusted to a positive value  so as to saturate the measured  CDM relic density, and an integrated luminosity of 100~fb$^{-1}$ is assumed. The solid (red)  line is for the signal with no requirement on $b$-tagging, the dashed  (black) line is with the requirement of at least one tagged $b$-jet,  and the dotted (blue) line is with at least two tagged $b$-jets. The
  signal is observable if the statistical significance is above the  horizontal line at $N_{\rm signal}/\sqrt{N_{\rm back}}=5$. } 
 \end{center}
 \end{figure}

\subsubsection{High ${\bf M_2}$ dark matter model}

As a final example, we consider the LHC reach in the HM2DM model, where agreement with (\ref{wmap}) is obtained by raising $|M_2({\rm GUT})|$ from its canonical value of $m_{1/2}$ in models with gaugino mass unification, so that the lightest neutralino is MHDM. Since the LSP
contains a substantial higgsino component, it is again reasonable to expect that $b$-jet tagging may increase the SUSY reach of the LHC.\\
As we have already seen in other examples, the increased reach from $b$-jet tagging depends on the value of the squark mass relative to
$m_{\tg}$. This led us to consider two model lines with, ({\it a})~$m_0=m_{1/2}$, and ({\it b})~$m_0=M_2({\rm GUT})$, for both of which we take $\tan\beta=10$, $\mu >0$ and $A_0=0$. Since the correct relic density is obtained by {\it raising} $M_2$, model-line ({\it b}) which gives heavier squarks than model line ({\it a}) will give a smaller reach as measured in terms of $m_{\tg}$. The {\it increase} in the reach from $b$-jet tagging will, however, be larger for model line ({\it b})
since squark contributions to sparticle production are kinematically suppressed.\\ 

The statistical significance of the SUSY signal in the HM2DM model is shown for the two model lines in the two frames of Fig.~\ref{reach:hm2dm}. Indeed we see that while the reach in the left frame for $m_0=m_{1/2}$ extends to $m_{\tg} \leq 2.5$~TeV (as compared to 2.1~TeV in the right
frame), there is very little gain in the reach from $b$-jet tagging in this case where squark and gluino masses are comparable. This is in contrast to the gain in reach of $\sim 8$\% for the case of heavier squarks in the right hand frame.\\ 
\section{Is direct detection of third generation squarks possible?} \label{sec:third}
Establishing that any new physics signals at the LHC arise from supersymmetry will require the identification of several superpartners.
In models where the third generation is significantly lighter than the other generations, it is natural to ask whether it is possible to detect signals from the {\it direct production} of third generation
squarks. As already mentioned, their detection as secondaries from production and subsequent decays of gluinos is possible if the gluino itself is not very heavy \cite{nojiri}. Our goal, therefore, is to examine whether the signal from the direct production of third generation squarks can be separated both from SM backgrounds, as well as from production of other SUSY particles. Clearly, this is a model-dependent question, since the SUSY ``contamination'' to the third generation signal will depend strongly on the masses of the other squarks and the gluino. In this section, we will study this issue within the context of the inverted mass hierarchy model with $\mu < 0$, that
we have used as our canonical test case.\\ 

\begin{figure}[htdp]
\begin{center}
\includegraphics[width=10cm]{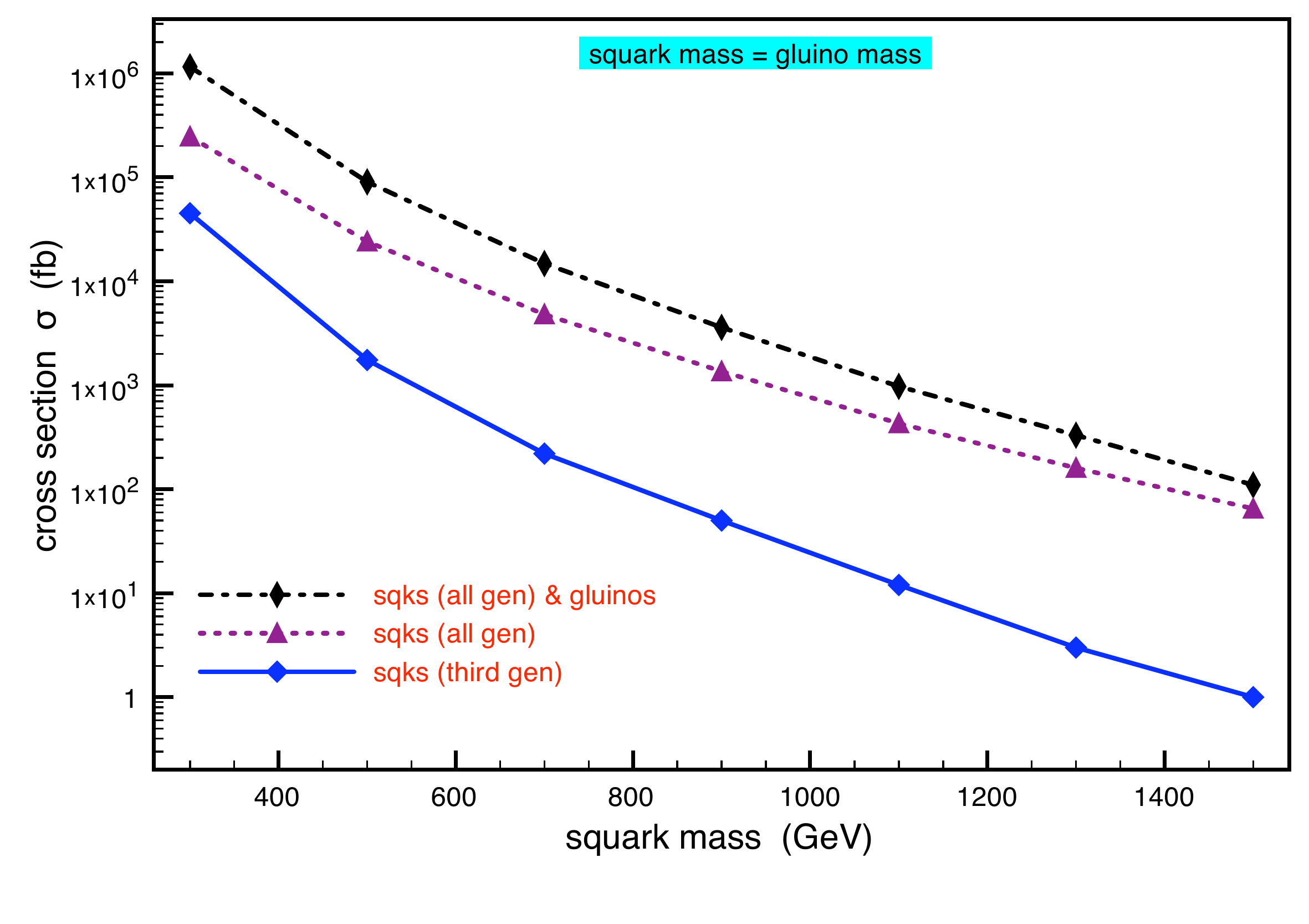}
 \end{center}
 \caption{\label{fig: sigtp1}Plot of cross section vs squark mass for third generation squarks as solid (blue) curve, all generation squarks as dotted (purple) curve, all generation squarks and gluinos  as dashed-dotted (curve) curve, with $m_{\tq}=m_{\tg}$.}
 \end{figure}

Since there are essentially no third generation quarks in the proton, the cross section for third generation squarks falls rapidly with the squark mass, and the signal becomes rapidly rate-limited. 
We show this with the solid (blue) curve  in  Fig.~\ref{fig: sigtp1} where we plot the cross section in fb vs. the squark mass in GeV units,  and contrast the results for the third generation squarks with those from either all generation squarks only, or from gluinos plus all generation squarks for   $m_{\tq}=m_{\tg}$. Here we  show the cross sections corresponding to production of all generation squarks as the dotted (purple) curve, and all generation squarks and gluinos  as the dashed-dotted (black) curve. We see that the cross section for third generation squarks is a subdominant part of the SUSY cross section, and further that it is very low, approaching 1 fb. \\
Therefore, we confine ourselves to the signal from third generation squarks with masses around 300--500~GeV, where the signal is likely to be the largest.
To unequivocally separate out the third generation signal, we must use cuts that are hard enough to reduce the SM backgrounds to
acceptable levels, yet not so hard as to enhance the ``contamination'' from heavier sparticles that, though they are produced with (much) smaller cross sections than third generation squarks, would pass these hard cuts with much larger efficiency.\\
 
Since third generation sfermions decay preferentially to third generation fermions (we focus on the case where $\tst_1 \to b\tw_1$ is accessible), we study the signal with at least one tagged $b$-jet.  We found, however, that even the softest set of cuts in both Table~\ref{tab:cuts1} and Table~\ref{tab:cuts2} that we actually use for our analysis of the SUSY $b$-tagged signal, are too hard for the purpose of extracting the signal
from third generation squarks.\\
We, therefore, returned to our basic cuts, 
$$ \eslt > 100~{\rm GeV}, E_T(j_1, j_2)> 100~{\rm GeV}$$ and augmented these with the requirements, $$E_T(j_3, j_4) > 100~{\rm GeV}, \ S_T \ge 0.1, \ n_b\ge 1, $$ and ran the third generation signal through the analysis cuts in Table~\ref{tab:3rdgen} to extract the optimal $N_{\rm signal}/N_{\rm back}$ ratio  (where the background includes the SM and the SUSY contamination as we discussed).  These cuts, which are applied ``from below'', primarily serve to control the SM background which is very large after just the basic cuts (see Table~\ref{tab:bkg}), but reduced by the additional requirements of a tagged $b$-jet and two additional 100~GeV jets.\\ 

\begin{table}[htb]
\begin{center}
\begin{tabular}{lc}
\hline \hline
Variable & Values \\
\hline
 $n_b\ge$  &   $1$ \\
$\eslt$ (GeV) $\ge$ & $100, 150, 200, 250$ \\
$[E_T(j_1),E_T(j_2)] $ (GeV) $\ge$ & $(100, 100), (200, 100), (200, 150)$ \\
&$(300, 100), (300, 150), (300, 200), $ \\
& $(400, 100), (400, 150), (400, 200)$ \\
$E_T(b_1)$ (GeV) $\ge$   & $40, 100, 200, 300, 400$ \\
$m_{\rm eff}$~(GeV) $\ge$ & $500, 600, ..., 1500$ \\
$n_j\ge$ & $4, 5, 6, 7 $\\
\hline \hline
\end{tabular}
\end{center}
\caption{The set of cuts examined for the extraction of the third generation squark signal at the LHC. See the text for the additional cuts we have imposed.}  
\label{tab:3rdgen}
\end{table}

We show the results of our analysis in Table~\ref{tab:thirdgenfinal}. The parameters are shown in the first four rows of the Table, while the next few rows show representative sparticle masses. The first two cases are along the $\mu< 0$ model line that we had introduced previously. In the first two cases $B(\tst_1\to b\tw_1)=1$, while in Case~3, $B(\tst_1\to b\tw_1)= 0.74$, with the remainder being made up by the decay $\tst_1\to t\tz_1$. The next several rows list the optimized choice of cuts from the $4\times 9\times 5\times 11\times 4$ possibilities in Table~\ref{tab:3rdgen}, along with
the cross sections for ({\it i})~the third generation signal, ({\it ii})~the SM background, and ({\it iii})~the ``SUSY contamination'' defined as the SUSY signal from production of sparticles other than
third generation squarks, after these cuts.\\
We see from these cross sections that both the event rates and the statistical significance of the third generation signal (even with the SUSY contamination included in the background) is very large. The problem, however, is that the signal to background ratio is smaller than 0.1, if the SUSY contamination is included in the background, and fails to satisfy our observability criterion.\footnote{Many authors do not impose such a requirement on the
observability of the signal. We believe that some requirement on the $N_{\rm signal}/N_{\rm back}$
 ratio is necessary since otherwise a signal with 5K events, above a background of 1M events would be considered significant. This would be indeed be the case if the background were known to a very high precision; however a systematic uncertainty of 0.5\% on the background could clearly wipe out the signal, at least if the signal is extracted by subtracting the theoretically calculated background! In the case at hand, where the SUSY model is not {\it a priori} known, and has to be arrived at using the same data, it is clear that subtraction of the SUSY contamination will suffer from considerable uncertainty until the data and theory both become mature enough for such a subtraction to be
carried out. While our criterion requiring $N_{\rm signal}/N_{\rm back}>0.25$ is admittedly
arbitrary, we believe that it is necessary to impose some lower limit on the signal to background ratio for a semi-realistic assessment.} We can, however, reduce the SUSY contamination (primarily from heavier sparticles) relative to the third generation signal by requiring that {\it the signal is not too hard.} Toward this end, we impose an {\it upper limit}, $m_{\rm eff} < 1000$~GeV, which efficiently reduces the contamination from heavy sparticles with correspondingly modest reduction of the cross sections from the softer third generation and SM processes. The corresponding cross sections after this cut are shown on the next three rows of the Table, while the last row shows the final two
signal to total background ratio that we are able to obtain, along with the statistical significance of the third generation signal with an integrated luminosity of 100~fb$^{-1}$.\\

Several comments about the Table are worth noting. 

\begin{itemize}
\item We see from the Table that before the cut restricting the value of  $m_{\rm eff}$ from above, the background was dominated by SUSY  contamination. In contrast, after this cut, the dominant source to the  background comes from SM processes. 

\item With the cuts that we have devised, the event rates for the third  generation signal as well as its statistical significance are  large. For reasons already discussed, we do not, however, believe that
  it will be easy to unequivocally ascertain the direct production of  third generation squarks in the signal. For this to be unambiguously  possible, it will be necessary to have an understanding of the
  contributions from other SUSY sources to the event rate after our  cuts. This may well be possible because with hard cuts it should be  possible to isolate the signal from heavy squarks and gluinos where  contamination from both SM and the lighter third generation squarks is  small. Just how well it will be possible to extrapolate this measured  signal into ``softer kinematic regions'' will determine the precision  with which the SUSY contamination can be subtracted. This issue is  beyond the scope of the present analysis  
\begin{table}[htb]
\begin{center}
\begin{tabular}{lcccc}
\hline
 &CASE 1 & CASE 2  & CASE 3  \\
\hline
$m_{16}$ (GeV) &717 & 854   & 739  \\
$m_{1/2} $ (GeV) & 306 & 355  & 361  \\
$A_{0}$  (GeV) & -1434 & -1708 & -1478 \\
tan$\beta$&47 &47 & 47\\ \hline
$\mu$ (GeV) & -372 & -428& -477\\
$m_{\tg}$ (GeV)  & 764 & 879  & 886 \\
$m_{\tu_R}$ (GeV) & 966 & 1127 & 1070 \\
$m_{\tst_1}$ (GeV) & 274 & 316 & 460 \\
$m_{\tb_1}$ (GeV)& 442 & 559 &  400 \\
$m_{\tw_1}$ (GeV) & 236   & 279 & 287 \\ \hline
$\eslt$ (GeV) $>$ & 150 & 100 & 150 \\ 
$[E_T(j_1), E_T(j_2)]$ (GeV) $>$ & 100, 100 & 100, 100 & 200, 100 \\ 
$E_T(b_1)$ (GeV) $>$ & 40 & 40 & 40 \\
$m_{\rm eff}$ (GeV) $>$& 500 & 500 & 600 \\ 
$n_j\ge$ & 5 & 6 & 4 \\
$\sigma_{{\rm 3rd \ gen.}} $ (fb) & 120.2 & 74.1 & 80.6 \\ 
$\sigma_{\rm SUSY \ cont.} $ (fb) & 1176.3 & 590.6 & 828.9 \\ 
$\sigma_{\rm SM}$ (fb) & 432.6 & 454.1 & 580.4 \\ \hline 
$m_{{\rm eff}}$ (GeV) $<$  & 1000 & 
1000 & 1000 \\ 
$ \sigma_{{\rm 3rd \ gen.}} $ (fb) & 47.2 & 30.9 & 20.5 \\
$\sigma_{\rm SUSY \ cont.}$ (fb) & 109.5 & 42.0 & 40.0\\ 
$\sigma_{\rm SM}$ (fb) & 141.7 & 180.6 & 155.1\\
$\sigma_{{\rm 3rd \ gen.}}/\sigma_{\rm tot. \ bkg}$ & 0.188 & 0.14 & 0.105\\
$N_{\rm signal}/\sqrt{N_{\rm back}}$ & 29.8 & 20.7 & 14.7 \\
\end{tabular}
\end{center}
\caption{The optimized cuts, along with cross sections for the signal  from direct production of light third generation squarks, for Standard  Model background, and for other SUSY contamination in the third  generation signal (discussed in the text). The first four  rows specify the input parameters for our three case studies while the next  six rows specify $\mu$ and selected sparticle masses. The next several  rows detail the choice of cuts from the set in Table~\ref{tab:3rdgen}
  chosen to ameliorate the softer Standard Model background, along with  cross sections for the third generation signal, for contamination to  this signal from other SUSY sparticles,  and for Standard Model  background after these cuts. In the last six rows we show the cut
  ``from above'' discussed in the text along with  our results for the various cross sections, the signal to  total background ratio (including SUSY contamination) and the  statistical significance of the signal.}
\label{tab:thirdgenfinal}
\end{table}
%

\item We examined additional cuts on $E_T(j_1, j_2)$ and $n_j$ to see if  we could raise the signal to background ratio. We found that a small  increase ($\sim 10$\%) may indeed be possible by restricting $n_j$  from above to be smaller than 8 or 9. Since our calculation of the  background with high jet multiplicity is carried out only in the  shower approximation, we did not feel that our estimate of this  improvement is reliable, and choose not to include it in the Table. 

\item We stress again that the SUSY contamination is model-dependent. We  can see from the Table that if gluinos and other squarks are indeed  decoupled at the LHC, and only third generation squarks are light,  their signal should be readily observable in all three cases. 

\end{itemize}
\section{Top tagging and the reach of the LHC}\label{sec:toptag}

We have seen that requiring a $b$-tagged jet reduces the SM background relative to the SUSY signal in a wide variety of models, and so increases the SUSY reach of the LHC. This then raises the question whether it is possible to further increase this reach by requiring a top-tagged jet, since the mechanisms that serve to enhance the decays of SUSY particles to $b$-quarks frequently tend to enhance decays to the entire third generation. SM backgrounds to $\eslt$ events with $t$-quarks should, of course, be smaller than those for events with $b$-quarks. In this section, we study the prospects for top tagging, once again using the inverted mass hierarchy model line (\ref{imhneg}) to guide our thinking. \\

Top tagging in SUSY events has been suggested previously for the reconstruction of SUSY events, assuming that $\tst_1$ or $\tb_1$ are light enough so that $\tg \to t\tst_1 \to t b\tw_1$ and/or $b\tb_1 \to b t\tw_1$ occur with large branching fractions \cite{nojiri}. It was shown that for $m_{\tg}\sim 700$~GeV, for which the SUSY event rate is very large, partial reconstruction of SUSY events with gluinos decaying to third generation squarks was possible at the LHC.\\

We follow the approach developed in this study to reconstruct the top quark via its hadronic decay mode. In a sample of multi-jet + $\eslt$ events with at least one tagged $b$-jet, we identified a
hadronically decaying top by first identifying all pairs of jets (constructed from those jets that are not tagged as a $b$-jets) as a hadronically decaying $W$ if $|m_{jj}-M_W| \leq 15$~GeV. We then
pair each such $W$ with the tagged $b$-jet(s) and identify any combination as a top if $|m_{bW}-m_t|\leq 30$~GeV. If we can reconstruct such a ``top'', we defined the event to be a top-tagged event.  The efficiency for tagging tops in this way turns out to be small.\footnote{In a simulated sample of about 90K $t\bar{t}$ pairs with a hard scattering $E_T$ between 50--400~GeV, we found only 6,255 top tags even with $\epsilon_b=1$. To understand this large loss of efficiency we
note that first, leptonically decaying tops (branching fraction of $\sim 1/3$) are clearly not identified. Second, $b$-jets are within their fiducial region ($E_{Tj}>40$~GeV, $|\eta_j|\leq 1.5$, with a $B$-hadron with $p_T(B)\geq 15$~GeV within a cone of $\Delta R=0.5$ of the jet axis) only about 5/8 of the time. Third, it is necessary for the top with the $b$-jet inside the fiducial region to decay hadronically in order to make the top mass window, since the wrong combination mostly falls outside. Finally, if the jets from the $W$ from the top with the tagged $b$ merge or radiate a separate jet at a large angle, this $W$ is lost, and hence the top, is not tagged. We have checked with our synthetic top sample that the choice of mass bins of $\pm 15$~GeV about $M_W$ and $\pm 30$~GeV about $m_t$ suggested in Ref.~\cite{nojiri} does not lead to loss of signal from events where the top decays hadronically into well separated jets: most of the loss in
efficiency comes from the other factors detailed above.}
\\
For our examination of the impact of top tagging on the SUSY reach of the LHC, we have chosen the $SO(10)$ model line (\ref{imhneg}) with $\mu < 0$ as a test case. In this case, since other squarks are heavy, the gluino mainly decays with roughly equal likelihood via $\tg \to \tst_1
t$ and $\tg \to \tb_1 b$, where subsequent decays of the third generation squarks can lead to yet more top quarks in SUSY events. As for the case of $b$-jet tagging, we have run the SUSY sample
through a set of cuts shown in Table~\ref{tab:topcuts} to optimize our top-tagged signal relative to SM background. Because of the small efficiency for top-tagging we cannot, however, afford a large reduction of the signal from multiple cuts. We have, therefore, restricted our optimization to cuts on just the three variables $\eslt$, $m_{\rm eff}$ and $n_j$, imposing the basic requirements on the signal as discussed in Sec.~\ref{sec:sim}. The results of our SUSY reach analysis with top-tagging are summarized in Table~\ref{tab:toptag}. \\
\begin{table}[htb]
\begin{center}
\begin{tabular}{lcc}
\hline \hline
Variable & Values \\
\hline
 $n_b\ge$  &   $1$ \\
$\eslt ({\rm GeV}) \ge$ & $300, 400,..., 900$ \\
$m_{\rm eff} ({\rm GeV}) \ge$ & $800, 900,..., 2000$ \\
$n_j\ge$ & $3, 4,..., 8$ \\
\hline \hline
\end{tabular}
\end{center}
\caption{The complete set of cuts examined for extraction of the   SUSY signal with tagged $t$-jets. In addition to the basic cuts detailed  in the text, we require that $S_T \ge 0.1$.  }
\label{tab:topcuts}
\end{table}
Here, we show the optimized statistical significance of the SUSY signal for three cases in the vicinity of the ultimate reach using this technique. In this table, we show representative sparticle masses along with branching fractions for sparticle decays that lead to top quark production in SUSY cascades. We then detail the final choice of cuts that optimizes the top-tagged SUSY signal. We also show the top-tagged signal cross section after these cuts along with the corresponding SM
background, and the statistical significance of the top-tagged signal achieved in cases 1 and 2; for case~3, the signal is not observable by our criteria. Finally, in the last two rows we show the corresponding statistical significance using $b$-jet tagging discussed in Sec.~\ref{sec:btaggs}. We see from the Table that while top tagging allows an LHC reach for $m_{\tg}$ just above 1600~GeV, {\it the top-tagged  rate becomes too low} for heavier gluinos. In contrast, $b$-jet tagging yields a statistical significance in excess of 50 close to the top-tagged reach. We thus conclude that while top-tagging can be used as a diagnostic tool, or even for reconstruction of SUSY events
\cite{nojiri} in favourable cases, it will not extend the SUSY reach of the LHC. \\
\begin{table}[htb]
\begin{center}
\begin{tabular}{lcccc}
\hline \hline
& CASE 1  & CASE 2  & CASE 3  \\
\hline
$m_{16}$~(GeV) & 1650 & 1770   & 1820  \\
$m_{\tg}$ (GeV)  & 1522 & 1614  & 1661 \\
$m_{\tu_R}$~(GeV) & 2108 & 2255 & 2319 \\
$m_{\tst_1}$~(GeV) & 714 & 766 & 792 \\
$m_{\tb_1}$~(GeV) & 744 & 842 &  876 \\
$m_{\tw_1}$~(GeV) & 533 & 570 &589 \\
$m_{\tz_1}$~(GeV) & 279 & 299 & 309 \\
$B(\tst_1 \to t\tz_i)$ & 0.64 & 0.69 & 0.70 \\
$B(\tb_1 \to t\tw_1)$ & 0.37 & 0.31 & 0.30 \\ \hline
$\eslt $ (GeV) $\ge$ &  300 & 500 & n/a \\
$m_{\rm eff}$ (GeV $\ge$ & 1700 & 800 & n/a \\
$n_j\ge$ & 8  &  3  & n/a \\ \hline
$\sigma_{\rm SUSY}$ (fb) & 0.138 & 0.108 &  n/a  \\  
$\sigma_{\rm back}$ (fb) & 0.0117 & 0.0306 &  n/a \\ \hline
$N_{\rm SUSY}/\sqrt{N_{\rm back}}$ &  & & \smallskip\\
top tag  & 12.7 & 6.14 & 0.00 \\
$1b$ & 62.8 & 52.5 & 44.7  \\
$2b$ & 93.5 & 64.0 & 46.4 \\
\hline \hline
\end{tabular}
\end{center}
\caption{A comparison of the statistical significance of the LHC signal using top-tagging described in the text,  for three different cases along the $SO(10)$ model line (\ref{imhneg}), with other parameters as fixed by Eq.~(\ref{above}).  The first few lines show the value of $m_{16}$ along with sample particle masses and branching fractions. The next three lines show the choice of cuts for
the variables in Table~\ref{tab:topcuts} that maximizes the statistical significance of the top-tagged signal. The signal and SM background cross sections for these cuts are shown on the next two lines for the cut choice that leads to an observable signal with the greatest
statistical significance. The last three rows compare the statistical significance of the signal using
top-tagging with that obtained using $b$-jet tagging discussed inSec.~\ref{sec:btaggs}.}
\label{tab:toptag}
\end{table}

\section{Charm-jet tagging} \label{sec:ctag}

Charm jet tagging offers a different possibility for enhancing the SUSY signal, especially in the case where a light top squark dominantly decays via $\tst_1 \to c\tz_1$. Charm jets may be tagged via the detection of a soft muon within the jet. Muons inside jets also arise from semi-leptonic decays of $b$-quarks and from accidental overlaps of unrelated muons with jets. Since $m_b$ is significantly larger than $m_c$, the variables $|{\vec{p}}_T^{\; \rm rel}|\equiv |{\vec{p}}_T(\mu)\times{\hat{p}}_j|$ and $\Delta R(\mu,j)\equiv \sqrt{\Delta\phi(\mu,j)^2+\Delta\eta(\mu,j)^2}$ can serve to distinguish muon-tagged $c$-jets from correspondingly tagged $b$-jets or accidental
overlap of an unrelated muon with jets.\\
Charm jet tagging with soft muons was first examined in Ref.~\cite{bst} as a way of enhancing the $t$-squark signal from $p\bar{p} \to \tst\tst X \to cc+\eslt +X$ production at Run I of the Fermilab Tevatron, but was found to have a reach smaller than the reach obtained via the conventional $\eslt$ analysis  {\it because the muon-tagged signal was severely   rate-limited}. It was, however, subsequently shown that using soft muons to tag the $c$-jet indeed enhances the top squark reach \cite{sender} but only  for an integrated luminosity larger than $\sim$~1 fb$^{-1}$, available today after the upgrade of the Main Injector.\\ 

These considerations led us to examine whether charm tagging may be similarly used at the LHC, at least for the case where $\tst \to c\tz_1$. Since the goal is to separate the charm jets from the decay of $\tst_1$ from other SUSY sources (which are frequently rich in $b$-jets), it is crucial to be able to separate the $c$ and $b$ jets with at least moderate efficiency and purity.  Following
Ref.~\cite{sender}, we examined many strategies to obtain this separation in the plane formed by the variables $|{\vec{p}}_T^{\; \rm rel}|$ and $\Delta R(\mu,j)$ but without any success. The difference between the situation at the Fermilab Tevatron, where this strategy appears to be moderately successful, and the LHC is the kinematics of the events. In contrast to the Tevatron, where jets with $E_T > 25$~GeV are readily detectable, at the LHC we have required $E_T(j)>50$~GeV in order not to be overwhelmed by mini-jet production. For this harder jet kinematics, the difference between $m_b$ and $m_c$ appears to be too small to yield significant
separation between $c$- and $b$-jets that are not vertex-tagged. The larger contamination from $b$-jets at the LHC only exacerbates this situation.\\
Before closing this section, we also mention one other (also unsuccessful) strategy that we tried for $c$-tagging. The idea was to utilize the difference in the distributions of $z\equiv E_{\mu}/E_c$ for
muons of a fixed sign of the charge from $b$ or $\bar{c}$ decays. While the expected distributions from the quark decays are indeed significantly different, this strategy also fails because these quarks hadronize before they decay, and the $z$-distributions of the muons from the
corresponding bottom or charm meson decays are essentially the same.\\ 

\section{Summary}\label{sec:resumen}
\begin{figure}[htdp]
\begin{center}
\includegraphics[width=10cm]{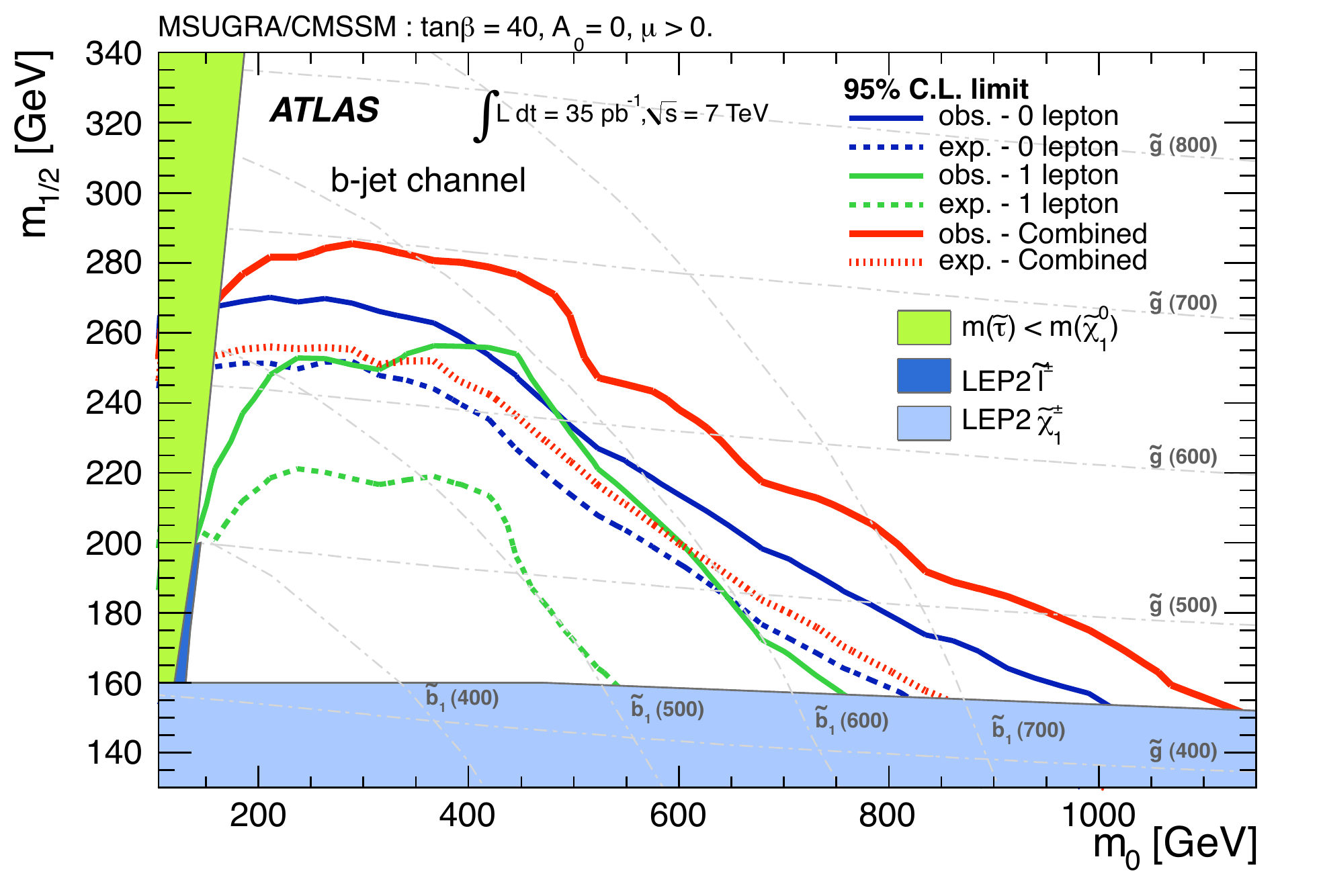}
 \end{center}
 \caption{\label{fig:batlas}Plot of $m_{1/2}$ vs $m_{0}$ for mSUGRA/CMSSM models, with an integrated luminosity of 35 pb$^{-1}$ at the LHC with $\sqrt{s}=7$. The $b$=jet channel is used to establish a region of exclusion in this parameter space. }
 \end{figure}

Summarizing, we have found that the use of $b$-tagging enhances the SUSY reach of the LHC by up to 20$\%$ in a variety of well-motivated models, with the largest increase in reach occurring in models where $m_{\tq} \gg m_{\tg}$. We note that the LHC has used $b$ tagging to establish regions of exclusion after 35 pb$^{-1}$ of data analyzed in the  mSUGRA/CMSSM models, as shown in Fig.~\ref{fig:batlas}, from the ATLAS Collaboration \cite{bjetatlas}. We have also examined $t$-tagging, since this would have the potential to reduce SM backgrounds more effectively than with $b$-tagging, but due to low efficiencies we did not obtain an increase in reach over our efforts with  $b$-tagging. We also attempted to separate the signal of third generation squarks from both the SM background and the SUSY signal from all other sources. Although this can be readily achieved with respect to the SM background, it proved to be more difficult to discriminate between the third generation squarks and all the other SUSY sources.




\chapter{Studying Neutralinos Bottom-Up at the LHC }
\label{chap:zslep}

\section{\textbf{Introduction and Goals}}

\begin{figure}[ht]
\begin{center}
\includegraphics[width=3.3cm]{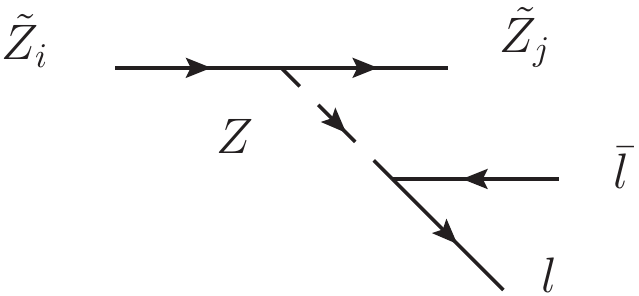}
\includegraphics[width=3.3cm]{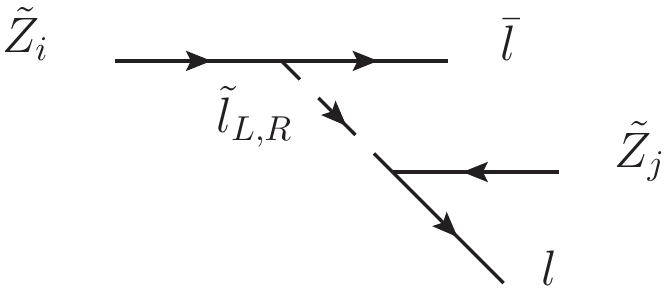}
\includegraphics[width=3.3cm]{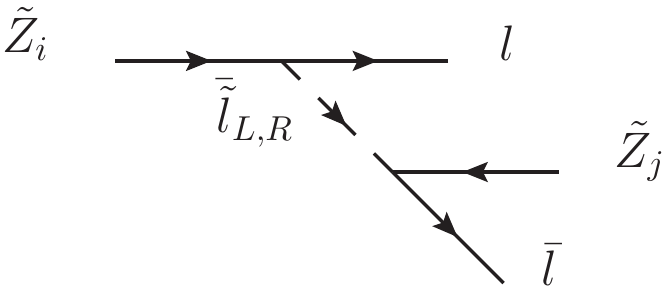}
 \end{center}
 \caption{\label{zslep}3-body decay of the neutralino via virtual $Z$-boson exchange, and by virtual $\tl_{L,R}$ exchange }
 \end{figure}

\begin{figure}[htdp]
\begin{center}
\includegraphics[width=10cm]{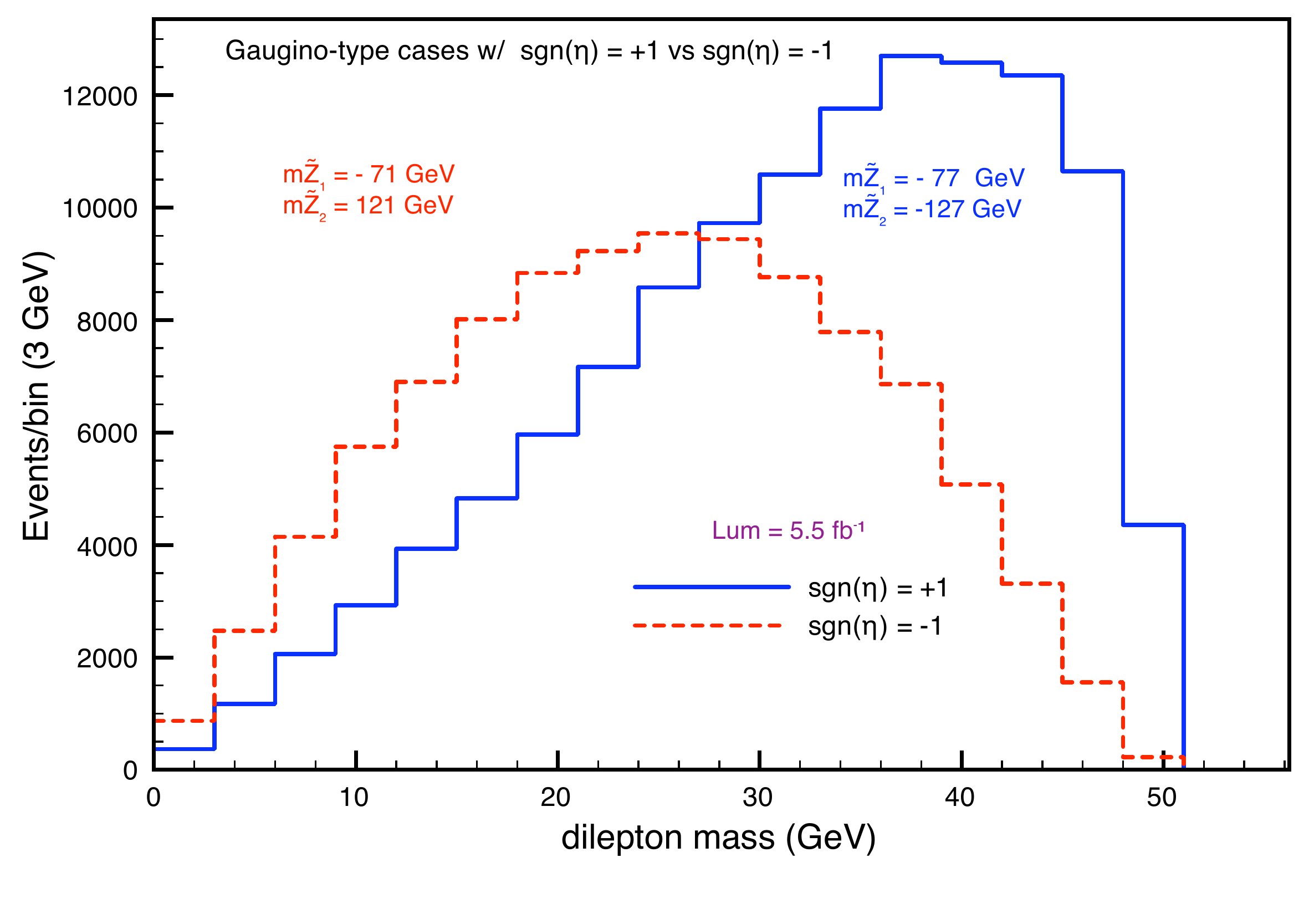}
 \end{center}
 \caption{\label{sgneta}The $m_{ll}$ distribution from the decay $\tz_{2} \rightarrow \tz_{1} l \bar{l}$ at $\sqrt{s}=14$ TeV $pp$ collider for MSSM parameters in (\ref{eqn: gau50}) for the solid (blue) histogram. For the dotted (red) histogram the sign of $M_{1}$ is reversed, representing the cases for $\eta = +1$ and $\eta = -1$ respectively.}
 \end{figure}

In this chapter we turn to the last of our projects introduced in Sec.~\ref{sec:goals}. Since heavier neutralinos are expected to be copiously produced via the cascade decays of squarks and gluinos at the LHC, we examine what we can deduce about their properties from a study of SUSY events.  We take a bottom-up approach and focus on the dilepton mass ($m_{ll}$) distribution of lepton pairs (e or $\mu$) produced via the decay $\tz_{i} \rightarrow \tz_{f} l \bar{l}$, assuming that the two body decays of neutralinos are kinematically forbidden. Since the leptons from neutralino decays always lead to opposite sign (\textbf{OS}), same flavor (\textbf{SF}) lepton pairs, i.~e. $e^{+}e^{-}$ and $ \mu^{+}\mu^{-}$ pairs, the OS,SF dilepton mass distribution will play the central role in our analysis.\\
We pick the $m_{ll}$ distribution for our analysis for several reasons. Gluino and squark cascade decay signals in a wide variety of models exhibit readily observable rates of dilepton production. We can efficiently suppress the SM background to multijet, dilepton events with $\eslt$ from gluinos and squarks with simple cuts on just jets in the event so as to minimize the loss of information contained in the $m_{ll}$ distribution. Cuts on the lepton $p_{T}$ or $\eslt$ would impact the overall shape of the distribution making a fit to the Theoretical expectation difficult. 
OS, SF dileptons can also come from production of chargino pairs which typically lead to 
 $e^{-}$$e^{+}$,  $\mu^{-}\mu^{+}$, $e^{+}\mu^{-}$, $e^{-}\mu^{+}$ at equal rates. Since neutralinos   
 \underline{\it always} decay to OS, SF dileptons, we can use the readily constructible distribution $N( e^{-}+e^{+}+\mu^{-}\mu^{+}-e^{+}\mu^{-}-e^{-}\mu^{+})$ to statistically remove the "chargino contamination" to the neutralino signal. Modulo cuts, unlike energy or angular distributions, the $m_{ll}$ distribution is a Lorentz invariant, unaffected by the boost of the parent neutralino and so it is straightforward to extract  from the data.\\
The kinematic mass edge of the $m_{ll}$ distribution has been examined in many studies to extract $m_{\tz_{i}}-m_{\tz_{f}}$, e.~g. $m_{\tz_{2}}-m_{\tz_{1}}$. This as a good starting point for an attempt at reconstructing the masses of the neutralinos involved in the decay. We ask whether the $m_{ll}$ distribution can provide us with more information than just this mass difference. For the case of $\tz_{2}$ decaying to $\tz_{1}$, more information would be possible if we could also extract  $m_{\tz_{2}}+m_{\tz_{1}}$ which would inmediately give us a measurement of the neutralino masses involved in the decay. \\
 Other information we could extract from the $m_{ll}$ distribution can be the relative sign of the mass eigenvalues. Previous studies \cite{kitano} have focused on identifying this feature as a means of distinguishing between a higgsino-like versus a gaugino-like neutralino, which, we argue, is incorrect. We also ask whether it is possible to extract whether the $Z$-exchange or the slepton $\tl_{L,R}$-exchange contributions (shown in Fig.~\ref{zslep}) dominate the neutralino decay amplitude. Most ambitiously we ask whether we can extract the parameters of the neutralino mass matrix that determines the masses and mixing angles of the neutralinos as we detail next.
Recall that, the neutral gauginos ($\lambda_{3}$ and $\lambda_{0}$) and higgsinos ($\psi_{h_{u}^0}$ and $\psi_{h_{d}^0}$) are not physical particles with definite mass, but that these will mix to form the neutral mass eigenstates, the neutralinos. The neutralino Lagrangian density can be written as:\\
\[
\lagr_{neutralino}  = - \frac{1}{2}  \, \overline{\Psi} \, M_{neutral} \, \Psi
\]
where \\
\[
M_{neutral} = 
  \left[
     \begin{array}{cccc}
     0 & \mu &-\frac{gv_{u}}{\sqrt{2}} & \frac{g^{'}v_{u}}{\sqrt{2}}\\
     \mu & 0 & \frac{gv_{d}}{\sqrt{2}} & -\frac{g^{'}v_{d}}{\sqrt{2}}\\
     -\frac{gv_{u}}{\sqrt{2}}& \frac{gv_{d}}{\sqrt{2}} & M_{2} & 0\\
     \frac{g^{'}v_{u}}{\sqrt{2}} & -\frac{g^{'}v_{d}}{\sqrt{2}} & 0 & M_{1}\\
     \end{array}
  \right ]
  \]
 and\\
 \[
 \Psi =
  \left (
    \begin{array}{c}
     \psi_{h_{u}^0}\\
     \psi_{h_{d}^0}\\
     \lambda_{3} \\
     \lambda_{0} \\
     \end{array}
     \right )
     \]      
The entries in the mass matrix for the neutralinos, arise from the following sources \cite{wss}: 
\begin{enumerate}
 \item The higgsino mass term $\mu$ from the superpotential. 
 \item $M_{1}$ and $M_{2}$  from  soft  SUSY breaking contributions from the gaugino masses.
 \item The remaining off-diagonal terms arise from gaugino-higgsino-Higgs boson interactions, when the Higgs boson fields  develop VEV's $v_{u}= \langle h_{u} \rangle$ and $v_{d}= \langle h_{d} \rangle$ because electroweak symmetry is spontaneously broken. 
 \end{enumerate} 
Before proceeding further we spell out the assumptions and ground rules for our bottom-up study.     
In order to keep the study as bottom-up as possible, we avoid specific constrained models, such as mSUGRA, where all MSSM masses and couplings are determined by a handful of parameters.
\bi
\item We assume the MSSM particle content; the shape of the $m_{ll}$ distribution is then completely fixed by the neutralino parameters,
\[
 \langle M_{1}, M_{2}, \mu, tan\beta, m_{\tl_{L}}, m_{\tl_{R}} \rangle
 \]
as shown in Sec.~\ref{sec: formula1}.
\item We will use only the \underline{shape} of the  $m_{ll}$ distribution as the normalization depends on gluino and squark properties, decay branching fractions.
\item Assume that the two body decay channels for $\tz_{2} \rightarrow \tz_{1} Z$ and $\tz_{2} \rightarrow \tl l $ are forbidden,\footnote{The first case leads to an $m_{ll}$ distribution sharply peaked at $M_{Z}$. For the second, mSUGRA case studies have shown that the $m_{ll}$ distribution can be used to exclude off-shell slepton decays if for the ÔÔtest pointÕÕ the decay $\tz_{2}\rightarrow \tl l $ is accessible \cite{matchev1}. We make no representation as to whether or not this is possible, and if so, whether the conclusions of \cite{matchev1} extend to other models, and conservately regard the absence of two body neutralino decays as an assumption in this study.} so our neutralino decays via the three-body mode, as shown in Fig.~\ref{zslep}.
\item We make a working technical assumption, for simplicity, that $m_{\tl_{L}} = m_{\tl_{R}}$.
\ei
We focus only upon what can be inferred from this data alone without combining it with information about other sparticles since that is likely to introduce other model dependence.\\
Also we carry out our analysis for the design energy  $\sqrt{s} = 14$ TeV of the LHC. We begin our phenomenological study by considering the $m_{ll}$ distribution for the decay $\tz_{2} \rightarrow \tz_{1} l \bar{l}$ where the neutralinos are gaugino-like as is typical in many models. For our case study we use the following parameters:
\be
< M_{1}, M_{2}, \mu, tan\beta, m_{\tilde{l}_{L}}, m_{\tilde{l}_{R}}>=<77, 127, -911, 10, 211, 211>
\label{eqn: gau50}
\ee
where all mass parameters are in GeV units, and  the gluino mass at $450$ GeV and squark masses at $400$ GeV, beyond the range of the Fermilab Tevatron. This is shown by the solid (blue) histogram in Fig.~\ref{sgneta}. We see that there is a sharp endpoint at  $m_{\tz_{2}}-m_{\tz_{1}}=50$ GeV. The rates shown correspond to an integrated luminosity of 5.5 fb$^{-1}$. We introduce,
\be
 \eta_{ij} = \frac{sgn(m_{\tilde{Z_{i}}})}{sgn(m_{\tilde{Z_{f}}})}
 \ee
with the convention that $\eta_{ij}$ = +1 for masses of equal sign , and $\eta_{ij}$ = -1 for masses of opposite sign, then we can see from Figs.~\ref{sgneta} that there is a difference in shape for the dilepton distributions corresponding to the different values of $\eta$. \\
The dotted (red) histogram in Fig.~\ref{sgneta} shows the same distribution for the case where the sign of $M_{1}$ is reversed for its value in (\ref{eqn: gau50}) with all other parameters held fixed. In this case the mass eigenvalue of $\tz_{1}$ flips its sign so that the relative sign $\eta_{12}$ goes from $+1$ to $-1$ for the dotted one. We see that the mass distribution is much softer in the $\eta_{12}=-1$ case. The difference in shape of the $m_{ll}$ distribution due to the  relative sign of the mass eigenvalues has been mentioned in the literature, \cite{kitano,XT1}. \\
The first of these references \cite{kitano} attributed the differences to the gaugino-like versus higgsino-like nature of the neutralinos, while the second reference \cite{XT1} more appropiately to the relative sign of the mass eigenvalues. It would thus seem that we can rather easily extract the relative sign of the neutralino mass eigenvalues, but we will revisit this below.\\
The remainder of this Chapter delves into details of information about neutralinos that can be extracted from the dilepton mass distribution and whether the ultimate goal of reconstructing the mass matrix is possible.\\
 \section{\textbf{The dilepton mass distribution in the MSSM}}
 \label{sec: formula1}
\subsection{Decay formula}
Let $v_i^{(j)}$ be the neutralino eigenvector components as defined in \cite{wss}, and $g'$ and  $g$ the hypercharge $U(1)$ and $SU(2)$ coupling constants. The five different coupling constants in the decay process $Z_2\to Z_1 l^+ l^-$ are:
\begin{eqnarray}
A_{\tz_1} &=& -\frac{g v_3^{(1)} + g' v_4^{(1)}}{\sqrt{2}}
\nonumber \\
A_{\tz_2} &=& -\frac{g v_3^{(2)} + g' v_4^{(2)}}{\sqrt{2}}
\nonumber \\
B_{\tz_1} &=& -\sqrt{2} g'v_4^{(1)}
\nonumber \\
B_{\tz_2} &=& -\sqrt{2} g' v_4^{(2)}
\nonumber \\
w_{12} &=& \frac{\sqrt{g^2 + g^{'2}}}{4}\left(v_1^{(1)} v_1^{(2)} -
  v_2^{(1)}v_2^{(2)}\right).
 \label{eq: def}
\end{eqnarray}
Let us define the following functions.\\
\begin{eqnarray}
\lambda(x, y, z) &=& x^2 + y^2 + z^2 - 2 x y - 2 y z - 2 z x
\nonumber \\
\Delta(m_{\tilde{l}}, m_{\tz_1}, m_{\tz_2}) &=& 2 m_{\tilde{l}}^2-m_{\tz_1}^2-
m_{\tz_2}^2
\end{eqnarray}

The contribution due to the $Z$-mediated process is,
\begin{eqnarray}
\frac{d\Gamma_Z}{d m_{ll}} =
\frac{4\pi^2 w_{12}^2 g^2 (1 + (-1 + 4\sin^2\theta_W)^2)}
{12 (2\pi)^5 \cos^2\theta_W}
\frac{m_{ll}\sqrt{\lambda(m_{ll}^2, m_{\tz_1}^2, m_{\tz_2}^2)}}
{(m_{ll}^2 - m_Z^2)^2 4 m_{\tz_2}^3}\times
\nonumber \\
((m_{\tz_2}^2 - m_{\tz_1}^2)^2 - 2m_{ll}^4 + m_{ll}^2 (m_{\tz_2}^2 + m_{\tz_1}^2) +
          6 a m_{\tz_1} m_{\tz_2} m_{ll}^2),
\label{eqn: Zmed}
\end{eqnarray}
where, $a$ is the relative sign of $\tz_1$,$\tz_2$ mass eigenvalues, i.e. 
$a=(-1)^{(\theta_1+\theta_2)}$. 

Let us define the function,
\begin{eqnarray}
F_{\tilde{l}}(m_{\tilde{l}})  =
-\sqrt{\lambda(m_{ll}^2, m_{\tz_1}^2, m_{\tz_2}^2)} 
- 4\sqrt{\lambda(m_{ll}^2, m_{\tz_1}^2, m_{\tz_2}^2)}\times
\nonumber\\
\frac{(m_{\tz_2}^2 - m_{\tilde{l}}^2)(m_{\tz_1}^2 - m_{\tilde{l}}^2)}
{(\Delta(m_{\tilde{l}}, m_{\tz_1}, m_{\tz_2}) + m_{ll}^2)^2 
- \lambda(m_{ll}^2, m_{\tz_1}^2, m_{\tz_2}^2)} 
\nonumber \\
+ \left(\Delta(m_{\tilde{l}}, m_{\tz_1}, m_{\tz_2}) + 
  \frac{2 a m_{\tz_1} m_{\tz_2} m_{ll}^2}{\Delta(m_{\tilde{l}}, m_{\tz_1}, m_{\tz_2}) 
+ m_{ll}^2}\right)\times
\nonumber\\
 \ln\frac{\Delta(m_{\tilde{l}}, m_{\tz_1}, m_{\tz_2}) + m_{ll}^2 + 
 \sqrt{\lambda(m_{ll}^2, m_{\tz_1}^2, m_{\tz_2}^2)}}
{\Delta(m_{\tilde{l}}, m_{\tz_1}, 
   m_{\tz_2}) + m_{ll}^2 - \sqrt{\lambda(m_{ll}^2, m_{\tz_1}^2, m_{\tz_2}^2)}}
\nonumber \\
\end{eqnarray}
The contribution from the pure slepton-mediated  decay is,
\begin{equation}
\frac{d\Gamma_{\tilde{l}\tilde{l}}}{d m_{ll}} =
\frac{1}{32\pi^3}\frac{m_{ll}}{4 m_{\tz_2}^3}
\left( A_{\tz_1}^2 A_{\tz_2}^2 F_{\tilde{l}}(m_{\tilde{l}_L}) + 
B_{\tz_1}^2 B_{\tz_2}^2 F_{\tilde{l}}(m_{\tilde{l}_R})\right),
\label{eqn: slep}
\end{equation}
where $m_{\tilde{l}_L},m_{\tilde{l}_R}$ are left- and right-slepton 
masses respectively.
Finally, define
\begin{eqnarray}
F_{\tilde{l}Z}(m_{\tilde{l}}) =
 \frac{1}{2}(-\Delta(m_{\tilde{l}}, m_{\tz_1}, m_{\tz_2}) + m_{ll}^2)
  \sqrt{\lambda(m_{ll}^2, m_{\tz_1}^2, m_{\tz_2}^2)}
\nonumber \\
+ ((m_{\tilde{l}}^2 - m_{\tz_1}^2)(m_{\tilde{l}}^2 - m_{\tz_2}^2) - a m_{\tz_1}
 m_{\tz_2}m_{ll}^2)\times
 \nonumber\\
 \ln\frac{\Delta(m_{\tilde{l}}, m_{\tz_1}, m_{\tz_2}) + m_{ll}^2 +
  \sqrt{\lambda(m_{ll}^2, m_{\tz_1}^2,
 m_{\tz_2}^2)}}{\Delta(m_{\tilde{l}}, 
m_{\tz_1},
m_{\tz_2}) + m_{ll}^2 - \sqrt{\lambda(m_{ll}^2, m_{\tz_1}^2, m_{\tz_2}^2)}}
\end{eqnarray}
The mixed $Z$ and slepton contribution, {\it i.e.} the cross term,  is
\begin{eqnarray}
\frac{d\Gamma_{\tilde{l}Z}}{d m_{ll}} =
 -\frac{g m_{ll}}{32 \pi^3 m_{\tz_2}^3 (m_{ll}^2 - m_Z^2)\cos\theta_W} \times
\nonumber \\
 (A_{\tz_1} A_{\tz_2} w_{12} (1 - 2\sin^2\theta_W) F_{\tilde{l}Z}(m_{\tilde{l}_L}) 
\nonumber \\
 + B_{\tz_1} B_{\tz_2} w_{12} 2 \sin^2\theta_W F_{\tilde{l}Z}(m_{\tilde{l}_R}))
 \label{eqn: xterm}
\end{eqnarray}
Finally, the decay width formula for the process $\tz_2\to \tz_1 l^+ l^-$ is
\begin{equation}
\frac{d\Gamma}{d m_{ll}} = 
  \frac{d\Gamma_Z}{d m_{ll}} + \frac{d\Gamma_{\tilde{l}\tilde{l}}}{d m_{ll}} + 
    \frac{d\Gamma_{\tilde{l}Z}}{d m_{ll}}
\label{eqn: fctn}
\end{equation}
Clearly, the functional form of the $Z$ and slepton contributions are completely different-looking. Note also, that each of the components in (\ref{eqn: fctn}) contain the ÔÔ$a$ÕÕ term, where $a$ is the relative    sign of $\tz_{1}, \tz_{2}$ mass eigenvalues, i.e. $a= \pm 1.$ It is this ÔÔ$a$ÕÕ term that causes the distribution to shift inward for the case $a=-1$ as seen in Fig.~\ref{sgneta}.\\
 \section{A Closer Look at the Test Case} 
\begin{figure}[htbp]
\begin{center}
\includegraphics[width=10cm]{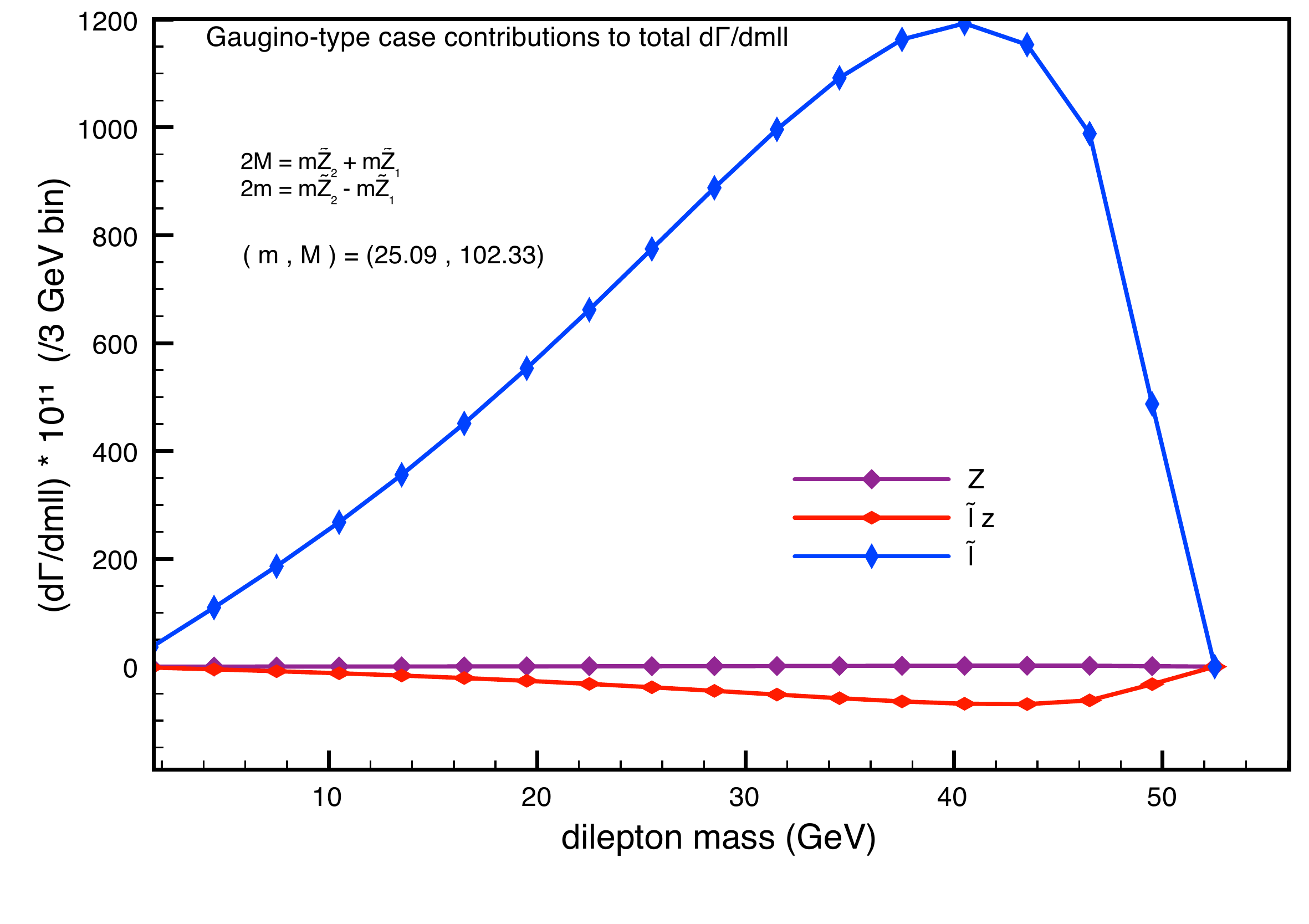}
 \end{center}
 \caption{\label{frame1}Different contributions to d$\Gamma$/dm$_{ll}$ for a gaugino-like test case from the components of the decay formula as in (\ref{eqn: fctn}). }
 \end{figure}

\begin{figure}[htbp]
\begin{center}
\includegraphics[width=10cm]{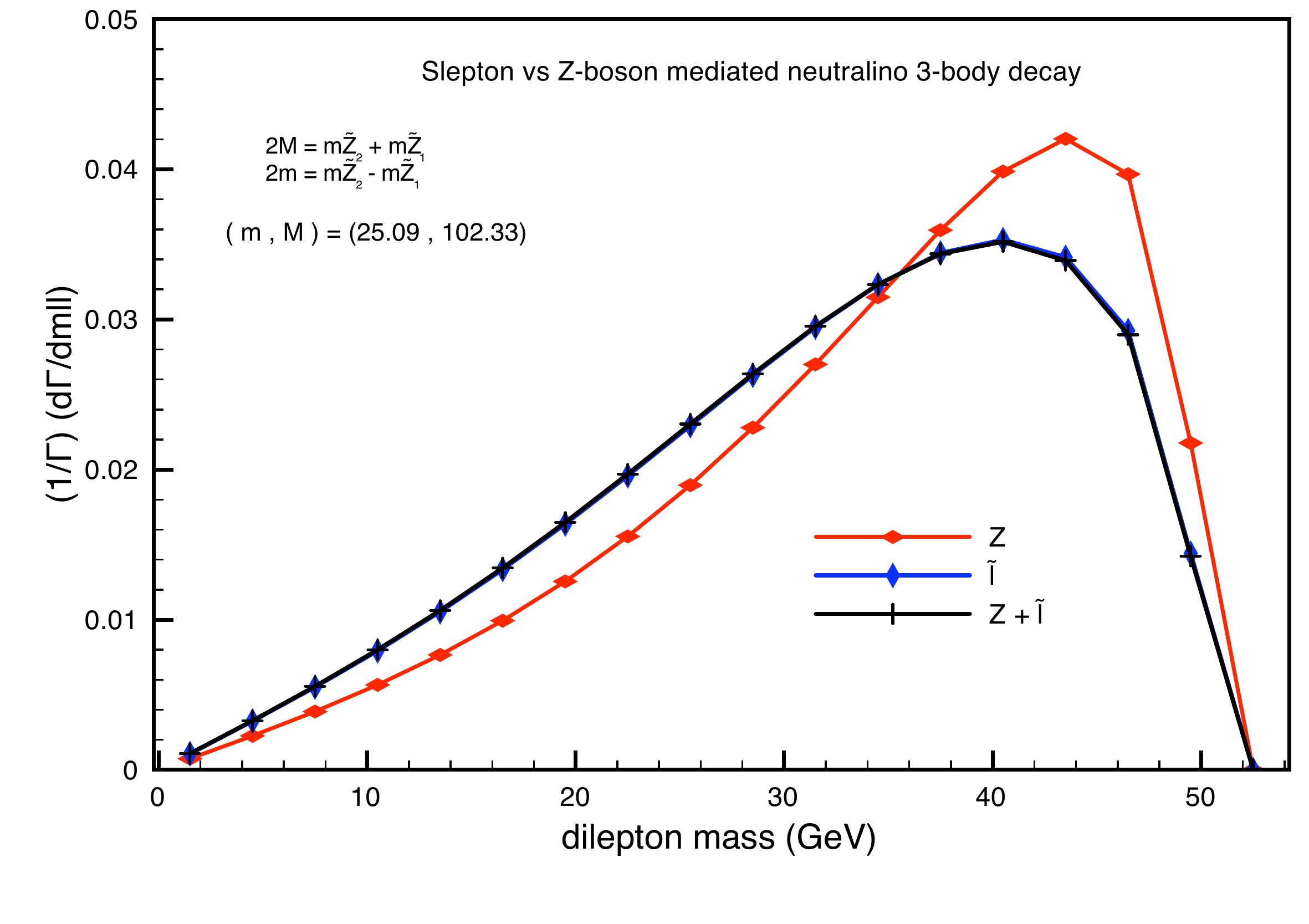}
 \end{center}
 \caption{\label{frame2}Different contributions to d$\Gamma$/dm$_{ll}$, normalized, for a gaugino-like test case with same MSSM inputs. }
 \end{figure}

\begin{figure}[htbp]
\begin{center}
\includegraphics[width=10cm]{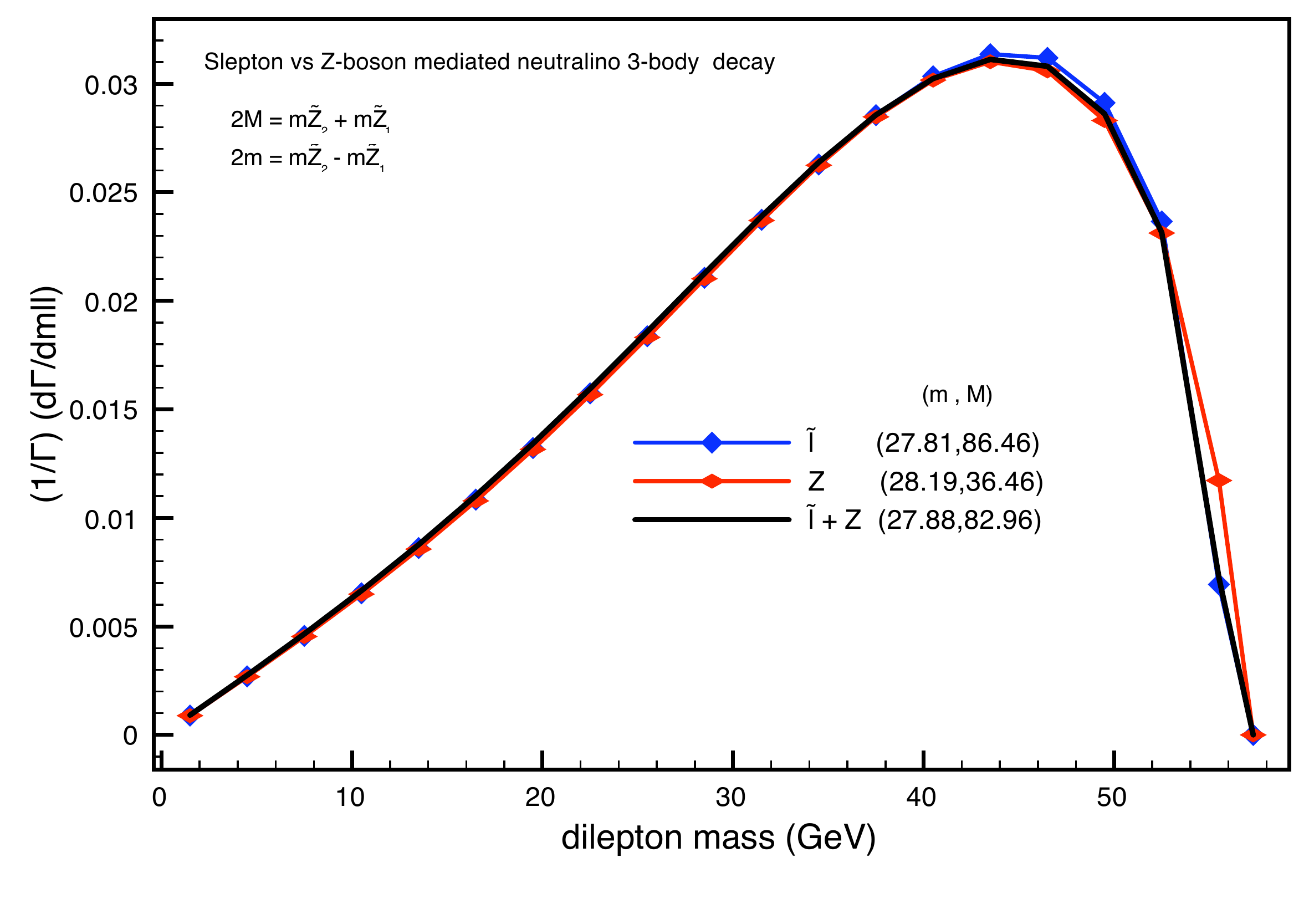}
 \end{center}
 \caption{ \label{fzvsfslep}Normalized $m_{ll}$ distributions showing the equivalence between $Z$-boson exchange and $\tilde{l}_{L,R}$ exchange. The values of the input MSSM parameters $M_{1}$ and $M_{2}$ are different for the respective curves, and the corresponding values of the parameters ($m, M$) are shown. }
 \end{figure}

 We continue our examination of our test case of eq.~(\ref{eqn: gau56}) and ask whether it is possible to infer whether the Z- or slepton-mediated amplitudes dominate the decay. Toward this end, we show the slepton-mediated contribution of eq.~(\ref{eqn: slep}), the $Z$-mediated contribution eq.~(\ref{eqn: Zmed}) and the interference term eq.~(\ref{eqn: xterm}) in Fig.~\ref{frame1}. As expected for this gaugino-like case, the slepton exchange process dominates over the $Z$ exchange, with a small contribution from the $Z$, $\tilde{l}$ cross-term. In this plot, we have used $\frac{d\Gamma}{dm_{ll}}$ which shows the relative strengths of the different contributions to the $m_{ll}$ distribution. In order to study whether it is possible to distinguish slepton and $Z$-mediated decays $\underline{directly}$ from the shape of the $m_{ll}$ histogram we show the corresponding normalized contributions $\frac{1}{\Gamma}\frac{d\Gamma}{dm_{ll}}$ in Fig.~\ref{frame2}.
The lighter curve shows the shape of the $Z$ mediated contribution while the darker curve shows that of the slepton mediated contribution. We can see a clear difference between the $Z$ and the slepton contribution, where the latter matches exactly the shape of the total $m_{ll}$ distribution. It seems that we should be able to distinguish between the two exchanges just from their shapes.\\
The situation is, however, not quite so simple because this conclusion presumes that we know the input model parameters. The data, however, only gives us a reliable measurement of $m_{\tz_2}-m_{\tz_1}$. In order to infer that the shape of the $m_{ll}$ distribution definitely picks out the slepton-mediated contribution, we have to ensure that this shape cannot be reproduced by the $Z$-contribution with the same values of $2m=m_{\tz_2}-m_{\tz_1}$ for all values of $2M=m_{\tz_2}+m_{\tz_1}$. Toward this end, we examine our original test point corresponding to gaugino-like neutralinos given by the parameter set,
\be
< M_{1}, M_{2}, \mu, tan\beta, m_{\tilde{l}_{L}}, m_{\tilde{l}_{R}}>=<55, 111, -911, 10, 234, 213>
\label{eqn: gau56}
\ee
that has an endpoint at 56 GeV. The decay is, once again, slepton dominated and the corresponding  $\frac{1}{\Gamma}\frac{d\Gamma}{dm_{ll}}$ distribution is shown in Fig.~\ref{fzvsfslep}. As expected the shape from the slepton mediated contribution also coincides with the total. The problem, however, is that this shape is also the same as that obtained assuming the decay occurs via the $Z$ mediated amplitude but for $(m, M)=(56.38, 72.92)$ as shown by the grey line in the figure which has the same endpoint. It is clear that without  
any additional information, we would not be able to certify whether the shape of the $m_{ll}$ histogram  was the result of a $Z$ dominated, or a slepton dominated situation.\\
This observation motivated to analyze our theoretical results further. We accomplish this in two ways. \\
First, we look closer at our decay formula, but this time in terms of the new variables we have introduced, with the assumption that $m^2 \ll M^2$. Second, we look at our Lagrangian terms directly, for the $Z$ and slepton contributions, applying a Fierz transformation, assuming that momentum transfer is much smaller than the virtual exchange masses. From this analysis we hope to explain the observed similarity in the two contributions, observed in Fig.~\ref{fzvsfslep}. The results of these analysis are discussed next.\\ 
\subsection{\label{closerlook}A Closer Look at the Decay Formula}
Consider the following transformations, in terms of new variables,$m$ and $M$ where,\\
\be
m_{\tilde{Z}_{2}} = \frac{M+m}{2}  ,   m_{\tilde{Z}_{1}}= \frac{M-m}{2}
\label{Mm1}
\ee
or,
\be
M = m_{\tilde{Z}_{2}}+m_{\tilde{Z}_{1}} , m=m_{\tilde{Z}_{2}}-m_{\tilde{Z}_{1}}
\label{mM2}
\ee
Now apply this transformation to $\frac{d\Gamma_{\tilde{l}\tilde{l}}}{d m_{ll}}$ from the decay formula result. The question is, can we extract some information regarding $M$ from the decay process. If the answer is affirmative, then together with the mass endpoint value, which corresponds to $m$, we would know the values for m$_{\tilde{Z}_{2}}$ and m$_{\tilde{Z}_{1}}$. Rewriting our formula for $\frac{d\Gamma}{dm_{ll}}$ and replacing the $Z$ propagator with $\frac{1}{M_{Z^2}}$ we find to leading order that
\[
\lambda = (M^2 - m_{ll}^2)(m^2-m_{ll}^2) 
\]
\[
\lambda^{\frac{1}{2}}= M(m^2-m_{ll}^2)^{\frac{1}{2}}\times [1-\frac{m_{ll}^2}{2M^2}] 
 \equiv \lambda_{0}^{\frac{1}{2}}\,[1-\frac{m_{ll}^2}{2M^2} ]
\]
\[
\Delta = 2m_{\tilde{l}}^2 - {\frac{1}{2}}M^2 - {\frac{1}{2}}m^2 = \Delta_{0}[1-(\frac{{\frac{1}{2}}m^2-m_{ll}^2)}{\Delta_{0}}]  
\]
\[
 \Delta_{0}\equiv 2m_{\tilde{l}}^2-\frac{1}{2}M^2
\]
The M completely factors out in this approximation and each of the three contributions to $\frac{d\Gamma}{dm_{ll}}$ in eq.~(\ref{eqn: fctn}) is proportional to
\begin{eqnarray}
\lambda_{0}^{\frac{1}{2}}\,M^2 [ (\frac{1}{3}+a)m_{ll}^2+ \frac{2}{3}m^2]
\label{mM3}
\end{eqnarray}
We see that the shape does not depend on M when $m^2 \ll M^2, M_{Z}^2$ and $m_{\tz_2}^2 \ll m_{\tl}^2$. We can understand this more simply using Fierz transformations as we see in the next section.\\
 \section{\label{fierzdecay}Fierz Transformation Applied to Decay Formula}
The $Z$ and slepton propagators in Fig.~\ref{zslep} can be shrunk to a point so that the decays occur as a four point interaction if  $m^2\ll M_{\tz_2}^2$, and the slepton is much heavier than $\tz_{2}$.
We define 
\[
W_{12}=(-i)^{\theta_1} (i)^{\theta_2} w_{12}
\]
where $w_{12}$ is defined in eq.~(\ref{eq: def}). We can write the Lagrangian for the $Z$ mediated exchange as:
\be
\lagr \propto W_{12} \bar{\tz_1} \gamma_{\mu} (\gamma_5)^{\theta} \tz_2 Z_{\mu} \bar{l} (a+b\gamma_5) l
\label{eq: zlagr}
\ee
where $\theta = \theta_1 + \theta_2 -1$ and $a=\frac{1}{4}\,(3tan\theta_{W}-cot\theta_{W})$ and $b=\frac{1}{4}\,(tan\theta_{W}+cot\theta_{W})$.\\
We define 
\[
A_{\tz_i}^l = (-i)^{\theta_{i} -1} A_{\tz_i}  ,   B_{\tz_i}^l = -(-i)^{\theta_{i} -1} B_{\tz_i}
\]
where $A_{\tz_i}$ and $B_{\tz_i}$ have been defined in eq.~(\ref{eq: def}). Then we can write the Lagrangian for the left and right slepton exchange as,
\be
 \lagr_{\tl}= iA_{ \tz_{1}}^{l} \, \tl_{L}^{\dag} \bar{\tz}_{1}\,\frac{1-\gamma_5}{2} \,l + iB_{ \tz_{2}}^{l} \, \tl_{R}^{\dag} \frac{1+\gamma_5}{2} l + h.c.
 \label{fierz14}
 \ee
 The $\tl_{L}$ mediated amplitude comes from a four point interaction
 \begin{eqnarray}
 (\bar{l}\, \frac{1+\gamma_5}{2})_{a} \tilde{Z}_{2a} \cdot \bar{\tilde{Z}}_{1b} (\frac{1-\gamma_5}{2}\, l)_{b}
 \label{fierz11}
 \end{eqnarray}
We can then write
 \[
 \tilde{Z}_{2a} \cdot \bar{\tilde{Z}}_{1b} = \Sigma_{A} \, C_{A}\, \Gamma^{A}_{ab} , with \,
  C_{A}= -\frac{1}{4} \bar{\tilde{Z}}_{1b} \, \Gamma^{A} \, \tilde{Z}_{2a}
  \]
 where A runs over the 16 Dirac bilinears which span the set of 4$\times$4 complex matrices. These bilinears are classified according to their properties under Lorentz transformations, and consist of the following:\\
 \begin{eqnarray}
 \{\Gamma^{A}\}= \{ \mathcal{I} , \gamma_{5} , \gamma^{\mu},\gamma_{5}\gamma^{\mu} , \sigma^{\mu\nu} \}  \; \;\; ; \; \; \mu,\nu = 0,1,2,3
 \label{base0}
 \end{eqnarray}
 Substituting back into eq.~(\ref{fierz11}) we obtain,
  \begin{eqnarray}
  -\frac{1}{4} \Sigma_{A}\,  \bar{\tilde{Z}}_{1} \, \Gamma^{A} \, \tilde{Z}_{2} \cdot \bar{l} \frac{1+\gamma_5}{2} \, \Gamma^{A} \, \frac{1-\gamma_5}{2} l
  \label{fierz12}
  \end{eqnarray}
from which only $\Gamma^{A} = \gamma^{\mu}$ and $\Gamma^{A} = i\gamma^{\mu}\gamma_{5}$ will contribute non-zero values so that
  \begin{itemize}
  \item for the $\tl_{L}$ exchange amplitude  arises from,
  \begin{eqnarray}
  -\frac{1}{2}\bar{\tz_{1}}\, \gamma^{\mu} \frac{1+\gamma_5}{2} \tz_{2}\cdot \bar{l} \gamma_{\mu} \frac{1-\gamma_5}{2} l
  \label{fierz13a}
  \end{eqnarray}
  \item while  the $\tl_{R}$ amplitude arises from,
  \begin{eqnarray}
  -\frac{1}{2}\bar{\tz_{1}}\, \gamma^{\mu} \frac{1-\gamma_5}{2} \tz_{2}\cdot \bar{l} \gamma_{\mu} \frac{1+\gamma_5}{2} l
  \label{fierz13b}
  \end{eqnarray} 
  \end{itemize}
From eq.~(\ref{fierz14}) the $\tl_{L}$ exchange amplitude will have the form
\be
 \frac{ A_{ \tz_{1}}^{l} A_{ \tz_{2}}^{l\ast}}{m_{\tilde{l}_{L}}^2} \bar{\tz}_{1}\,\frac{1-\gamma_5}{2} l \bar{l}\,\frac{1+\gamma_5}{2} \,\tilde{Z}_{2} + \frac{ A_{ \tilde{Z}_{2}}^{l} A_{ \tilde{Z}_{1}}^{l\ast}}{m_{\tilde{l}_{L}}^2}\, \bar{\tilde{Z}}_{2}\,\frac{1-\gamma_5}{2} l \bar{l}\,\frac{1+\gamma_5}{2} \,\tilde{Z}_{1}
\label{fierz18}
 \ee
Fierz transforming the above, we obtain
\be
 \frac{ A_{ \tz_{1}}^{l} iA_{ \tz_{2}}^{l\ast}}{m_{\tilde{l}_{L}}^2} \bar{\tz}_{1}\,\frac{1+\gamma_5}{2} \tz_{2}  \bar{l}\,\frac{1-\gamma_5}{2} \ l + h.c.
 \label{fierz21}
 \ee
 but, with $ A_{ \tz_{1}}^{l} A_{ \tz_{2}}^{l\ast}=A_{ \tz_{1}}(-i)^{\theta_{1} - 1} A_{ \tz_{2}} (i)^{\theta_{2} -1}$ and after rearranging terms  (\ref{fierz21}) becomes
\begin{eqnarray}
-\frac{1}{2}\,\frac{A_{ \tz_{1}}A_{ \tz_{2}}}{m_{\tilde{l}_{L}}^2}\,(i)^{\theta_{2}-\theta_{1}} \bar{l}\,\gamma^{\mu}\,\frac{1-\gamma_5}{2} \,l \,\bar{\tilde{Z}}_{1}\gamma^{\mu}\times \nonumber \\\frac{[1-(-1)^{\theta_{1}+\theta_{2}}]+\gamma_{5}[1+(-1)^{\theta_{2}+\theta_{1}}]}{2}\,\tilde{Z}_{2}
\label{fierz19}
\end{eqnarray}
Adopting the following convention,
for $\theta_{1}+\theta_{2}$ even, choose $\theta$ odd, and viceversa, then we see that when $\theta$ is odd the neutralino current exhibits an axial vector nature, while for $\theta$ odd it exhibits a vector nature.Then, we can write ~(\ref{fierz19}) as,\\
\begin{eqnarray}
 -\frac{1}{2}\,\frac{[A_{ \tilde{Z}_{1}}^{l}][A_{ \tilde{Z}_{2}}^{l}]}{m_{\tilde{l}_{L}}^2}\,(i)^{\theta_{2}-\theta_{1}} \bar{l}\,\gamma^{\mu}\,P_{L}\,l\, \bar{\tilde{Z}}_{1}\gamma_{\mu}\,(\gamma_{5})^{\theta}\,\tilde{Z}_{2}
 \label{fierz20}
 \end{eqnarray}
 Summarizing, we have,\\
 \begin{itemize}
 \item from the Z graph,\\
 \[
 \bar{\tilde{Z}}_{1}\,\gamma_{\mu}\,(\gamma_{5})^{\theta}\,\tilde{Z}_{2}\,\bar{l}\,\gamma^{\mu}\,(a+b\gamma_{5})\,l
 \]
 \item and from the $\tilde{l}_{L}$ and $\tilde{l}_{R}$ graphs\\
 \[
 \bar{\tilde{Z}}_{1}\,\gamma_{\mu}\,(\gamma_{5})^{\theta}\,\tilde{Z}_{2}\,\bar{l}\,\gamma^{\mu}\,\frac{1\mp \gamma_{5}}{2}\,l
 \]
 \end{itemize}
 From the Z graph, we can actually write $\bar{l}\,\gamma^{\mu}\,(a+b\gamma_{5})\,l$ as, \\$\bar{l}\,\gamma^{\mu}\,[(a+b)P_{R} + (a-b)P_{L}] \,l$ which looks exactly like the  $\tilde{l}_{L}$ and $\tilde{l}_{R}$ contributions.\\ 
 We see that (~\ref{fierz20}) has the same form as (~\ref{eq: zlagr}). Also, the contribution from the right slepton would have the same structure. In our decay formula, we have used the massless lepton limit, so chirality is conserved, making $\tilde{l}_{L}$ and $\tilde{l}_{R}$ different final states which cannot interfere, explaining the absence of crossterms.\\
\section{\textbf{SM background subtraction}}
\begin{figure}[htbp]
\begin{center}
\includegraphics[width=10cm]{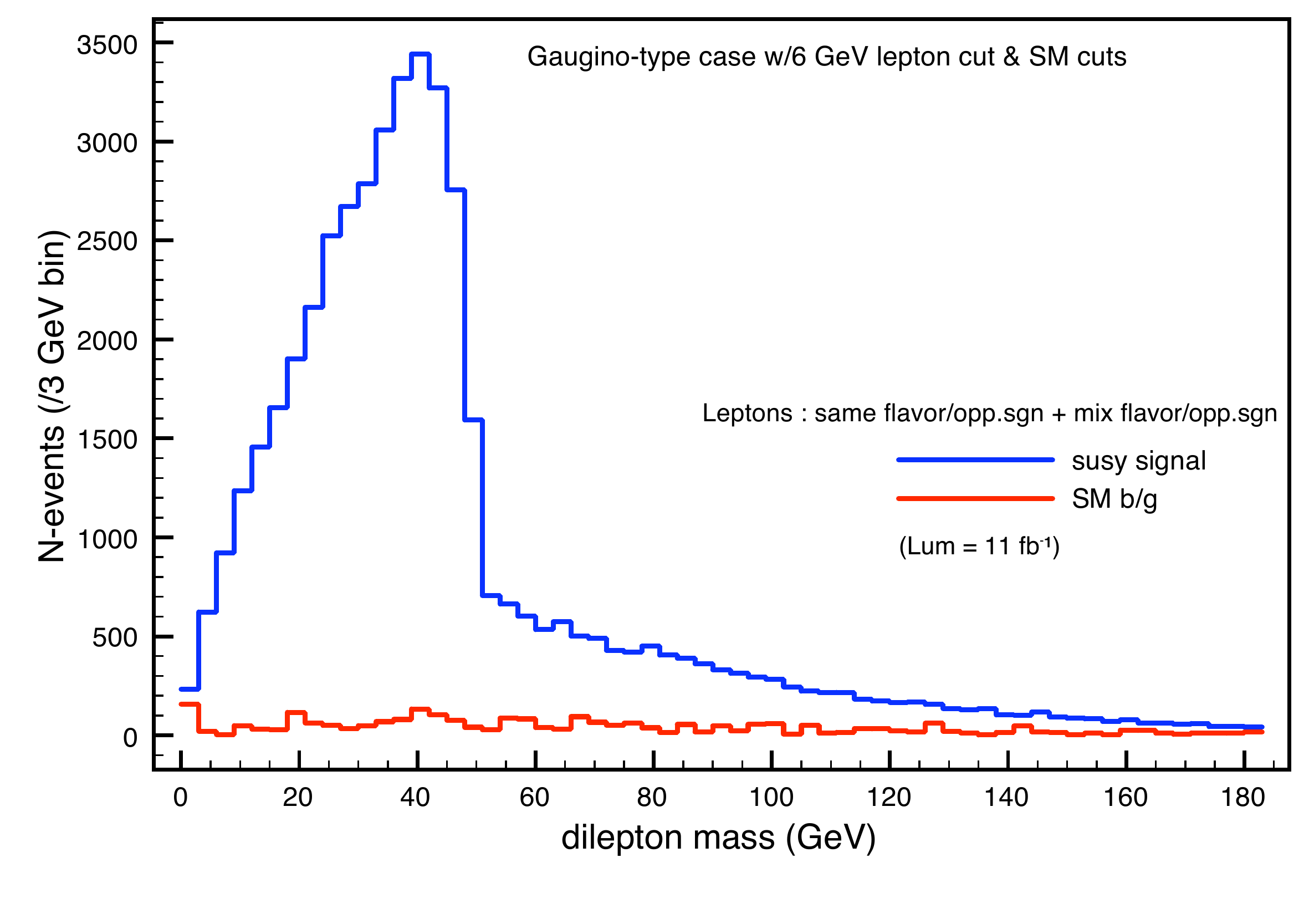}
 \end{center}
 \caption{ \label{histogauS}Histogram showing total $m_{ll}$ distribution of OS, SF dilepton events  from all SUSY events (larger solid blue), together with that from SM sources (smaller solid red) at a $\sqrt{s}=14$ TeV $pp$ collider after the cuts discussed in the text. For MSSM input parameters given by $\langle M_1, M_2, \mu, tan(\beta), m_{\tl}, m_{\tg}, m_{\tq} \rangle$ the gaugino-like case shown has MSSM parameters $\langle 77, 127, -911, 10, 211, 441, 441 \rangle$.}
 \end{figure}

\begin{figure}[htbp]
\begin{center}
\includegraphics[width=10cm]{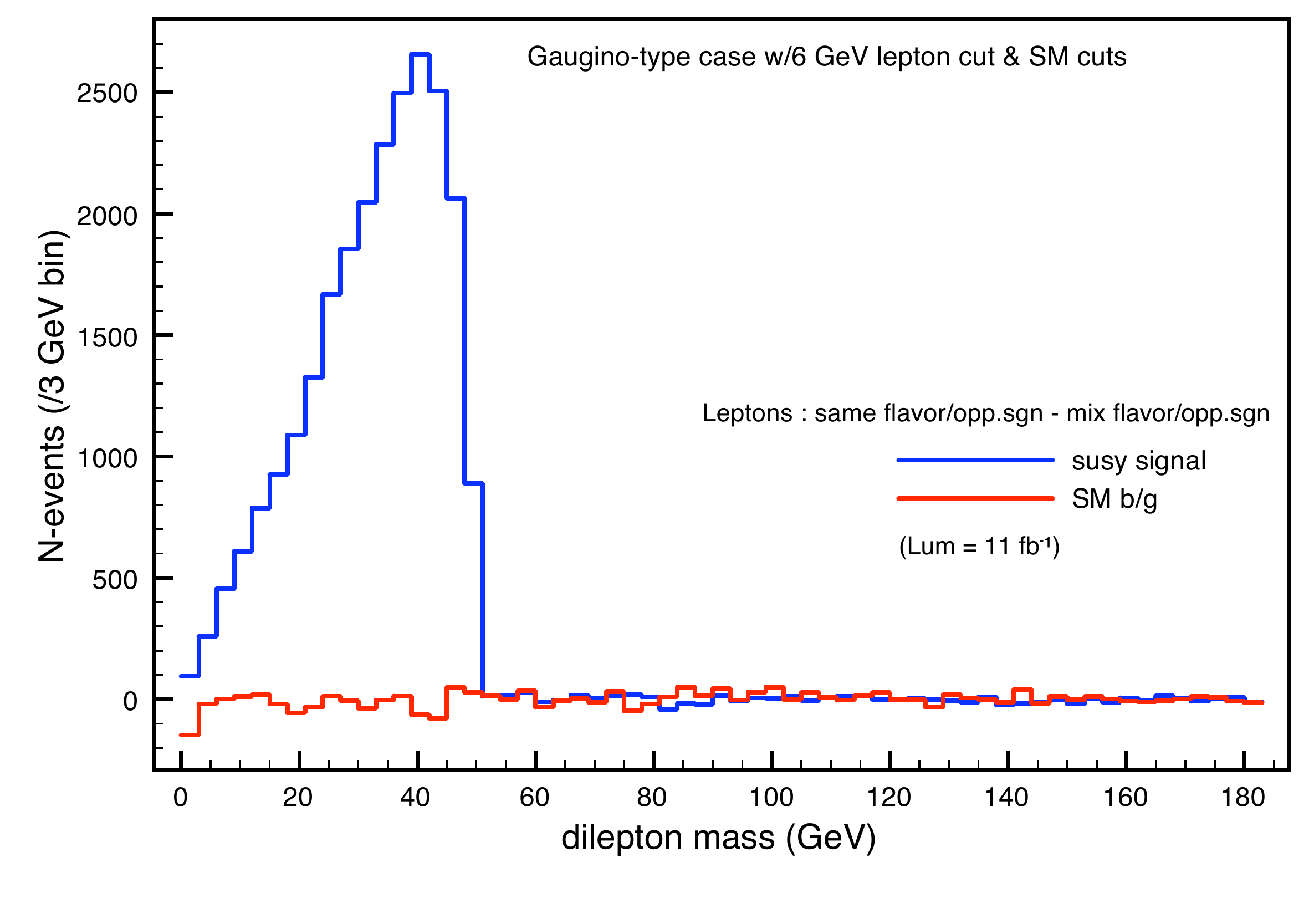}
 \end{center}
 \caption{\label{histogauD}Histogram showing the $m_{ll}$ distribution of OS, SF dilepton events  from all SUSY events (larger solid blue) from all SUSY events after flavour subtraction, together with that from SM sources (smaller solid red) also after flavour subtraction, at a $\sqrt{s}=14$ TeV $pp$ collider after the cuts discussed in the text.. For MSSM input parameters given by $\langle M_1, M_2, \mu, tan(\beta), m_{\tl}, m_{\tg}, m_{\tq} \rangle$ the gaugino-like case shown has MSSM parameters $\langle 77, 127, -911, 10, 211, 441, 441 \rangle$.}
 \end{figure}

The real data will be contaminated by the SM background, so we still need to implement cuts to effectively remove it, while at the same time, make sure our cuts do not alter the shape of our $m_{ll}$ distribution significantly. Effectively this implies that we should focus on making cuts on jets in the event, avoiding significant cuts on $\eslt$ and of course on the leptons. We use the sample gaugino-like point with MSSM parameters given by eq.~(\ref{eqn: gau50}) and show the effectiveness of our cuts on the SM background. We require $n_{leptons}=2$, $n_{jets}\ge 4$, $\eslt \ge 50$ GeV and $h_{T} \ge 550$ GeV, where $h_{T}= \sum_{i} pTjet_{i}  i= 1,4$. The $m_{ll}$ distribution for the SUSY signal and the SM background is shown in Fig.~\ref{histogauS}.\footnote{We have checked that $\eslt \ge 50$ GeV cut does not significantly distort the shape of the $m_{ll}$ distribution.} We see that the cuts effectively eliminate SM backgrounds.\\
The question is, are there any sources of contamination to our signal still present at this point. Recall that  our focus is with leptons of same flavour and opposite charge, as expected from neutralino decays (e$^{-}$e$^{+}$ + $\mu^{-}\mu^{+}$) but there will be additional leptons from charginos in all combinations of flavor ($e^{-}$$e^{+}$, $\mu^{-}\mu^{+}$, $e^{+}\mu^{-}$, $e^{-}\mu^{+}$) in equal amounts (statistically). This is seen as the long tail beyond $m_{ll} = 50$ GeV, the kinematic endpoint of the distribution for the neutralino decays. In order to remove the contamination from charginos, we take the leptons of opposite flavor ($e^{+}\mu^{-}$+$e^{-}\mu^{+}$) and subtract them from the total, thus, statistically eliminating the contribution from the charginos to the dilepton output. We show the flavour subtracted dilepton distribution with the corresponding SM background after cuts in Fig.~\ref{histogauD}.
We see that the tail beyond the kinematic endpoint is very efficiently eliminated.
At this point, we are in a position to begin our analysis of the selected case studies, and we present  our results in the next section.\\
\section{\textbf{Results}}\label{sec: gauhiggs}
To conduct our analysis we chose three representative cases, according to the composition of the participating neutralinos. 
\begin{itemize}
\item Gaugino-like neutralinos.
\item Higgsino-like neutralinos.
\item Mixed higgsino-gaugino-like neutralinos. 
\end{itemize}
We chose these cases at the beginning of 2010, prior to the LHC commencing operations. All three cases have gluino and squark masses in the $400-450$ GeV range, and were considered as LEP 2 and Tevatron-safe cases. The mass values for both gluino and squark guaranteed us a significant cross section, which for initial integrated luminosity values circa 10 fb$^{-1}$ would provide us with statistically significant numbers of events per 3 GeV bin  in our $m_{ll}$ distributions. \\
The first two case studies are presented  in principle to confirm the accuracy of our theoretical findings presented in Secs.~\ref{closerlook} and ~\ref{fierzdecay} where we concluded that the shapes of the dilepton distributions did not suffice to uniquely fit the parameters. This led us to consider a third case where three neutralinos are kinematically accessible. Such a case is motivated by an interesting mechanism for obtaining CDM in the MSSM. The third case presented new challenges to our analysis, such as having three decay processes ($\tz_{3}\rightarrow\tz_{2}, \, \tz_{2}\rightarrow\tz_{1},\, $and$\, \tz_{3}\rightarrow\tz_{1}$) producing dilepton events for the $m_{ll}$ distribution, resulting in three different mass edges, and the LSP having a mixed higgsino-gaugino like nature. We focused on this case first, and as it produced some interesting results with regards to our objectives, we were motivated to pursue this study further. As we were concluding this study in April 2011, the first results were published by the LHC, summarizing the completed analysis of data corresponding to an integrated luminosity of 35 pb$^{-1}$. These results included a region of exclusion for SUSY in the $(m_0, m_{1/2})$ plane. Our case studies were right at the edge of the exclusion region at the time. Later, towards the end of August, beginning of September 2011, new data analysis representing integrated luminosities just in excess of 1 fb$^{-1}$ were published, and the new regions of exclusion for SUSY definitely excluded our case study points. We expect that our techniques may still be useful for heavier gluinos and squarks in certain types of models, such as those having a compressed mass spectrum. Other models, such as mSUGRA,  requiring gluinos and squarks to be at 1 Tev or more introduces a factor of 2-3 multiplying all the masses, so mass gaps will grow by these factors, and the separation between the neutralino masses will increase to a level which allows two-body decays, which will dominate, suppressing the three-body decays essential for our work. As we mentioned before, our case studies were designed to give significant statistics at initial luminosities for the LHC. If we raise the gluino and squark masses beyond 1 TeV the cross sections will reduce significantly (by a factor of about 10), so an integrated luminosity of 100 fb$^{-1}$, about a year of operation at design luminosity, would be needed in these higher mass cases.\\
\subsection{\textbf{CASE 1: dominant gaugino-like neutralinos}}
\begin{figure}[htdp]
\begin{center}
\includegraphics[width=10cm]{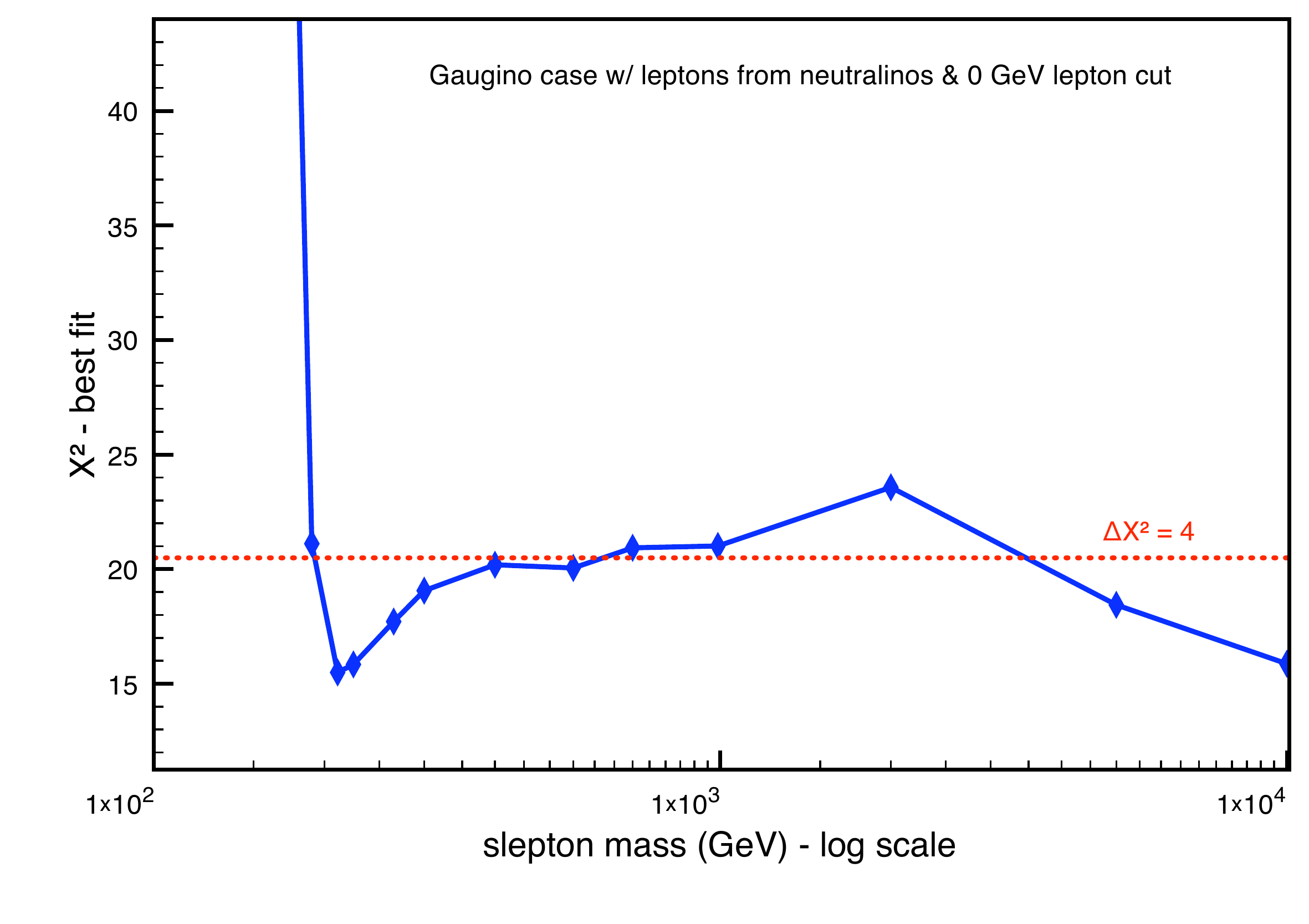}
 \end{center}
 \caption{\label{gaug00}$\chi^{2}$ best fit for a gaugino-type case with leptons from neutralinos w/ 00GeV lepton cut, $m_{ll}^{(00)}$, plotted  vs slepton mass marginalized over other MSSM parameters.}
 \end{figure}

As mentioned above, we examine this case to verify our somewhat pessimistic conclusions  obtained early in this chapter, as referenced at the beginning of this section. Specifically, as we try to fit our theoretical function to the data, will these fits be the result of the slepton mediated decay, or will it be the result of the Z-boson mediated decay masquerading as the slepton exchange. One way to verify this would be to choose the slepton mass to be very large (10 TeV) to guarantee that the slepton has effectively decoupled from the process, and check to see if we can obtain as good a fit as was obtained when the slepton was clearly involved in the process.\\
Before proceeding further, a brief summary of our fitting procedure is in order. Initially, we would like to generate an ideal, i.~e. not experimentally accessible, $m_{ll}$ distribution, with all dileptons being produced by neutralinos only, which we can do at the simulation level by identifying the parent of each lepton. In addition, we apply only a minimal cut of $\eslt = 5$ GeV, no jet cuts, and no isolation requirements for the leptons. We require exactly two such leptons per event, of \textbf{OS} and \textbf{SF}, and no $p_{T}$ cuts on the leptons. This will give us a quasi-pure distribution, which we will denote by $m_{ll}^{(00)}$, so that we can attempt to fit our theory function $(\frac{d\Gamma}{dm_{ll}})_{00}$,  to our expression in eq.~(\ref{eqn: fctn}). The shape is fit to the parameters,
\[
\langle M_{1}, M_{2}, tan(\beta), \mu, m_{\tl} \rangle
\]
and an additional parameter determining the normalization. We fit by looking for a set of MSSM input parameters for  $(\frac{d\Gamma}{dm_{ll}})_{00}$ and evaluating it at the midpoint of each 3 GeV bin, then calculating a total $\chi^2$ between the data value and the formula value for all the bins\footnote{For the $i^{th}$ bin $\chi_{i}^2=(formula\, value_{i} - data\, value_{i})^{2}/n_{i}$ where $n_i$ is the total number of dileptons in bin}, finally choosing the minimum of the $\chi^2$ values. The corresponding MSSM values plus the value for the overall constant will be our best fit values for the $m_{ll}^{(00)}$ data. For this first case, we used $\eta$ = +1, and the  MSSM input parameters\footnote{This is the same case used to test the chargino subtraction, and SM background elimination.} in eq.~(\ref{eqn: gau50}) where the theoretical endpoint is at 50 GeV.
This case provides some interesting results. The number of bins is 16, and here we only fit one theory decay formula instead of 3 as will be required in the mixed case, so the number of variables is 6, $\langle M_{1}, M_{2}, \mu, tan\beta , \tl \rangle$ and the overall constant, for a net number of degrees of freedom = 10.
Initially we attempted to identify a minimum by utilizing different programs available for this sort of procedure, Minuit (PAW) and ROOT, but it was difficult to achieve a convergent solution, and results depended strongly on the input of initial parameters. We proceeded by using Mathematica v8.1, creating a 5 parameter grid and then using a $\chi^{2}$ best fit. \\
We include here the main results for the gaugino-like point corresponding to the fit for the $m_{ll}^{(00)}$  quasi-pure case,  summarized in Table ~\ref{gaugpure00}   and Fig.~\ref{gaug00}.\\ 
\begin{table}[htdp]
\begin{center}
\begin{tabular}{|c||r| |r| |r| |r| |r|}
\hline \hline
$\chi^{2}$&$m_{\tilde{l}}$ GeV&$M_1$&$M_2$&$\mu$&$tan{\beta}$\\
\hline
98.41&160&60&110&-1425&05\\
21.11&190&79&129&-1300&10\\
15.49&211&95&145&-925&10\\
15.84&225&70&120&-1100&14\\
17.71&265&56&103&-1100&02\\
19.06&300&56&103&-1050&02\\
20.19&400&59&109&-1100&11\\
20.05&550&71&121&-1200&11\\
20.93&700&65&112&-1100&02\\
21.01&1TeV&83&130&-950&05\\
23.57&2 TeV&07&57&-1050&05\\
18.44&5 TeV&10&60&-1300&11\\
15.74&10 TeV&10&60&-1200&08\\
\hline \hline
\end{tabular}
\end{center}
\caption{$\chi^{2}$ results for the fit to the gaugino-like $m_{ll}^{(00)}$ case, corresponding to OS, SF dileptons from neutralinos w/(00) GeV lepton cuts.} 
\label{gaugpure00}
\end{table}%
The flat behaviour observed in the plot for higher values of m$_{\tilde{l}}$ has a distinguishing feature. In the present case, the MSSM parameters corresponding to the best-fit $\chi^{2}$ for the different values of $m_{\tilde{l}}$ remain low, all in a relatively narrow region until the jump at Super-TeV slepton masses. 
The $\Delta\chi^2$ remains smaller than 4 for a wide range of  $m_{\tilde{l}}$ and never exceeds 8. For very heavy slepton masses we see that the $\Delta\chi^2$ again attains a very low value compatible with a good fit, but for a very different set of gaugino masses. This is compatible with our finding earlier where we saw that it was not possible to distinguish between $Z$ and slepton mediated amplitudes.\\
\subsection{\textbf{CASE 2: dominant higgsino-like neutralino}}
This case illustrates clearly the theoretical results obtained previously in this chapter. The mass endpoint of 24 GeV is low enough that our condition of $m_{ll}^2 \ll M_{Z}^2$ is satisfied over the entire range of $m_{ll}$ values, as well as $m^2 \ll M^2$. This results in a complete insensitivity to the MSSM input parameters. When we create our grid over the MSSM parameters, we do so for a 3 GeV bin containing the value of $m$, so small perturbations from the flat minimum $\chi^2$ could occur for this reason.
The  MSSM inputs are the following:\\
\be
< M_{1}, M_{2}, \mu, tan\beta, m_{\tilde{l}_{L}}, m_{\tilde{l}_{R}}>=<-155, 170, 167, 10, 170, 170>
\label{eqn: higgsino}
\ee
where slepton masses are in GeV units, with $m_{\tz_{1}}=112.4$ GeV and $m_{\tz_{2}}=137.0$ GeV for a mass endpoint of 24.6 GeV. The gluino mass is 450 GeV and the squark masses are 400 GeV.We analyzed the  $m_{ll}^{(00)}$ case, and as expected, the $\chi^2$ behaviour was flat with respect to any of the MSSM parameters marginalized over the other parameters. Extraction of MSSM parameters is not possible. Though discouraging, we understand why.\\
\section{\textbf{CASE 3: mixed higgsino-gaugino type neutralino}}
\begin{figure}[htbp]
\begin{center}
\includegraphics[width=10cm]{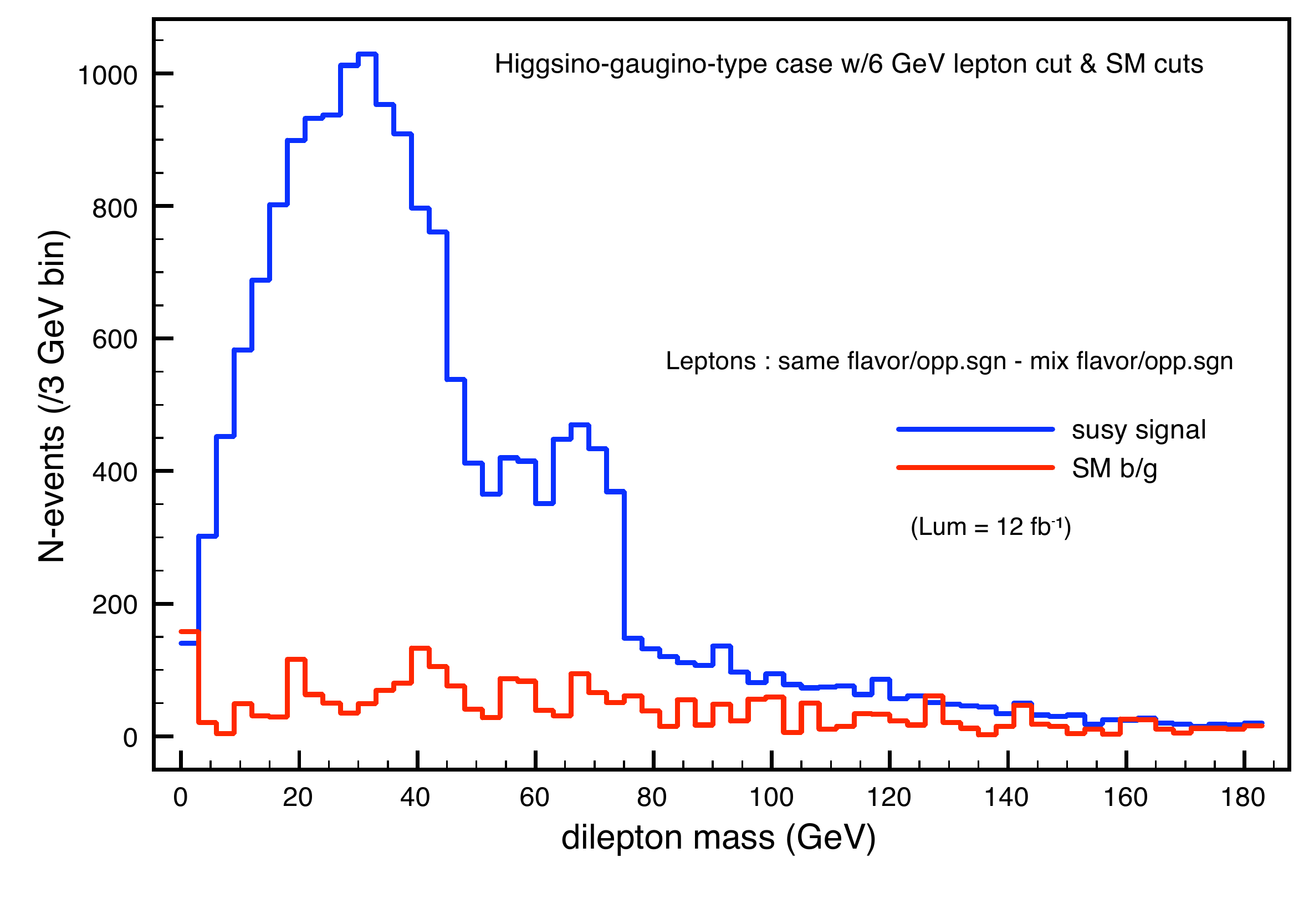}
 \end{center}
 \caption{\label{histomixS}Histogram showing total $m_{ll}$ distribution of OS, SF dilepton events  from all SUSY events (larger solid blue), together with that from SM sources (smaller solid red) at a $\sqrt{s}=14$ TeV $pp$ collider after the cuts discussed in the text. For MSSM input parameters given by $\langle M_1, M_2, \mu, tan(\beta), m_{\tl}, m_{\tg}, m_{\tq} \rangle$ the mixed higgsino-gaugino-like case shown has MSSM parameters $\langle -70, 400, 120, 10, 170, 450, 400 \rangle$.}
 \end{figure}

\begin{figure}[htbp]
\begin{center}
\includegraphics[width=10cm]{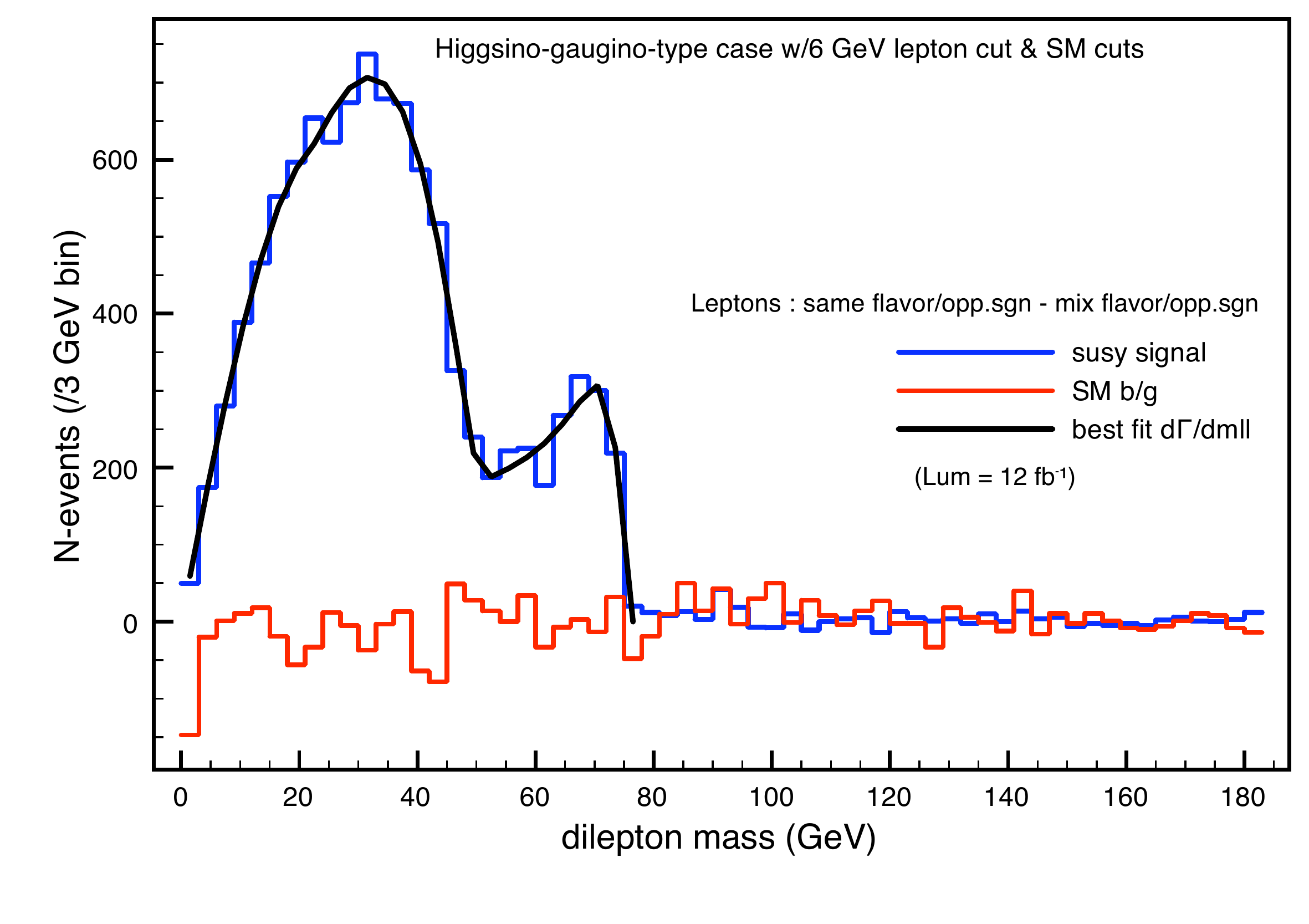}
 \end{center}
 \caption{\label{histomixD}Histogram showing the $m_{ll}$ distribution of OS, SF dilepton events  from all SUSY events (larger solid blue) from all SUSY events after flavour subtraction, together with that from SM sources (smaller solid red) also after flavour subtraction,  at a $\sqrt{s}=14$ TeV $pp$ collider after the cuts discussed in the text.. For MSSM input parameters given by $\langle M_1, M_2, \mu, tan(\beta), m_{\tl}, m_{\tg}, m_{\tq} \rangle$ the mixed higgsino-gaugino-like case shown has MSSM parameters $\langle -70, 400, 120, 10, 170, 450, 400 \rangle$. The solid (black)  curve shows the result of the fit discussed in Sec.~\ref{sec: qrc} of the text.}
 \end{figure}

There are SUSY models, such as the so-called High $M_{2}$ DM models introduced in the previous chapter \cite{XT1}., where  there is the possibility of a Dark Matter component with a mixed bino-higgsino structure. In this case there will be a visible \underline{double mass edge} from both $\tz{2}$ and $\tz_{3}$ decays, while the third edge though present is not manifest. This motivated our third case study, which provided us with a bigger challenge for the fit, but also with very positive results.\\
For this case. the gluino mass is taken at 450 GeV, while the squarks of all generations are at 400 GeV. The remaining MSSM input parameters are,
\be
< M_{1}, M_{2}, \mu, tan\beta, m_{\tilde{l}_{L}}, m_{\tilde{l}_{R}}>=<-70, 400, 120, 10, 170, 170>
\label{mixparams}
\ee
The resulting values for $\eta$ are,
\begin{center}
{$\langle  \eta_{21} ,  \eta_{31} ,   \eta_{32}  \rangle $ = $\langle -1 , +1 , -1 \rangle$}
\end{center}
where the slepton masses are in GeV units. In such a case, while attempting to fit our theory function, the fit will be sensitive to small changes in both $M_{1}$ and $\mu$ , and to a lesser degree to changes in $M_{2}$ , while exhibiting a relatively flat behaviour for changes in tan$\beta$.\\
The 2-body neutralino decays are kinematically suppressed, leaving dominant the 3-body decays involving $ \tilde{Z_{3}} \rightarrow  \tilde{Z_{1}}$ , $ \tilde{Z_{2}}\rightarrow  \tilde{Z_{1}}$, $ \tilde{Z_{3}}\rightarrow  \tilde{Z_{2}}$, through virtual   Z , $\tilde{l}\tilde{r}$-slepton exchanges, with the relevant mass gap endpoints at:
\begin{itemize}
\item $m_{\tz_{3}}- m_{\tz_{2}}$ = 25 GeV
\item $m_{\tz_{2}}- m_{\tz_{1}}$ = 50 GeV
\item $m_{\tz_{3}}- m_{\tz_{1}}$ = 75 GeV
\end{itemize}
The corresponding dilepton mass distribution after the analysis cuts detailed earlier is shown in Fig.~\ref{histomixS} for all SUSY sources together with the corresponding SM background. The mass edges at 50 GeV and 75 GeV are evident from the figure while the existence of the edge at 25  GeV can be inferred. The corresponding flavour subtracted distribution is shown in Fig.~\ref{histomixD} 
The latter two endpoints are easily distinguishable in the $m_{ll}$ distribution, but as shown previously, we require additional information, and regretfully we have confirmed that $m_{ \tz_{j}}+ m_{\tz_{i}}$ is not accessible in specified limiting scenarios analyzed previously. Our goal is to determine if with multiple endpoints and neutralinos participating, some information can be gained which could specify the values of the MSSM parameters of the neutralino mass matrix\\
\subsection{Quasi-Pure Case}\label{sec: qpc}
\begin{figure}[htdp]
\begin{center}
\includegraphics[width=10cm]{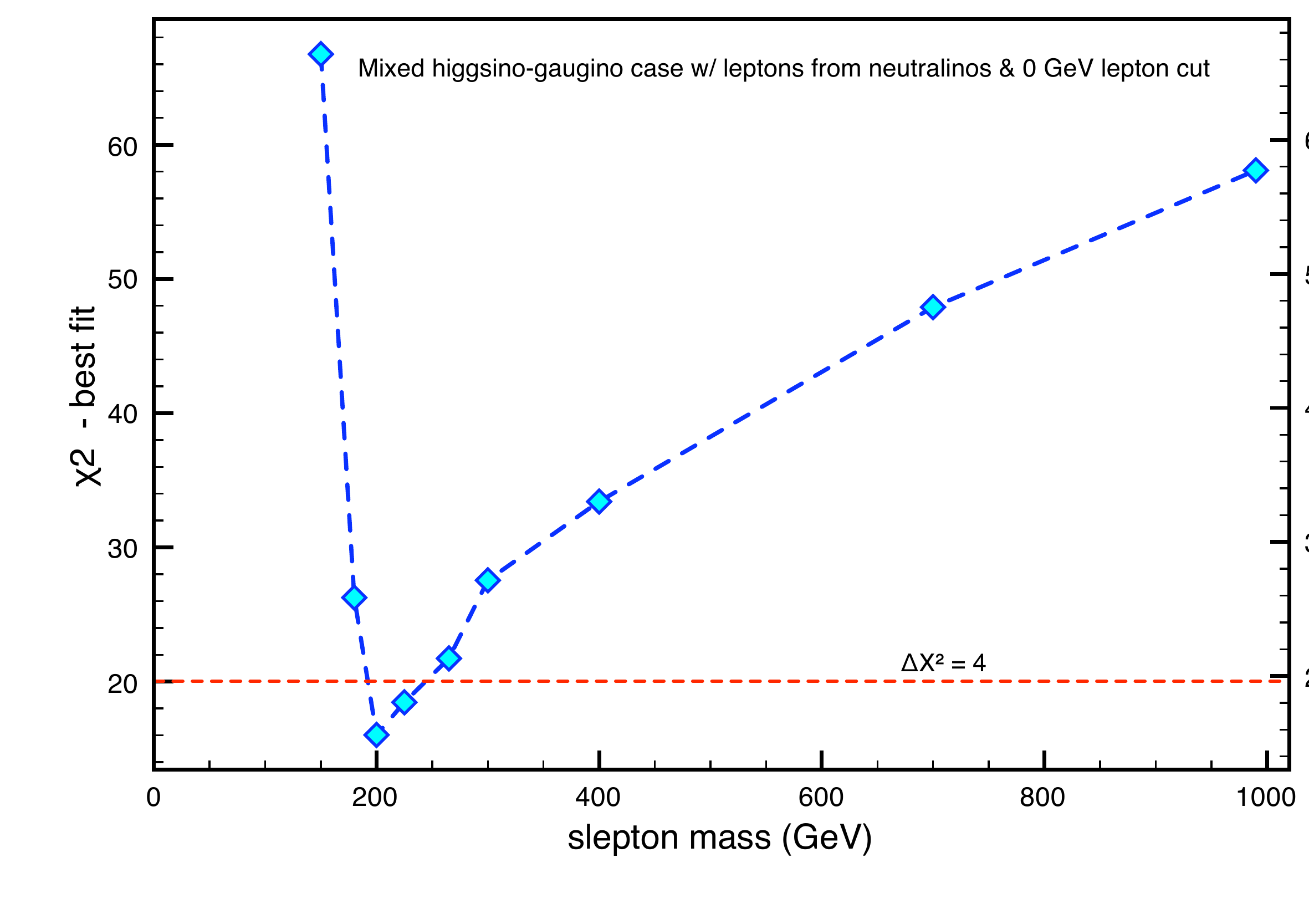}
 \end{center}
 \caption{\label{mixpure00}$\chi^{2}$ best fit to the $m_{ll}^{(00)}$ distribution for a mixed higgsino-gaugino-like case, corresponding to OS, SF leptons only from neutralinos w/(00) GeV lepton cut, plotted  vs slepton mass marginalized over other MSSM parameters.}
 \end{figure}

As mentioned before for the previous two case studies, we consider first the $m_{ll}^{(00)}$ distribution for this case. We perform a least $\chi^{2}$ over a grid of values for  $\langle m_{\tilde{l}} , M_{1}, M_{2}, \mu, tan{\beta} \rangle$, and three overall constants, each one multiplying the decay formula for the respective neutralino decay, up to its respective endpoint. This process is more complex, we have a total of 25 bins, and 8 fit parameters (the SUSY paameters above plus independent normalizations for each of the three decays that can contribute to the dilepton spectrum),  for a net of 17 degrees of freedom. The results from this analysis  are summarized in Table~\ref{pure00mix} for the  $m_{ll}^{(00)}$ case.
\begin{table}[htdp]
\begin{center}
\begin{tabular}{|c||r| |r| |r| |r| |r|}
\hline \hline
$\chi^{2}$&m$_{\tilde{l}}$ GeV&$M_{1}$&$M_{2}$&$\mu$&tan${\beta}$\\
\hline
66.75&150&-44&365&91&17\\
26.26&180&-69&430&116&15\\
16.03&200&-82&490&126&27\\
18.46&225&-106&520&151&19\\
21.73&265&-147&550&194&13\\
27.55&300&-169&640&212&31\\
33.42&400&-264&740&307&31\\
42.12&550&-409&870&452&29\\
47.90&700&-565&1010&610&20\\
58.09&1TeV&-850&1300&895&20\\
\hline \hline
\end{tabular}
\end{center}
\caption{$\chi^{2}$ results for the mixed higgsino-gaugino-like case of OS, SF  leptons directly from neutralinos w/${00}$ GeV  lepton cuts corresponding to the fit to the $m_{ll}^{(00)}$ distribution.}
\label{pure00mix}
\end{table}%
We have also plotted the $\chi^2$ results for the $m_{ll}^{(00)}$  case  in  Fig~\ref{mixpure00}  where we have plotted $\chi^{2}$  vs  m$_{\tilde{l}}$ marginalized over the other MSSM parameters.\\
The plot for the $m_{ll}^{(00)}$ case exhibits a clear minimum with a  $\chi^{2}$ fit of $\approx$~1 per degree of freedom (25 bins ; 8 fit variables) for a m$_{\tilde{l}}$ $\approx$ 200 GeV. We see the  $\chi^{2}$ clearly rising on both sides of the minimum value. The ÔÔbest fitÕÕ values are,
\[
\langle m_{\tilde{l}} , M_{1}, M_{2}, \mu, tan{\beta} \rangle=\langle 200, -82, 490, 126, 27\rangle
\]
to be compared with the input values in eq.~(\ref{mixparams}) above. The reason that we are able to obtain an unambiguous fit for this case is that the ambiguity that we had for the single endpoint case is removed because of the contribution of the extra neutralinos. \\
\subsection{Quasi-Real Case}\label{sec: qrc}
Up to now, our analysis has been idealized in that we have pretended that experiments can identify arbitrarily soft leptons and further identify the parent of the lepton (since we have retained leptons from the neutralinos only). The latter issue can be addressed by considering the flavour subtracted spectrum as we have already discussed. Before proceeding with results, we first discuss how we handle the issue of lepton cuts.  
\subsubsection{R-function}
\begin{figure}[htbp]
\begin{center}
\includegraphics[width=10cm]{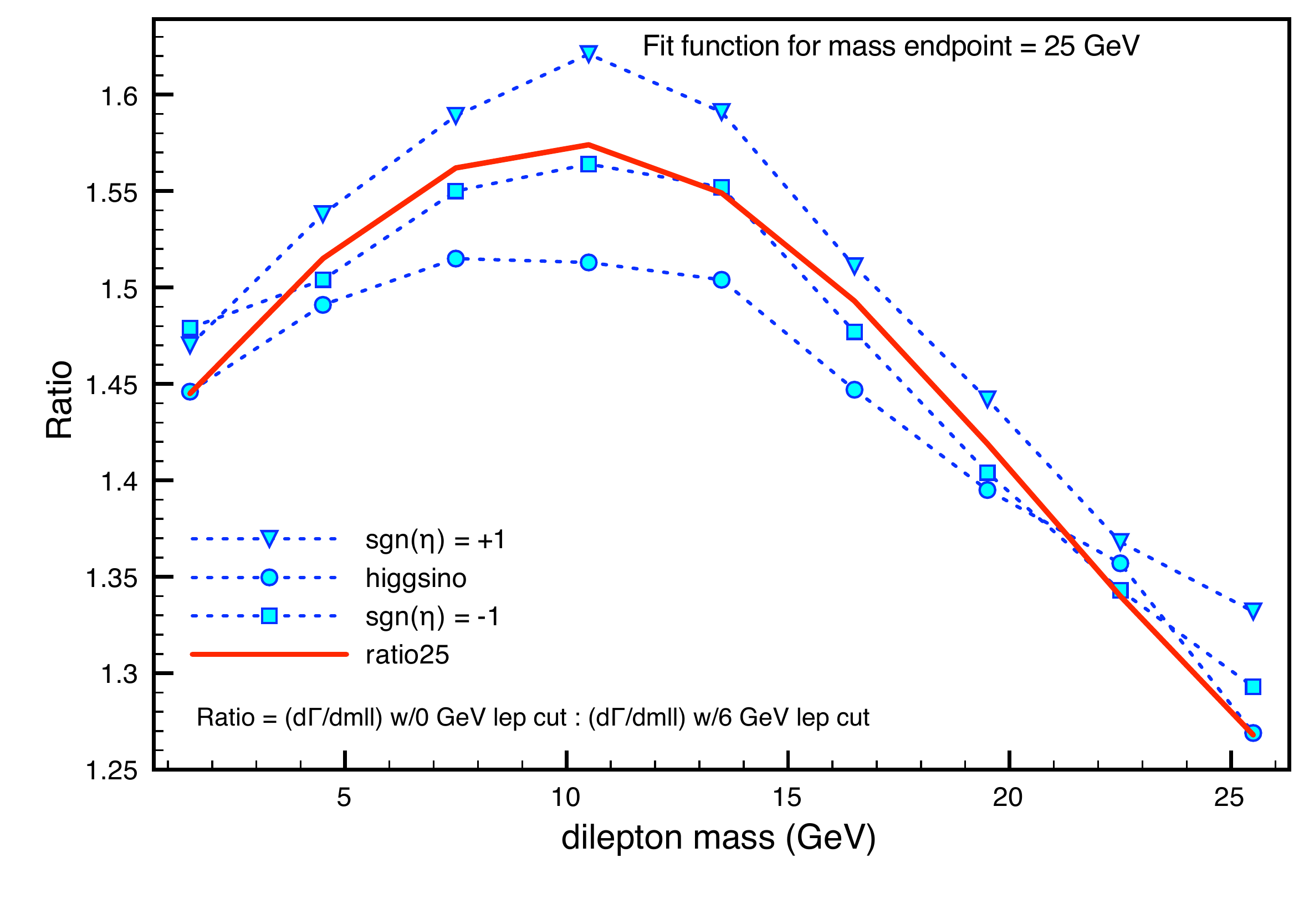}
 \end{center}
 \caption{\label{vec25}Comparison of function $R_{25}$ (solid red curve) to corresponding data sets (dotted blue curves). For MSSM input parameters given by $\langle M_1, M_2, \mu, tan(\beta), m_{\tl}, m_{\tg}, m_{\tq} \rangle$ with all masses in GeV units, the case with $sgn(\eta)=+1$ corresponds to $\langle 101, 126, -711, 10, 175, 441, 441  \rangle$. The case with $sgn(\eta)=-1$ has $\langle 101, -130, -711, 10, 175, 441, 441  \rangle$ and the higgsino-like case has $\langle -155, 170, 167, 10, 170, 450, 400  \rangle$ }
 \end{figure}

\begin{figure}[htbp]
\begin{center}
\includegraphics[width=10cm]{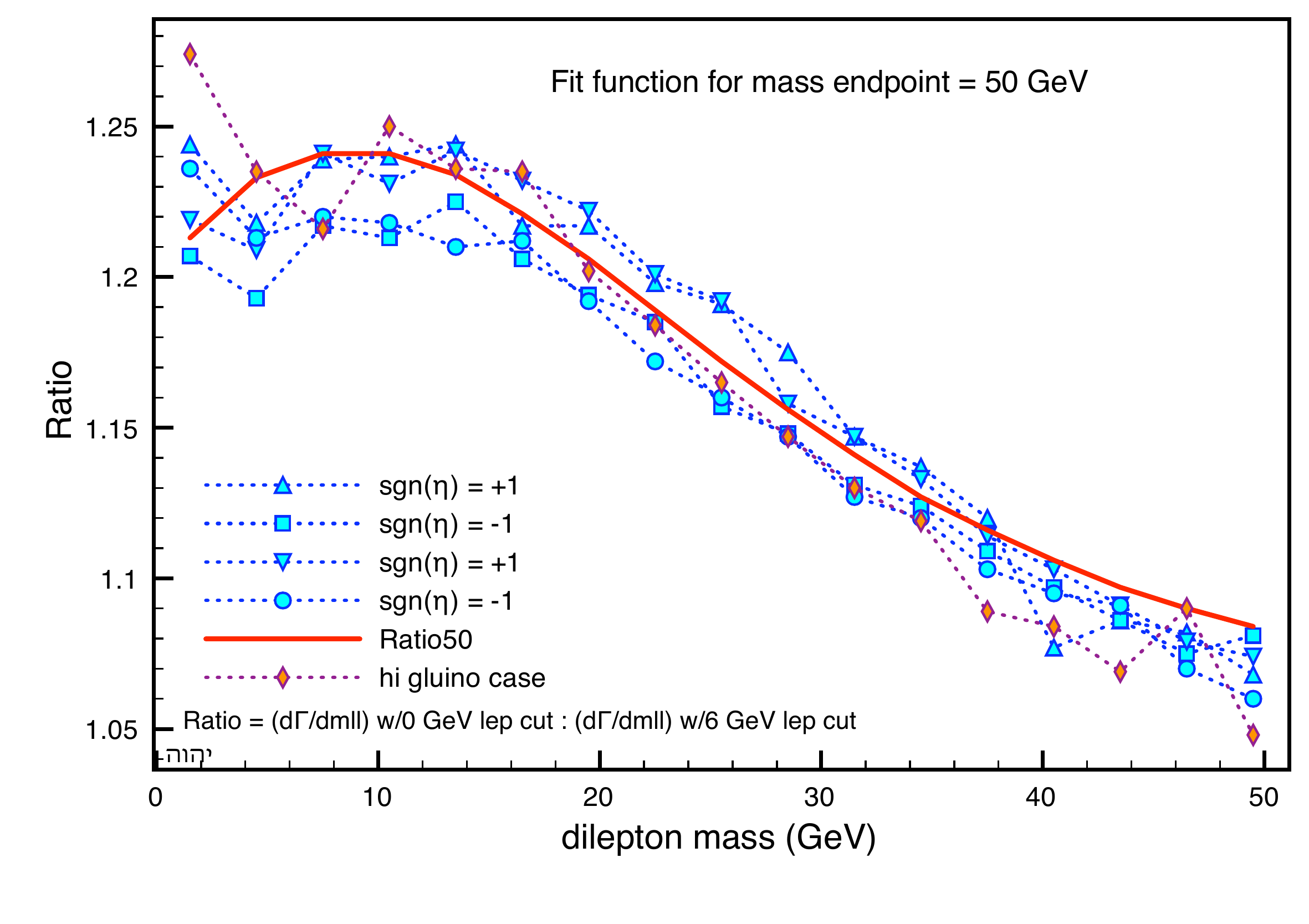}
 \end{center}
 \caption{\label{vec50}Comparison of function $R_{50}$ (solid red curve) to corresponding data sets (dotted blue curves) and the high gluino mass data set (broken red curve w/diamonds). For MSSM input parameters given by $\langle M_1, M_2, \mu, tan(\beta), m_{\tl}, m_{\tg}, m_{\tq} \rangle$ with all masses in GeV units, the first four cases correspond to MSSM vals $\langle 77, 127, -911, 10, 211, 441, 441  \rangle$,   $\langle 77, -130, -911, 10, 211, 441, 441  \rangle$, to $\langle 91, 141, -911, 10, 175, 441, 441  \rangle$, and $\langle 91, -144, -911, 10, 175, 441, 441  \rangle$, while the high gluino case corresponds to MSSM parameters $\langle -71, 121, -911, 10, 315, 900, 800  \rangle$.}
 \end{figure}

\begin{figure}[htbp]
\begin{center}
\includegraphics[width=10cm]{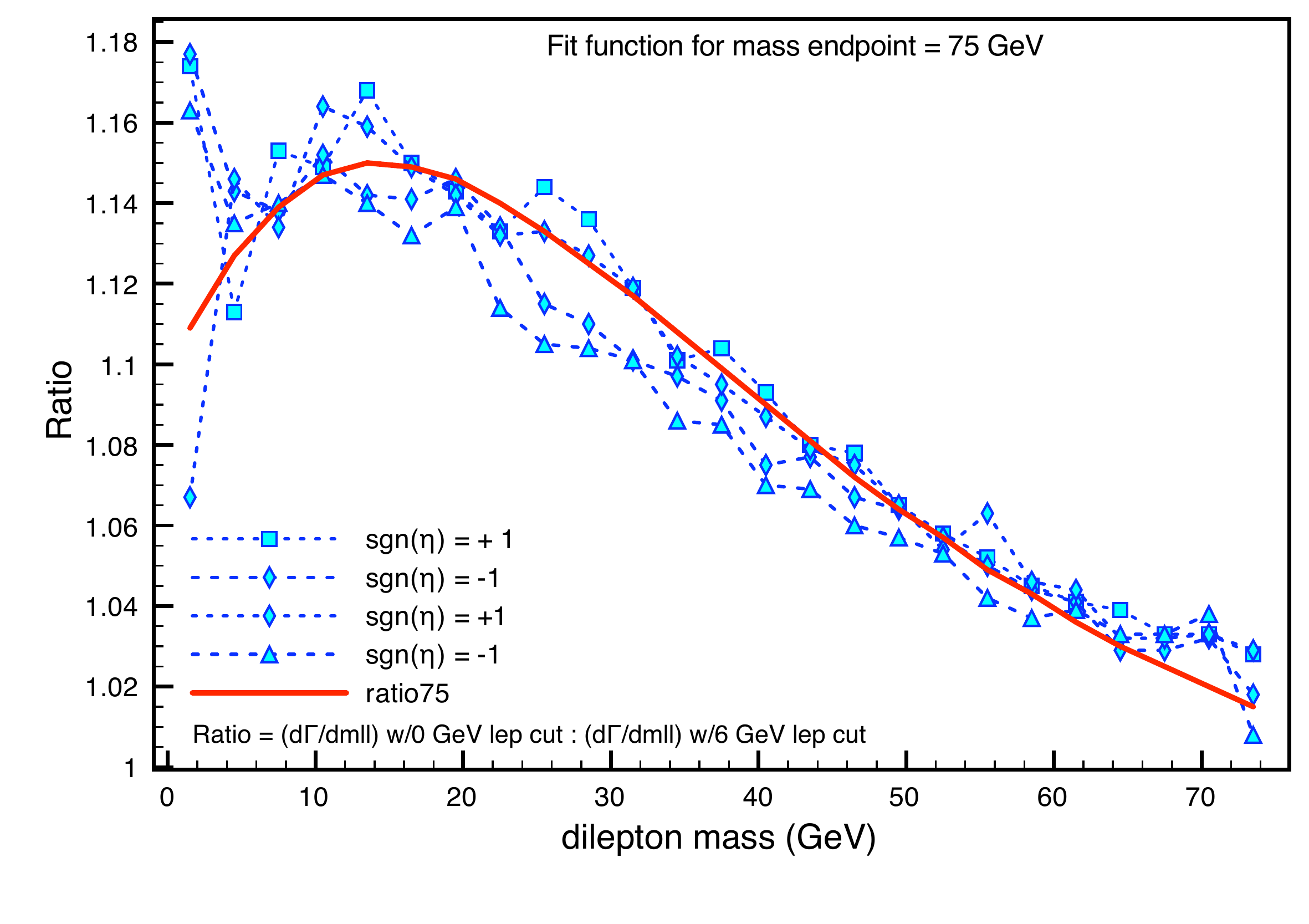}
 \end{center}
 \caption{\label{vec75}Comparison of function $R_{75}$ (solid red curve) to corresponding data sets (dotted blue curves).  For MSSM input parameters given by $\langle M_1, M_2, \mu, tan(\beta), m_{\tl}, m_{\tg}, m_{\tq} \rangle$ with all masses in GeV units, the four cases correspond to MSSM values of $\langle 69, 144, -944, 10, 315, 414, 441  \rangle$,   $\langle 69, -147, -944, 10, 315, 414, 441  \rangle$, to  $\langle 77, 152, -911, 10, 211, 441, 441  \rangle$, and $\langle 77, -155, -911, 10, 211, 441, 441  \rangle$.}
 \end{figure}

Our theoretical decay formula has no lepton cuts in it, while real data events will necessarily have cuts in the lepton $p_{T}$ because very soft electrons and muons  ($< 6$ GeV in our analysis) are not readily identifiable. We expect to cause the most distortion to the $m_{ll}$ distribution shape in the low $m_{ll}$ region. Also, we expect that most of the effect of a not so hard lepton cut will be from kinematics rather than from details of the matrix element. Since we want to fit our theoretical results to the actual distribution we need to find a way to make an appropiate correction to our formula that will simulate the effect of the lepton cuts. Ideally, when calculating the formula, changing our integration limits for $E_{T}(l)$ to include these cuts would work, but the integrals became too complex to be able to write an analytical expression similar to the one we obtained for our theory function without lepton cuts, eq.~(\ref{eqn: fctn}) or \textbf{$(d\Gamma/dm_{ll})_{00}$}, where the subscripts indicate a $0$ GeV cut on each of the two leptons.\\
In order to incorporate the effect of the lepton $p_{T}$ cut into our analysis, we need to include the effect of this cut in our theoretical fit function in eq.~(\ref{eqn: fctn}). Toward this end, we generate Monte-Carlo samples both with and without the lepton cut of $6$ GeV, and obtain the corresponding dilepton mass distributions, \textbf{$(d\Gamma/dm_{ll})_{66}$} and \textbf{$(d\Gamma/dm_{ll})_{00}$}. We define the ratio
\be
R= \frac{\textbf{$(d\Gamma/dm_{ll})_{00}$}}{\textbf{$(d\Gamma/dm_{ll})_{66}$}}.
\label{eqn: rfctn}
\ee
$R$ will be larger than unity. We expect $R$ is largest for small values of $m_{ll}$ and approaches close to unity if $m_{ll}$ is very large, for which the cut on the lepton has the smallest effect. Instead of writing R as a function of $m_{ll}$, we write it as a function of the scaled variable
\[
x\equiv \frac{m_{ll}}{m_{\tz_2}-m_{\tz_1}}
\]
so that the argument of $R$ runs between $0$ and $1$. Of course the form of $R$ will depend on the endpoint $m_{\tz_i}-m_{\tz_f}$. If, as we expect, the effect of the lepton cut is largely kinematic R will be roughly independent of other parameters as long as $m_{\tz_i}-m_{\tz_f}$ is held fixed.\\
To extract the $R$ functions for our mixed gaugino-higgsino like case with endpoints of $25$ GeV, $50$ GeV and $75$ GeV, we generate four sample points having only one endpoint at the specified value. We can easily generate these by allowing $\mu$ to take on relatively high values, so that M$_{2}$ - M$_{1}$ yields the desired mass gap. For the $25$ GeV endpoint we generated a gaugino like point for both the cases with $M_{1}=\pm $ and also used the higgsino point with parameters given by eq.~(\ref{eqn: higgsino}). For the $50$ GeV case 
We show the $R$-values that we obtain from our simulation in Fig.~\ref{vec25}, Fig.~\ref{vec50} and Fig.~\ref{vec75} for endpoints of $25$ GeV, $50$ GeV and $75$ GeV respectively.
We see that the scatter for the different models is indeed small, confirming that R is largely determined by kinematics of the decay. The solid (red) line is our analytical fit for the ratios $R_{25}$, $R_{50}$ and $R_{75}$ which are parametrized as
\begin{eqnarray}
R_{25}(x)-1=\frac{0.575}{[1+(x-0.4)^{2}]^{2.35}}\\
R_{50}(x)-1=\frac{0.725*(x+0.20)}{[1+(x+0.20)^{2}]^{4}}\\
R_{75}(x)-1=\frac{1.15*(x+0.15)}{[1+(x+0.15)^{2}]^{4}}
\end{eqnarray}
We can then write our prediction for the spectrum of $m_{ll}$ as
\be
 (\frac{d\Gamma}{dm_{ll}})_{66} =\frac{a}{R_{25}}(\frac{d\Gamma}{dm_{l\bar{l}}})_{\tilde{Z_{3}} \rightarrow  \tilde{Z_{2}}}+\frac{b}{R_{50}}(\frac{d\Gamma}{dm_{l\bar{l}}})_{\tilde{Z_{2}} \rightarrow  \tilde{Z_{1}}}+\frac{c}{R_{75}}
(\frac{d\Gamma}{dm_{l\bar{l}}})_{\tilde{Z_{3}} \rightarrow  \tilde{Z_{1}}}
\label{verafctn}
\ee
where $a, b, c$ are parameters determining the normalization, and each of the $(\frac{d\Gamma}{dm_{l\bar{l}}})_{\tilde{Z_{i}} \rightarrow  \tilde{Z_{f}}}$ correspond to  \textbf{$(d\Gamma/dm_{ll})_{00}$} in eq.~(\ref{eqn: fctn}).
We are now ready to present our results of the fit to the mixed point.
\subsubsection{Results for the quasi-real case}
\begin{figure}[htdp]
\begin{center}
\includegraphics[width=10cm]{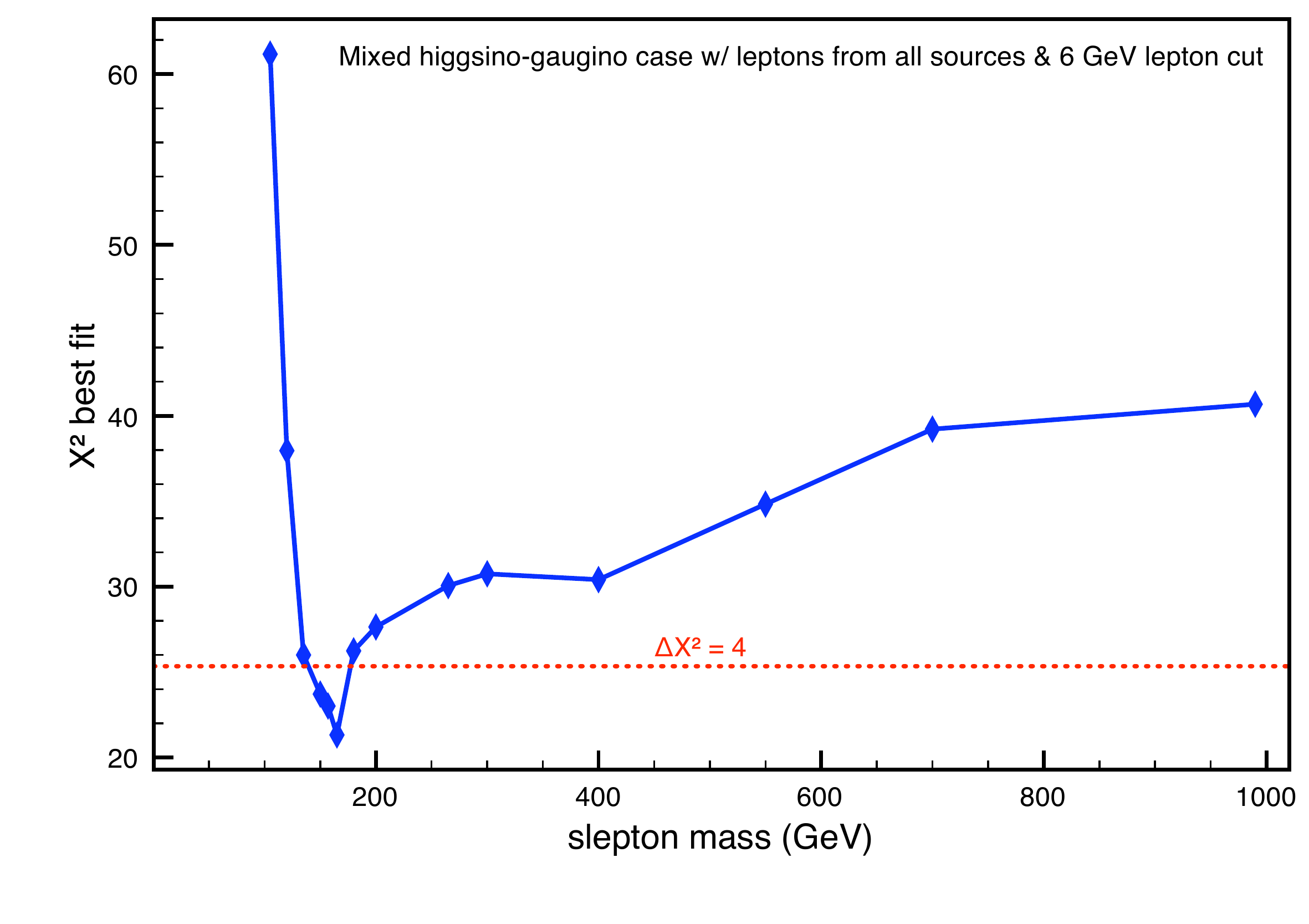}
 \end{center}
 \caption{\label{mixreal06}$\chi^{2}$ best fit to the $m_{ll}^{(66)}$ distribution for a mixed higgsino-gaugino-like case, corresponding leptons from all SUSY sources w/(06) GeV lepton cut after flavour subtraction and cuts discussed in text, plotted  vs slepton mass marginalized over other MSSM parameters. Results  correspond to the the quasi-real  $m_{ll}^{(66)}$ case discussed in Sec.~\ref{sec: qrc} of the text.}
 \end{figure}

Our idealized analysis of the mixed case of Sec.~\ref{sec: qpc}  suggests that it may be possible to make further progress even when realistic effects are incorporated into the analysis. We perform an 8-parameter fit to the flavour subtracted data in Fig.~\ref{histomixD}. The values of $\chi_{min}^2$ for each value of slepton mass are shown in Table~\ref{real06mix} where for each value of $m_{\tl}$ the fitted values of other MSSM parameters are shown.
\begin{table}[htdp]
\begin{center}
\begin{tabular}{|c||r| |r| |r| |r| |r|}
\hline \hline
$\chi^{2}$&m$_{\tilde{l}}$ GeV&$M_1$&$M_2$&$\mu$&tan${\beta}$\\
\hline
61.17&105&-02&210&61&09\\
37.96&120&-10&250&61&21\\
26.00&135&-30&350&75&32\\
23.71&150&-43&400&89&20\\
23.01&157&-46&420&92&20\\
21.32&165&-59&405&109&10\\
26.24&180&-65&500&110&23\\
27.64&200&-85&500&130&23\\
30.06&265&-140&600&185&20\\
30.75&300&-180&650&225&20\\
30.41&400&-270&750&315&20\\
34.83&550&-405&900&448&29\\
39.22&700&-271&600&342&02\\
40.68&1TeV&-460&800&530&02\\
\hline \hline
\end{tabular}
\end{center}
\caption{$\chi^{2}$ results for the case of leptons  from all SUSY sources w/${06}$ GeV  lepton cuts corresponding to the fit to the $m_{ll}^{(66)}$ distribution. This case corresponds to the quasi-real case in the text, Sec.~\ref{sec: qrc} where the cuts discussed in the text have been applied, together with flavour subtraction.}
\label{real06mix}
\end{table}%
The corresponding $\chi_{min}^2$ values\footnote{In calculating the $\chi_{i}^2$ corresponding to the $i^{th}$ bin, we divide by $n_{i}$, which corresponds to total number of dileptons in bin of OS, and both same and mixed flavour.} are shown in Fig.~\ref{mixreal06} where we have marginalized over other MSSM parameters.
We see that the slepton mass is fitted to be in the interval 150 GeV $\le m_{\tl} \le$ 175 GeV ($2\sigma$). Moreover, unlike the single mass edge cases in Sec.~\ref{sec: gauhiggs} the fit is unambiguous and allows us to extract the $\eta_{ij}$ values as well as the relative $Z$ and slepton mediated contributions in each case.
\subsection{Fitting Neutralino Mass Matrix Parameters Individually}
\begin{figure}[htbp]
\begin{center}
\includegraphics[width=5cm]{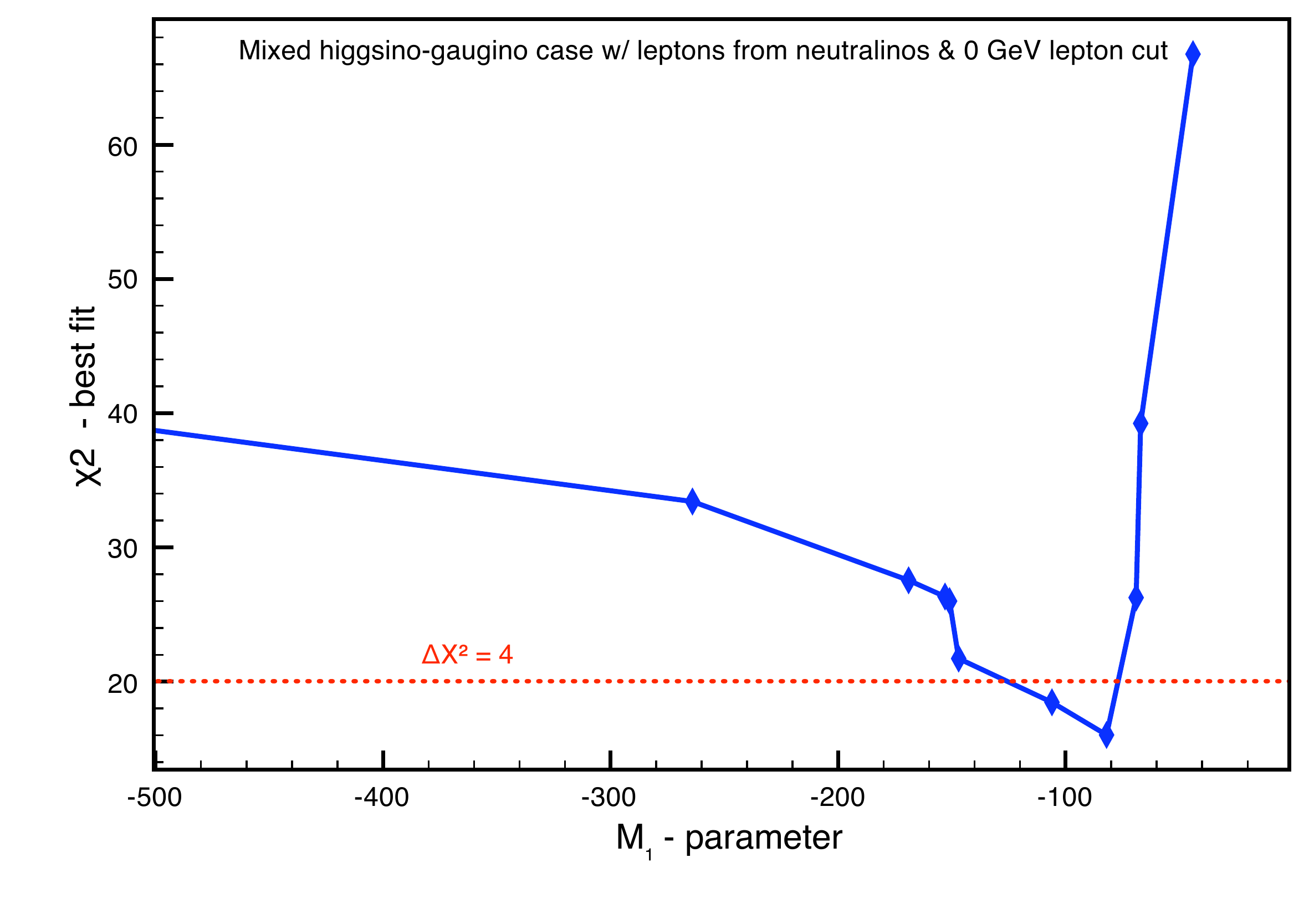}
\includegraphics[width=5cm]{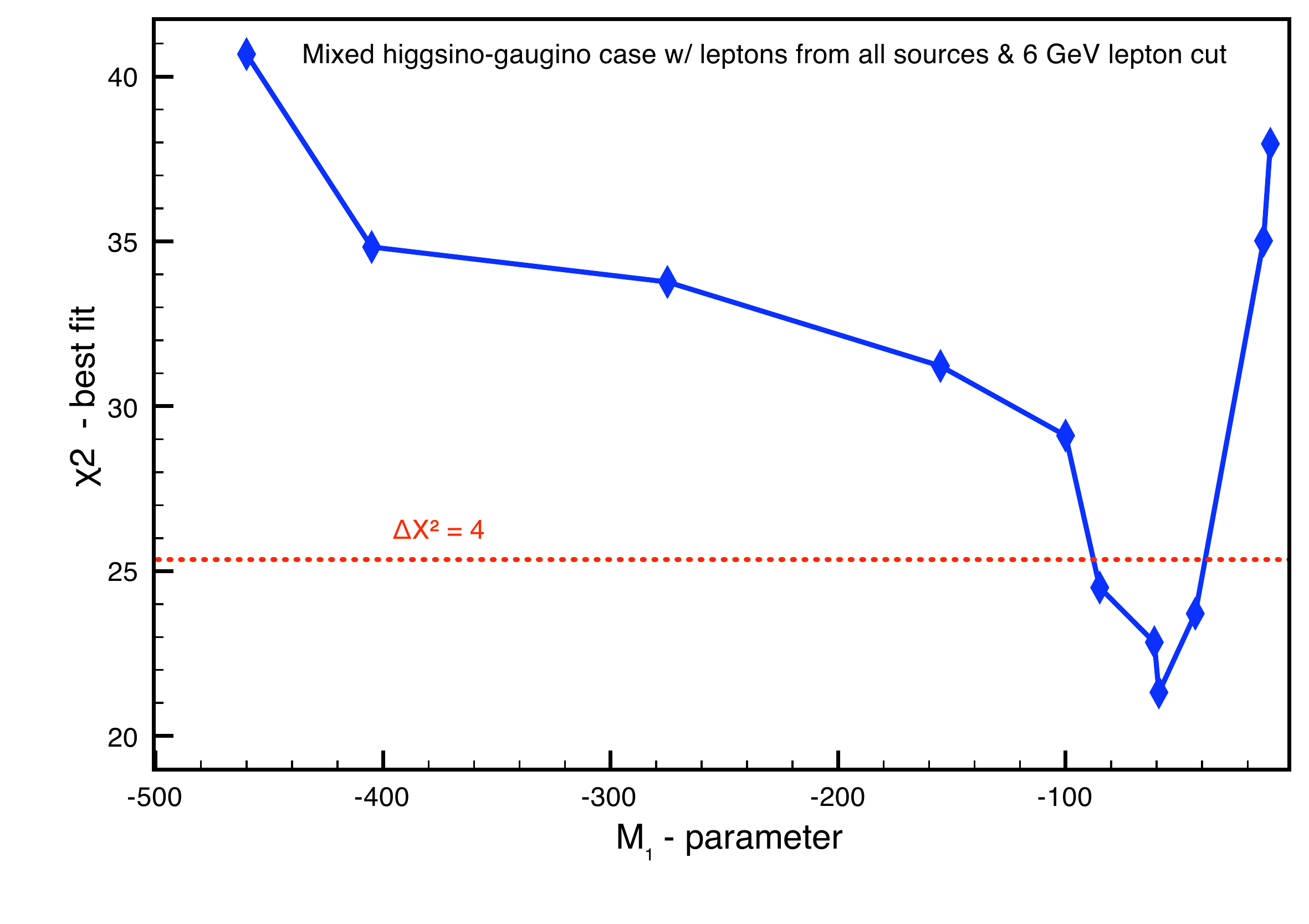}
 \end{center}
 \caption{\label{M100}$\chi^2$ best fit vs MSSM parameter $M_{1}$, marginalized over other MSSM parameters for both the  $m_{ll}^{(00)}$ case on the left,  and the $m_{ll}^{(66)}$ case on the right.}
 \end{figure}

\begin{figure}[htbp]
\begin{center}
\includegraphics[width=5cm]{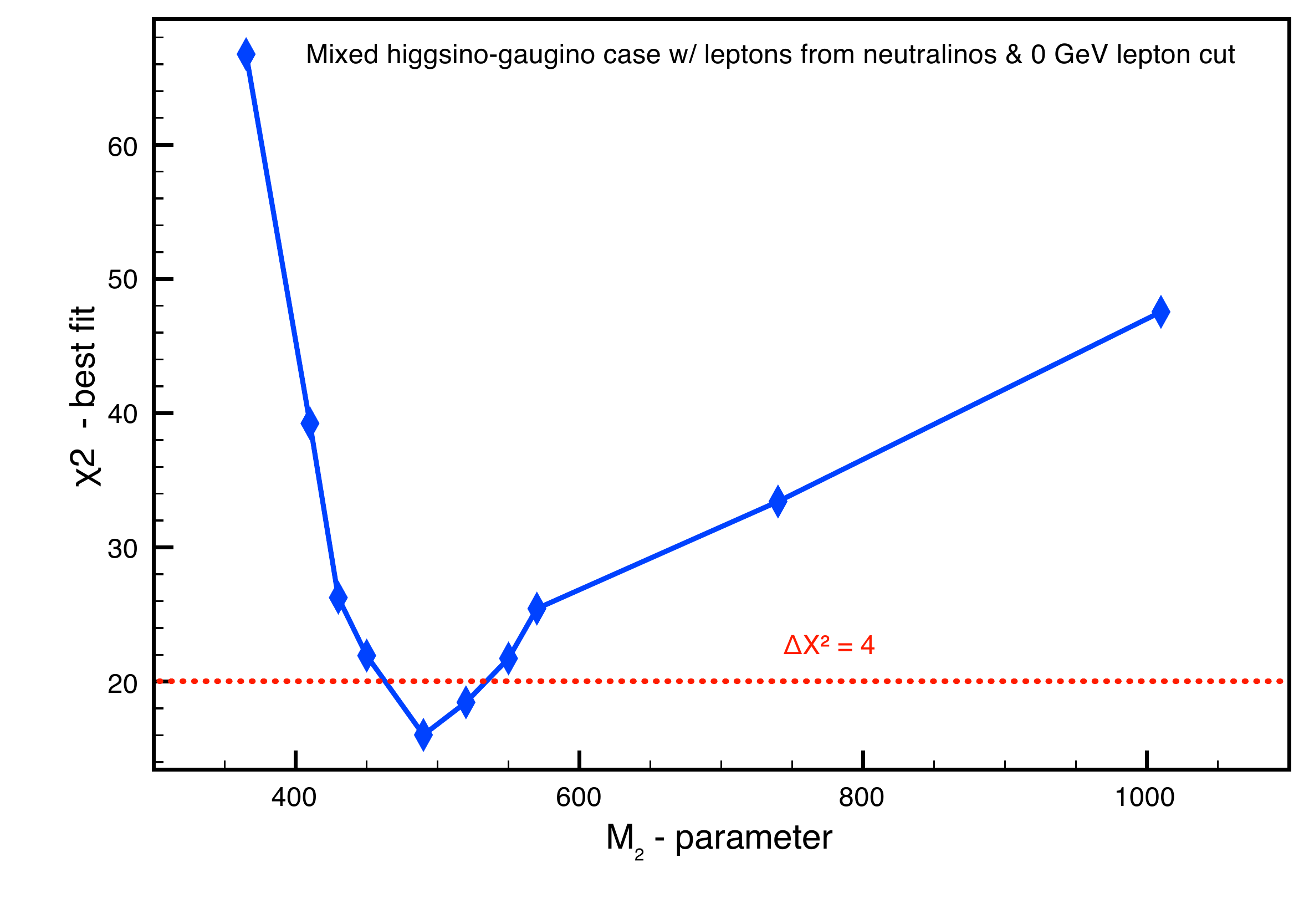}
\includegraphics[width=5cm]{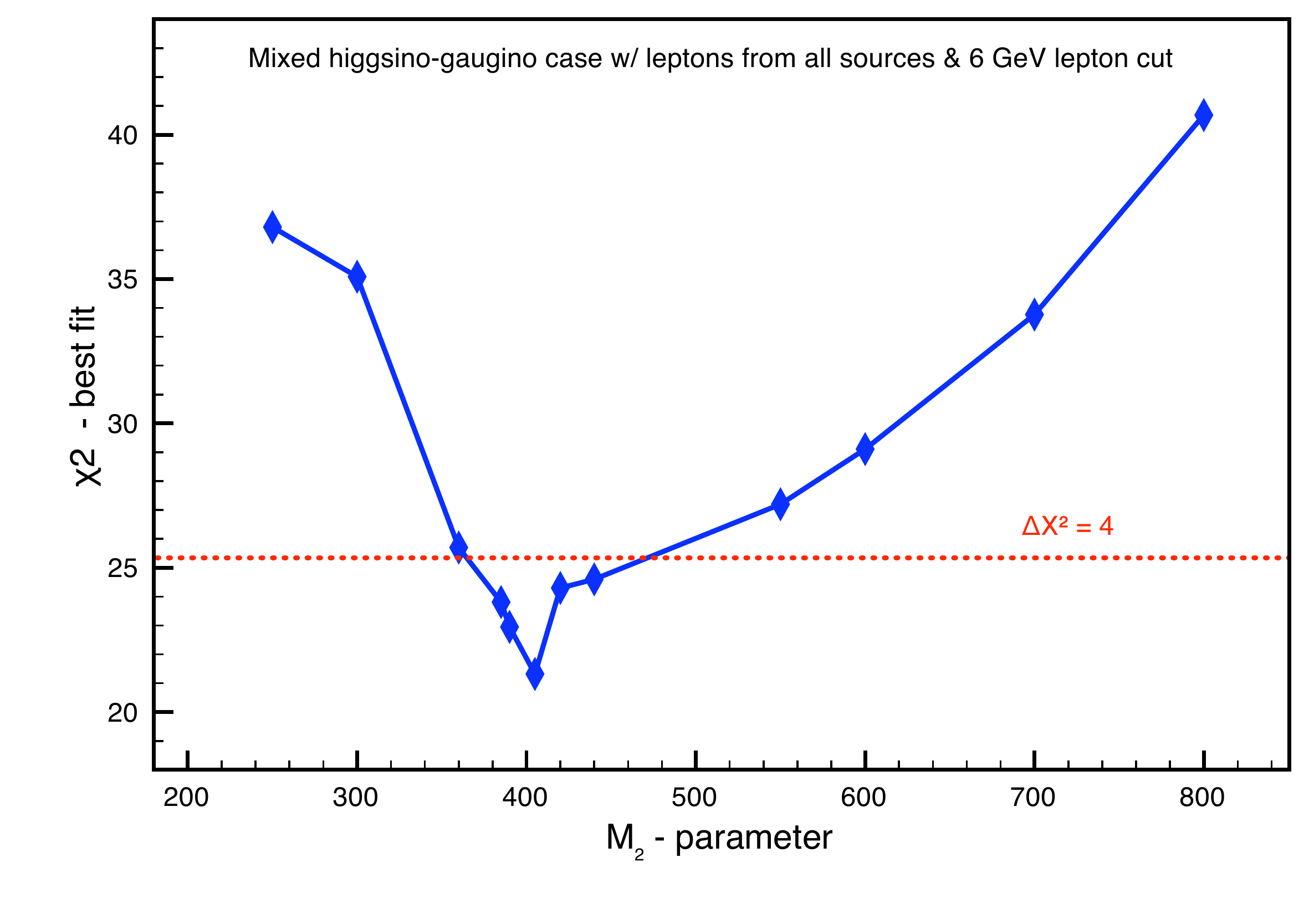}
 \end{center}
 \caption{\label{M200}$\chi^2$ best fit vs MSSM parameter $M_{2}$, marginalized over other MSSM paraameters for both the  $m_{ll}^{(00)}$ case on the left, and the $m_{ll}^{(66)}$ case on the right.}
 \end{figure}

\begin{figure}[htbp]
\begin{center}
\includegraphics[width=5cm]{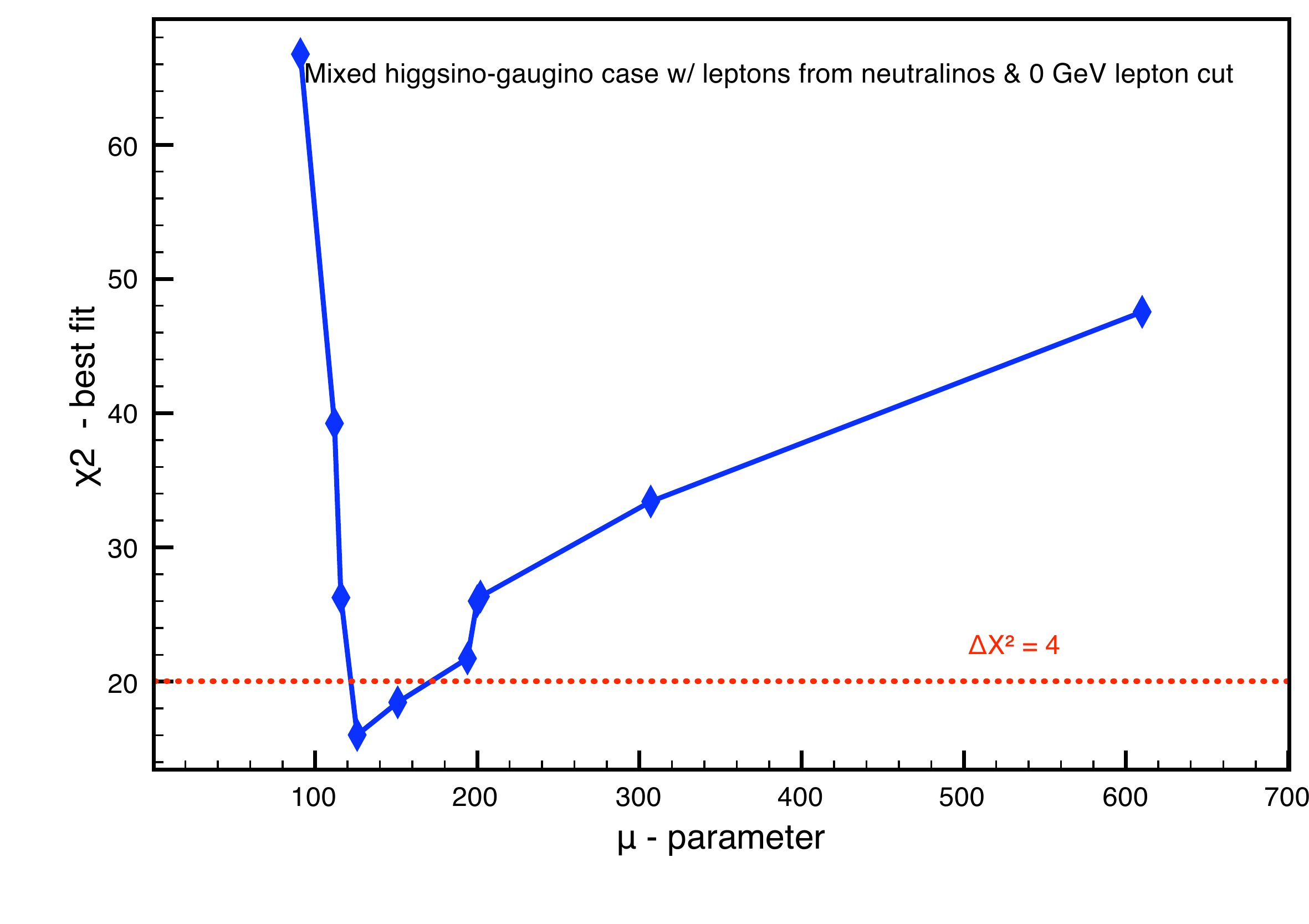}
\includegraphics[width=5cm]{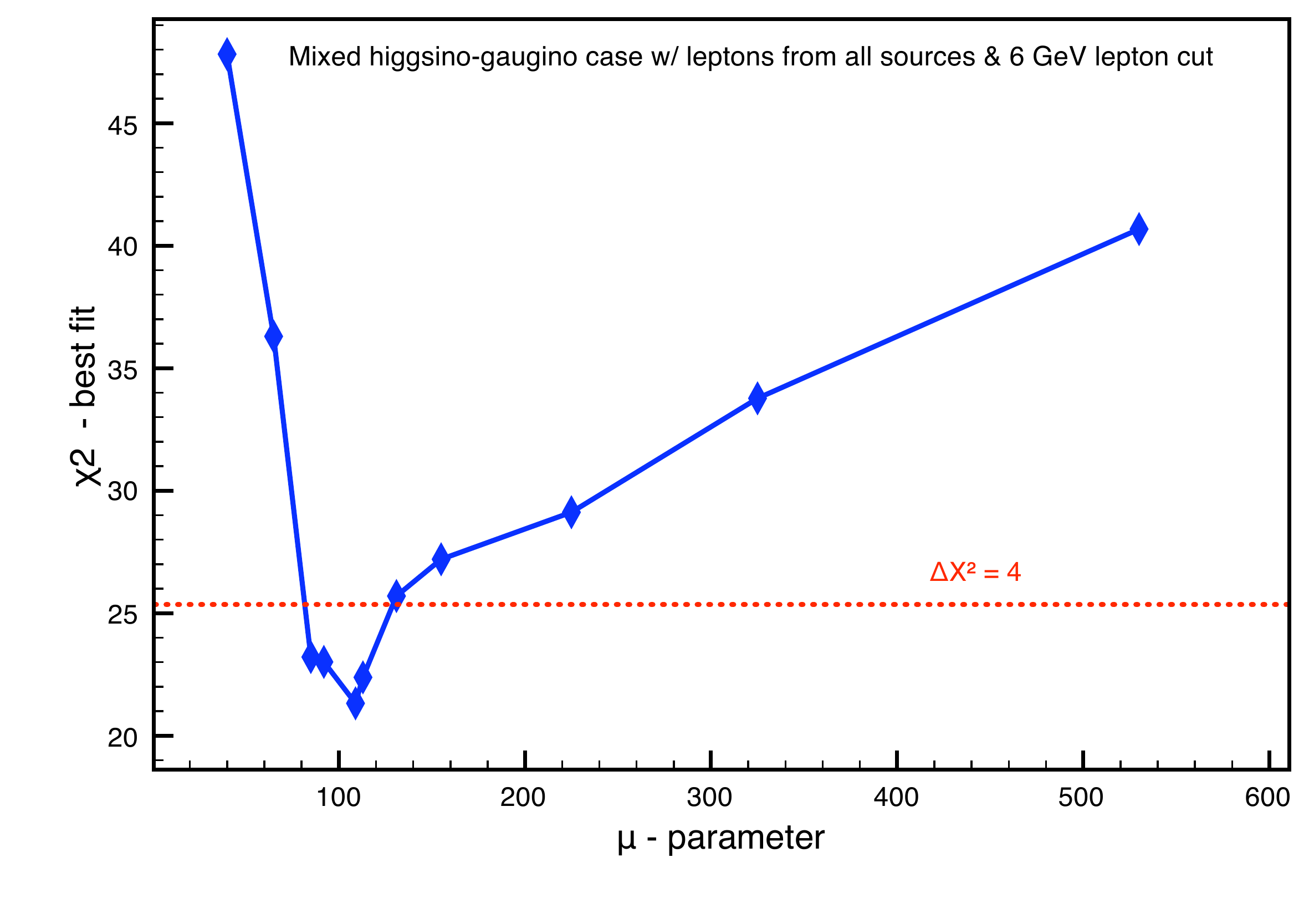}
 \end{center}
 \caption{\label{mu00}$\chi^2$ best fit vs MSSM parameter $\mu$, marginalized over other MSSM paraameters for both the  $m_{ll}^{(00)}$ case on the left, and the $m_{ll}^{(66)}$ case on the right.}
 \end{figure}

\begin{figure}[htbp]
\begin{center}
\includegraphics[width=5cm]{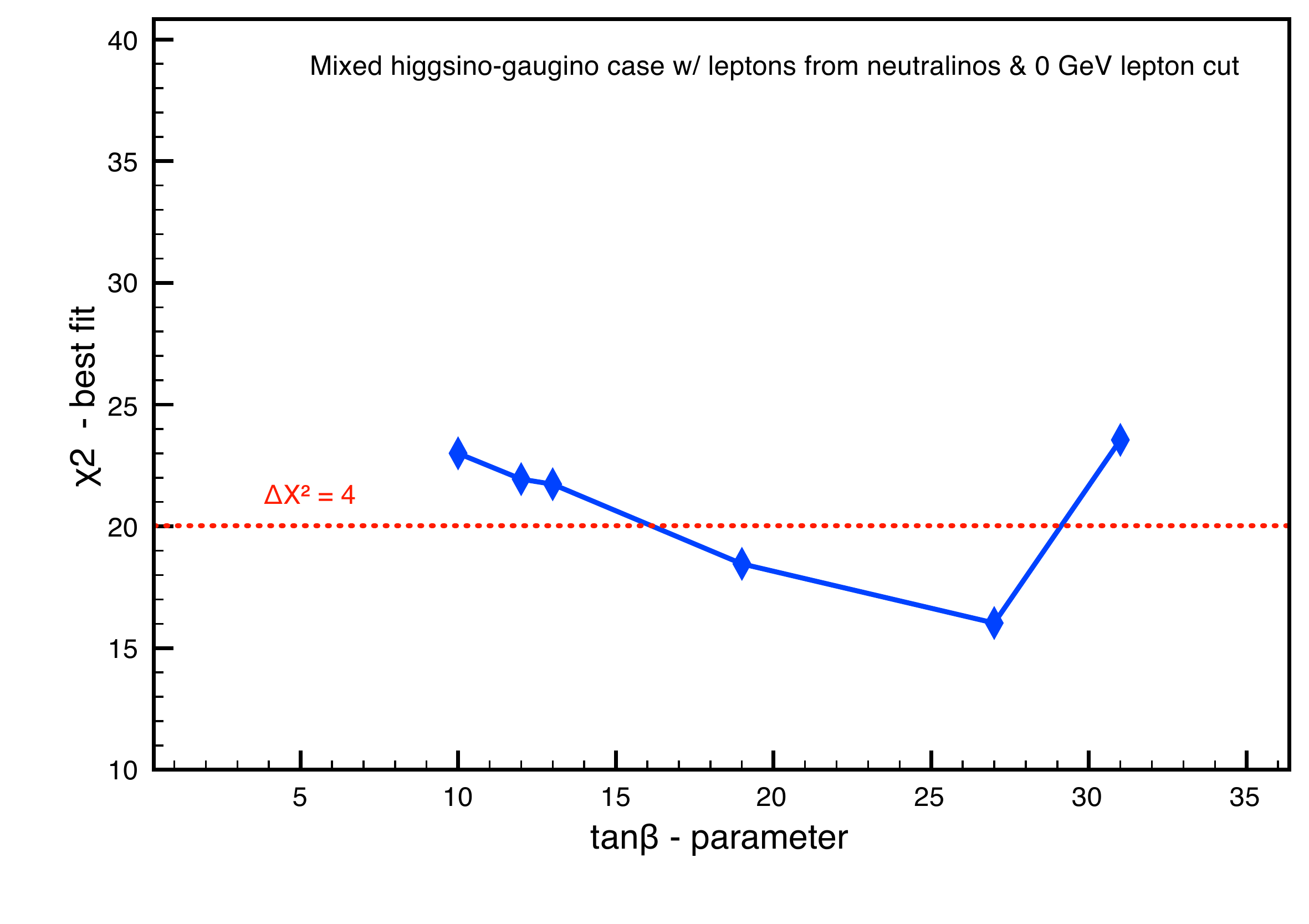}
\includegraphics[width=5cm]{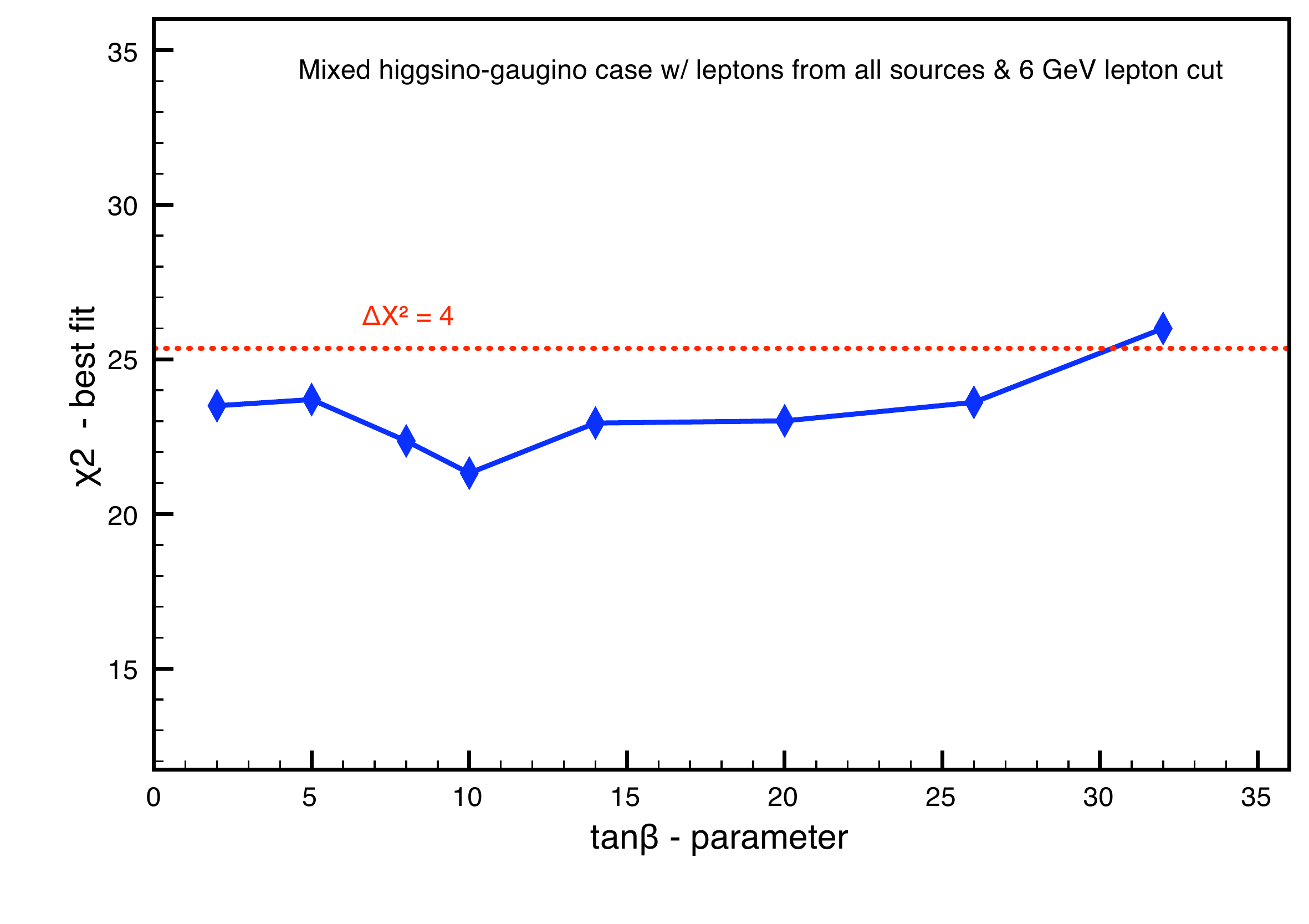}
 \end{center}
 \caption{\label{tanb00}$\chi^2$ best fit vs MSSM parameter $tan\beta$, marginalized over other MSSM paraameters for both the  $m_{ll}^{(00)}$ case on the left, and the $m_{ll}^{(66)}$ case on the right.}
 \end{figure}

We look at the $\chi^{2}$ sensitivity for each of the other MSSM input parameters. We present our results  for our two  previous cases, the $m_{ll}^{(00)}$ and the $m_{ll}^{(66)}$, as our most ideal and most real scenarios respectively. These results are contained in the accompanying figures. Fig.~\ref{M100}, Fig.~\ref{M200}, Fig.~\ref{mu00} and Fig.~\ref{tanb00} for $M_{1}$, $M_{2}$, $\mu$ and $tan(\beta)$ respectively. 
The results are as expected for $M_{1}$ and $\mu$, and surprisingly sensitive for $M_{2}$ but with a larger range of values. Also, as expected $tan\beta$ exhibits a flat behaviour with respect to $\chi^{2}$. for the quasi-realistic fits to the data, we find the fitted values at the ($2\sigma$) level are
\bi
\item $m_{\tl}=  139$ GeV to $180$ GeV
\item $M_{1}=    -95$  to $-58$ GeV             
\item $M_{2}=  360$ to $475$ GeV               
\item $\mu =  87$  to $127$ GeV
\ei
while $tan(\beta)$ remains undeterred, to be compared to the input values in eq.~(\ref{mixparams}). \\
Summarizing, we set out to obtain as much information as possible from the $m_{ll}$ distribution, and were surprised to learn of its relative insensitivity to $m_{\tz_i}+m_{\tz_f}$ for the case with a single mass edge. Another surprise was the ability of the Z-boson exchange to masquerade as a slepton exchange. This degeneracy may be resoluble from other data. On the other hand, the results for the favorable mixed gaugino-higgsino case with the double mass edge were positive, supporting our original goals and methodology, including our technique for simulating lepton $p_{T}$ cuts.\\


\chapter{Conclusions and Future Outlook}\label{chap:epilog}
In Chapt.~\ref{chap:btag} we investigated models with an inverted squark mass hierarchy to study how much $b$-tagging or $t$-tagging would increase the reach at the LHC. We also studied ways in which $b$-tagging would allow us to extract the signals for third generation squarks from both the SM background and the SUSY contamination from all other sources, which became an additional background to be eliminated. \\
Now that the LHC has analyzed data corresponding to an integrated luminosity of just in excess of 1.1 fb$^{-1}$, we find that the exclusion regions have pushed the gluino and squark masses above $1$ TeV. This makes models with an IMH appealing, because with high gluino and squark masses, it is still possible to obtain light third generation squarks between $200-400$ GeV. In addition, the value of $S$ which we use to quantify the degree of IMH inversion, would be higher than the values we used for our analysis. So, for a fixed light stop ($\tst_{1}$) mass, the value of $N_{SUSY}$ after cuts would be smaller for the higher mass gluinos and squarks, increasing the value of our observability criteria. This makes the study of this particular area of great importance given the results obtained by the LHC up to date. \\
In Chapt.~\ref{chap:zslep} our case studies are for points which are now excluded by the LHC. As mentioned before, it would be worth investigating whether the fitting techniques we used in our study may continue to be useful for higher values of gluino and squark mass as long as the neutralino mass endpoints remain at values which continue to suppress two=body decays, allowing three-body decays, via virtual $Z$ or slepton exchanges, to dominate. This will occur in models with a compressed spectrum, that have recently received some attention. We aim to pursue our study along these lines. \\



\include{papel}

\end{document}